\def\draftversion{false}

\RequirePackage{ifthen}
\ifthenelse{\equal{\draftversion}{true}}{
  \documentclass[prb,galley,showpacs,preprintnumbers,citeautoscript,
      amsmath,amssymb,longbibliography]{revtex4-1}
}{
  \documentclass[citeautoscript,floatfix,aps,prb,twocolumn,
      superscriptaddress,longbibliography]{revtex4-1}
}

\usepackage{amsmath,latexsym,psfrag}
\usepackage{latexsym}
\usepackage{psfrag}
\usepackage{graphicx}
\usepackage{epstopdf}
\usepackage{natbib}
\usepackage{array}
\usepackage{dcolumn}
\usepackage{bm}

\usepackage{soul}  

\usepackage[usenames,dvipsnames]{color}

\soulregister\cite7

\ifthenelse{\equal{\draftversion}{true}}{
  \usepackage{showlabels}
  \marginparwidth 2.7in
  \marginparsep 0.5in
  \newcounter{comm} 
  \def\commnext{\stepcounter{comm}}
  \def\commtext{{\bf\color{blue}[\arabic{comm}]}}
  \def\commmar{{\bf\color{blue}[\arabic{comm}]}}
  \def\msm#1{\commnext\marginpar{\small MS\commmar: #1}\commtext}

  \def\mlab#1{\marginpar{\small\bf #1}}
  
}{
  \def\dvm#1{}
  \def\cdm#1{}
  \def\msm#1{}
  \def\asm#1{}
  \def\miq#1{}
  \def\mlab#1{}
  
}

\newcommand{\vv}{V_{\rm H}}

\begin{document}

\title{ Novel mechanisms to enhance the capacitance 
        beyond the classical limits
        in capacitors with free-electron-like electrodes }

\author{ Javier Junquera }
\affiliation{ Departamento de Ciencias de la Tierra y
              F\'{\i}sica de la Materia Condensada, Universidad de Cantabria,
              Cantabria Campus Internacional,
              Avenida de los Castros s/n, 39005 Santander, Spain}
\author{ Pablo Garc\'{\i}a-Fern\'andez}
\affiliation{ Departamento de Ciencias de la Tierra y
              F\'{\i}sica de la Materia Condensada, Universidad de Cantabria,
              Cantabria Campus Internacional,
              Avenida de los Castros s/n, 39005 Santander, Spain}
\author{ Massimiliano Stengel}
\affiliation{ Institut de Ci\`encia de Materials de Barcelona (ICMAB-CSIC),
              08193 Bellaterra, Spain}
\affiliation{ICREA -- Instituci\'{o} Catalana de Recerca i Estudis Avan\c{c}ats,
             Pg. Lluis Companys, 23, 08010 Barcelona, Spain}              

\date{\today}

\begin{abstract}
 The so-called negative electron compressibility refers to the lowering 
 of the chemical potential of a metallic system when the carrier 
 density increases.
 This effect has often been invoked in the past to explain the 
 enhancement of the capacitance beyond the classical limits in 
 capacitors with two-dimensional electron gases as electrodes. 
 Based on experiments on strongly confined semiconductor quantum wells (QWs), 
 it has been traditionally ascribed to the electron exchange energy
 as the main driving force. 
 Recent research, however, has revealed that analogous effects can
 occur in other classes of materials systems, such as polar oxide interfaces,
 whose characteristics drastically depart from 
 those of the previously considered cases.
 To rationalize these new results, it is necessary to revisit
 the established theory of confined electron gases, and test whether 
 its conclusions are valid beyond the specifics of semiconductor-based QWs.
 Here we find, based on first-principles calculations of jellium slabs, that 
 one must indeed be very careful when extrapolating existing results
 to other realistic physical systems.
 In particular,
 we identify a number of additional, previously overlooked mechanisms 
 (e.g., related to the displacement of the electronic cloud and to the
 multiband structure of the delocalized gas), that enter into play
 and become new sources of negative capacitance in the weak-confinement regime. 
 Our detailed analysis of these emerging contributions, supported by analytic 
 models and multiple test cases, will provide a useful guidance in the ongoing
 quest for nanometric capacitors with enhanced performance.
\end{abstract}

\pacs{73.20-r, 73.40.Rw, 73.30.+y }

\maketitle

 \section{Introduction}
 \label{sec:intro}
 
 The constant drive for faster, smaller, cheaper and 
 more powerful micro-electronic
 devices has led to Moore's law,~\cite{Moore-65} the defining
 paradigm of the global semiconductor industry.
 Although its validity has firmly held since its formulation in the mid sixties,
 often disproving the recurring skepticisms that were raised over the years,
 the exponential~\footnote{
 According to Moore's law, the number of transistors
 that can be placed on an integrated circuit would approximately double 
 every eighteen months.} 
 improvement in computer power summarized in this empirical rule is now
 coming to an end.~\cite{Waldrop-16,Markov-14}
 The main culprit is the inevitable shrinking of the electronic components
 of the integrated circuits, whose size 
 is currently approaching the ultimate physical limit of a single atomic layer.
 Within this regime, traditional design principles have become 
 unreliable as quantum size effects start to kick in;
 moreover, Joule heating (e.g., related to parasitic tunneling currents)
 has become so dramatic as to seriously compromise the device  
 efficiency, or even its functionality.

 To illustrate why miniaturization is directly linked to
 energy waste, it is useful to consider
 the example of metal-oxide-semiconductor field-effect transistors (MOSFET), 
 an ubiquitous device in modern microprocessors.
 In MOSFETs, the gate metal electrode and the semiconducting channel 
 are separated by an insulating oxide, in such a way that
 the stack of the three materials forms a capacitor. 
 A voltage applied to the gate is then used to control the resistance of the
 channel,~\cite{Sapoval} and hence to amplify or switch electronic signals. 
 Now, a large capacitance $C$ is mandatory in order to operate the 
 transistor at low gate voltages. (The resistance of the channel depends
 on the ``free charge'' that is stored on the semiconductor side.)
 Recalling the textbook formula for $C$ (the subscript ``geom'' emphasizes
 that, for a given dielectric material, 
 its value only depends on the geometry of the parallel-plate capacitor), 
  \begin{equation}
     C = C_{\rm geom} 
    = \frac{\kappa A}{ 4 \pi d},
    \label{eq:cgeomdef}
 \end{equation}
 where $\kappa$ is the permittivity, $d$ is the thickness and 
 $A$ the surface area, one can see that miniaturization inevitably requires
 thinner dielectric layers if $C$ is to be kept constant upon a 
 reduction of $A$.
 Yet, when $d$ reaches the length scale of few (tens of) atomic layers, 
 tunneling currents become so large that further shrinking would be 
 impractical -- 
 this summarizes, in a nutshell, the conundrum that the semiconductor 
 industry is currently facing.

 To work around this issue, one would need to increase the capacitance 
 per unit area without further reducing the dielectric thickness,  
 a task in which both technology and
 fundamental research has invested tremendous efforts in the past few years.
 A particularly promising route revolves around the
 concept of ``negative capacitance''. 
 Briefly, it consists in
 connecting two or more capacitors [e.g., the usual dielectric film
 described by Eq.~(\ref{eq:cgeomdef}), plus an additional element 
 whose physical nature will
 be specified shortly] in series, in such a way that the overall capacitance 
 is \emph{larger} than the original value. 
 Of course, if we stick to ordinary device elements, this is impossible:
 elementary electrostatics dictates that the total capacitance 
 is always smaller than that of any individual capacitor in the series. 
 However, if one of the capacitances is \emph{negative},
 then the total capacitance can, in principle, be larger than that of the 
 constituents, i.e. $C$ could  be enhanced with respect to the classically 
 expected value, Eq.~(\ref{eq:cgeomdef}), without reducing 
 the geometric thickness, $d$.

 The obvious question, then, is how to realize a negative 
 capacitance in practice. 
 Different proposals have appeared in the literature during the last few years.
 They can be more easily understood if we notice that the inverse  
 of the capacitance can be written as the second derivative of the
 total energy with respect to the charge stored on the 
 plates.~\cite{Feynman-II}
 While a negative value may appear unphysical at first sight, as it indicates 
 a thermodynamic instability of the system, a ``negative capacitance''
 can nevertheless exist locally, in a composite device whose global capacitance
 is still positive.

 A first possibility has already been demostrated in 
 ferroelectric nanocapacitors.~\cite{Salahuddin-08,Cano-10,Gao-14,Zubko-16}
 The basic idea is that a ferroelectric material has a switchable
 spontaneous polarization ($P$), whose potential landscape,
 $E(P)$, can be described by a characteristic double-well curve. 
 In a vicinity of the centrosymmetric saddle-point configuration, 
 such a curve is convex (i.e., $d^2 E/dP^2<0$), indicating
 a polar instability; therefore, a ferroelectric film can in principle
 provide a negative contribution to $C$ when appropriately incorporated in
 a capacitor stack.~\footnote{In a capacitor, the charge 
 density stored on the plates is proportional to the normal component of
 the electric displacement field. 
 Since, in atomic units,  
 $Q/S = D/ (4\pi) = (E + 4 \pi P)/ (4 \pi)$, where
 $E$ is the normal component of the electric field, 
 and in a typical ferroelectric 
 material $4 \pi P \gg E$, then $Q/S \approx P$.}

 A second strategy
 puts, instead, the emphasis on the physics of the metallic 
 electrodes themselves.
 In particular, Kopp and Mannhart~\cite{Kopp-09} recently argued
 that the total capacitance, $C$, of a device depends
 on the quantum-mechanical nature of the electrodes via
\begin{equation}
    \frac{1}{C} = 
                  \frac{1}{C_{\rm geom}} + \frac{1}{Ae^{2}}\frac{d \mu}{dn}.
    \label{eq:invcap}
 \end{equation}
 Here $C_{\rm geom}$ is the classical capacitance of Eq.~(\ref{eq:cgeomdef}),
 $\mu$ is the chemical potential (Fermi level) 
 of the metallic plate, $e$ is the electron charge, 
 and $n$ is the electron density per surface area.
 The second term on the right-hand side is proportional to the
 so-called ``electron compressibility'', $d\mu / dn$, and encodes the 
 aforementioned electrode-dependent effects.
 Within specific conditions (dilute two-dimensional gases) 
 that can, in principle, be realized in MOSFETs,~\cite{Mannhart-10}
 an electron system can enter a regime of 
 ``negative electronic compressibility'', i.e. $d\mu / dn < 0$,
 and hence provide a negative contribution to the overall capacitance.
 
 Experimentally, this unconventional~\footnote{In the standard textbook picture,
 when an electron is added to 
 a metallic system it fills the lowest unoccupied energy state; 
 as a consequence, the chemical potential increases.}
 behavior was first detected 
 in semiconductor (GaAs/AlGaAs) quantum wells.~\cite{Ashoori-92,Eisenstein-94}
 The effect was ascribed to the quantum nature of the two-dimensional 
 electrodes: in the
 dilute limit, the response of confined electron gases is
 typically dominated by exchange effects and tend to yield a negative $d\mu/dn$.~\cite{Eisenstein-94}
 The interest in this effect has remained mostly academic until recently, 
 because the relative gain in capacitance that one expects for semiconductor-based 
 systems is too small for practical applications. 
 However, fundamental research in this area has regained momentum
 with the recent discovery of a very large capacitance enhancements 
 ($>$40\% with respect the geometrical value) in two-dimensional
 electron gas (2-DEGs) in oxide 
 nanostructures (LaAlO$_{3}$/SrTiO$_{3}$ interface).~\cite{Li-11,Tinkl-12}
 On one hand, it is very tempting to interpret these arresting new results as 
 a manifestation of the same physics as that observed and modeled by Eisenstein
 {\em et al.}~\cite{Eisenstein-94}
 On the other hand, it is important to keep in mind that a polar oxide interface
 drastically departs, both from the point of view of structural and electronic
 properties, from the (much simpler) case of a semiconductor quantum well.
 In order to avoid any uncontrolled extrapolation it is, therefore, necessary
 to critically assess, first of all, the generality of the existing negative 
 compressibility models, and verify whether their conclusions are general enough
 to encompass a wider range of materials and geometries.

 In this work we quantitatively evaluate, based on first-principles calculations
 of jellium slabs and on analytic derivations, all the major ingredients 
 that contribute to the electronic compressibility in ultrathin metallic
 electrodes.
 We devote special attention to the role played by electron confinement,
 which we identify as a key difference between the cases of the quasi-infinite
 square well (appropriate to deal with the case of a semiconductor quantum well,
 where the electrons are confined by large conduction band offsets),
 and that of the asymmetric wedge-like potentials
 (useful to describe polar interfaces).
 In the case of strong external confinement, our results show excellent 
 agreement with both the model of Eisenstein {\em et al.},~\cite{Eisenstein-94}
  and with the ideas that 
 Kopp and Mannhart~\cite{Kopp-09} brought forward to explain the origin of the negative 
 capacitance in ideal two-dimensional systems.
 However, when the confining potential is modified 
 (or removed altogether, leaving
 a jellium-like positive background to keep the electrons in place)
 we find that the properties of the system are affected rather dramatically,
 in some cases even reversing the expected qualitative trends.
 First we show that, in the weak confinement regime, the wavefunction
 response to an external field is much stronger than in a quasi 
 two-dimensional (2D)
 quantum well with essentially infinite potential barriers; this leads to
 a very strong charge-density contribution to the capacitance.
 This effect stems from the progressive displacement of the ``image charge''
 plane from the geometric center of the 2D-like electrode while the capacitor
 is progressively charged, and always provides a negative contribution to
 $C$. (In fact, we find that this contribution is well appreciable 
 even in the regime investigated in Ref.~\onlinecite{Eisenstein-94},
 although the 
 experimental set up therein was not specifically designed to detect
 it.)
 Second, we find that a more delocalized nature of the electron gas 
 facilitates population of higher subbands, and this introduces new
 contributions to the capacitance that go well beyond the assumptions of 
 Ref.~\onlinecite{Kopp-09}.
 In particular, when a new band starts to be populated we find a strong,
 abrupt drop in the electronic compressibility, 
 which can become even more negative
 than that of an ideal 2D system at comparable density.
 This effect originates from the orthogonality between the wavefunctions
 that belong to different energy bands, which allows for a more 
 efficient redistribution of the electron charge, and hence for
 a drastic reduction of the electrostatic (Hartree) contribution
 to the electronic compressibility.

 Our work, therefore, highlights two previously overlooked mechanisms 
 that can potentially lead to a negative contribution to the capacitance 
 in interacting electron systems.
 We note both effects are completely general,
 and applicable to a wide variety of physical systems, as we 
 illustrate via a variety of test cases. 
 As a matter of fact, the 
 ``image plane'' charge-density effect
 is not restricted to ultrathin quasi-2D
 metallic systems, but can be readily found in standard thick electrodes
 as well, where the concept of ``negative compressibility'' is not applicable.
 (The chemical potential is fixed by the Fermi level of the bulk metal.)
 At the same time, our work clearly shows that exceptional care is needed
 when dealing with two-dimensional electron systems: the underlying
 physical mechanisms at play 
 may be substantially different from case to case, and critically
 depend on the structural and electronic properties of the host material.

 Our work is organized as follows. In Sec.~\ref{sec:preliminaries} 
 we summarize the main features of our capacitor model.
 In Sec.~\ref{sec:invcapden} we provide a brief background on the physics of the inverse
 capacitance density and its relation with the negative electron 
 compressibility. We include the explicit formulation to compute it in the  
 cases where only one band is occupied (Sec.~\ref{sec:energydecom}),
 its relationship with perturbation theory (Sec.~\ref{sec:perturbation}),
 and the generalization to multiple occupied bands 
 (Sec.~\ref{sec:multiband}).
 The self-consistent numerical implementation of all the above is
 described in Sec.~\ref{sec:computationalmethod}. 
 The established theory of confined electron gases with frozen wave functions
 is revisited under the light of our prescription in 
 Sec.~\ref{sec:confinedgasesfrozen}, including 
 ideal two dimensional electron gases
 (Sec.~\ref{sec:ideal2D}) and the effects introduced by
 the finite thickness where the electron gas is confined 
 (Sec.~\ref{sec:finitethickness}).
 In Sec.~\ref{sec:confinedgasesscf} we demonstrate the importance of 
 relaxing the wave functions in three different regimes:
 when the quantum gas is confined by a strong external potential,
 typical case in semiconductor QWs (Sec.~\ref{sec:res-traditional-qw}), 
 weak confinement [jellium slab in the absence of a strong
 confinement potential; Sec.~\ref{sec:jellium_unconstrained}], 
 and asymmetric barrier [reminiscent of a polar interface;
 Sec.~\ref{sec:asymmetric}],
 highlighting in each case the connections to earlier works and 
 the two original mechanisms that we introduced above.
 Finally, in Sec.~\ref{sec:discussion} and Sec.~\ref{sec:conclusions} 
 we discuss some possible realizations of the ideas described here, 
 together with some limitations of our treatment that may motivate further work
 on this subject.  
  
 \section{Method}
 \label{sec:preliminaries}

\begin{figure}
    \begin{center}
       \includegraphics[width=1.\columnwidth]{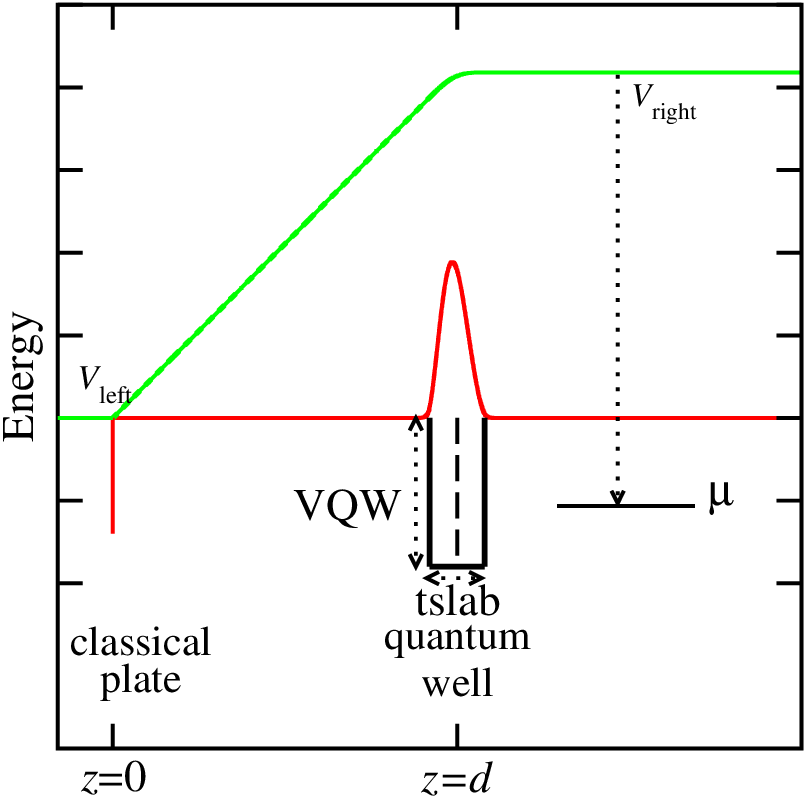}
       \caption{ (Color online) 
                 Schematic view of the computational setup used 
                 in our model.
                 The capacitor is made of a classical plate of zero thickness
                 where a negative background charge is located 
                 (solid vertical red line at the left),
                 and a quantum electrode, 
                 here represented as a quantum well of thickness 
                 $w$ where the electronic
                 charge (solid red line) is localized.
                 Green and black solid lines represent, respectively,
                 the electrostatic potential and 
                 a strong external potential,
                 typical in semiconductor QWs, that confines the quantum gas 
                 ($V_{\rm ext}$).
                 Vertical dashed line represents the geometrical center of the
                 quantum well, whose distance with respect to the classical
                 plate is $d$. 
               }
       \label{fig:cartoonmodel}
    \end{center}
 \end{figure}

 In Fig.~\ref{fig:cartoonmodel} we present a schematic version of the
 capacitor model that we shall use in this work.
 It consists in a classical electrode, 
 whose surface is located at
 $z = 0$, and of a quantum electrode, represented as 
 a quantum well of thickness $w$ and centered at $z=d$;
 the two are separated by a dielectric of
 permittivity $\kappa$.~\cite{same-kappa}
 In the remainder of this work, we shall assume that the system
 is infinitely extended in the $(x,y)$ plane, parallel to the
 electrode surface, and we shall indicate the perpendicular
 direction as $z$. Unless otherwise stated, we shall use atomic units 
 throughout ($\hbar = m_{\rm e} = a_{0} = c = e = 1$).
 The electronic charge is assumed to be positive, so the
 electrostatic energy and the electrostatic potential,
 as well as the electron density and the electronic charge density,
 amounts to the same value numerically.
 The in-plane electron density, $n$, can also be described
 by the parameter $r_{\rm s}$, measured in Bohr units, 
 which characterizes the average interparticle distance within the plane,
 \begin{equation}
    \frac{1}{n} = \pi r_{\rm s}^{2}.
    \label{eq:defnrs}
 \end{equation}

 \psfrag{tslab}[cc][cc]{$w$}
 \psfrag{VQW}[cc][cc]{$V_{\rm ext}$}

 Within the quantum electrode, we shall explicitly solve
 either the many-body or mean-field Schr{\"o}dinger equation, 
 in order to account for all relevant 
 electron-electron interactions at and beyond the Hartree description
 (further details of the numerical procedure 
 are provided in Sec.~\ref{sec:computationalmethod}).~\footnote{
 Note that in Ref.~\onlinecite{Eisenstein-94} a slightly more complicated 
 double-well setup was used, in order to facilitate the comparison
 with experimental measurements; the precise relationship between
 our model in Fig.~\ref{fig:cartoonmodel} and that of
 Ref.~\onlinecite{Eisenstein-94}
 will be discussed in Sec.~\ref{sec:confinedgasesscf}.}
 In particular, we shall write the total charge density as 
 \begin{equation}
   \rho_{\rm tot}(z) = ( n_{\rm jell} -n ) \delta(z) + \rho_{\rm el}(z) + 
      \rho_{\rm jell}(z),
   \label{eq:rhotot} 
 \end{equation}
 where the delta function at $z=0$ describes the classical plate, 
 $\rho_{\rm el}(z)$ is the electronic density of the quantum plate, and
 $\rho_{\rm jell}(z)$ accounts for the possible inclusion of a uniform jellium 
 background within the latter.
 The capacitor is overall neutral, so the integral of the total
 charge density written in Eq.~(\ref{eq:rhotot}) along the whole $z$ axis 
 vanishes; this implies that $\int dz \rho_{\rm jell}(z) = - n_{\rm jell}$.
 The electrostatic potential, $V_{\rm H}(z)$, is then calculated from
 $\rho_{\rm tot}$ via a one-dimensional Poisson equation,
 \begin{equation}
   \frac{ d^2 V_{\rm H}(z) }{d z^2} = -\frac{4 \pi}{\kappa} \rho_{\rm tot}(z).
   \label{eq:poisson}
 \end{equation}
 (We shall set the electrical boundary conditions in such a way that 
 the electrostatic potential is constant and equal to $V_{\rm left}$ for $z<0$.
 Due to the neutrality of the capacitor, the potential will be again constant
 for $z\gg d + w/2$ and equal to $V_{\rm right}$.)
 Since $\rho_{\rm tot}(z)$ differs from zero only in proximity of the 
 capacitor plates, the electric field equals 
 $\mathcal{E} = - d\vv(z)/dz = - 4 \pi n/\kappa$ 
 in the interior of the capacitor. 
 We shall indicate the center of the charge density 
 stored on the quantum electrode as the ``average charge plane'', 
 \begin{equation}
    \bar{z} = \frac{1}{n} \int dz \:\: z \rho_{\rm tot}(z). 
    \label{eq:averagez}
 \end{equation}
 It is an easy exercise to verify that, in full generality 
 and with the sign convention we have chosen,
 \begin{equation}
      \frac{4 \pi n \bar{z}(n) }{\kappa} = V_{\rm tot}(n),
    \label{eq:vtotbarz}
 \end{equation}
 where we 
 have defined $V_{\rm tot} = V_{\rm right} - V_{\rm left}$.
 
\subsection{Inverse capacitance density}
\label{sec:invcapden}

 Knowledge of the electrostatic potential offset, $V_{\rm tot}$, is not
 enough to compute the inverse capacitance density, 
 $\mathcal{C}^{-1} = C^{-1}A$.
 The latter is defined as the first derivative of the Fermi level offset,
 or equivalently as the second derivative (with respect to $n$) of 
 the \emph{total}
 energy of the capacitor per surface unit, $E_{\rm tot}$,  
 \begin{equation}
    \mathcal{C}^{-1} = \frac{d \mu_{\rm tot}}{d n} = 
                       \frac{d^2 E_{\rm tot}}{d n^2}.
    \label{eq:defcminusone}
 \end{equation}
 Here, similarly to the potential, we have defined 
 \begin{equation}
    \mu_{\rm tot} = \mu_{\rm right} - \mu_{\rm left}
    \label{eq:mutot}
 \end{equation}
 as the difference between the Fermi levels of the left 
 ($\mu_{\rm left}$) and right ($\mu_{\rm right}$) electrodes.
 Given the assumption of an ideal classical electrode on the left, we
 can assume $\mu_{\rm left}= V_{\rm left}$; thus, $\mu_{\rm tot}$ is the 
 Fermi level of the quantum electrode referred to $V_{\rm left}$.
 Clearly, $\mu_{\rm tot}$ is a geometry-dependent quantity: at a given $n$
 the thicker the capacitor, the larger the potential drop between
 the plates. 
 To have a geometry-independent definition of the chemical potential
 it is convenient to use $V_{\rm right}$ as a reference instead, and
 define
 \begin{equation}
    \mu = \mu_{\rm right} - V_{\rm right}.
    \label{eq:defmu}
 \end{equation}
 Then, by construction, we have
 \begin{equation}
    \mathcal{C}^{-1} = \frac{d V_{\rm tot}}{d n} + \frac{d \mu}{d n} = 
                       \frac{4 \pi z_{\rm im} }{\kappa} + \frac{d \mu}{d n},
    \label{eq:invc}
 \end{equation}
 where we have introduced the differential center of charge $z_{\rm im}$,
 also known as ``image charge'' plane,
 \begin{equation}
    z_{\rm im}(n) = \frac{d (n \bar{z})}{d n}.
    \label{eq:defzim}
 \end{equation}
 The relationship between $\bar{z}$ and $z_{\rm im}$ is in all 
 respects analogous to that 
 existing between average and instantaneous velocity,
 and reduces to an equality only in cases where the
 perturbation of the electronic wavefunction due to the 
 applied bias can be neglected.
 By replacing Eq.~(\ref{eq:averagez}) into Eq.~(\ref{eq:defzim}) one can trivially
 verify that 
 $z_{\rm im}$ is the first moment of charge-density variation with respect 
 to $n$; in other words, it corresponds to the average location where the free 
 electrons accumulate when an infinitesimal charge is added to the plate.
 $z_{\rm im}$ is thus uniquely given by the charge-density
 response of the system to an applied bias [recall Eq.~(\ref{eq:vtotbarz})].
 
 In Eq.(\ref{eq:invc}) we have achieved an intuitive partition of 
 $\mathcal{C}^{-1}$ into an ``electrostatic'' term 
 which is reminiscent of the geometric 
 contribution to the rhs of Eq.~(\ref{eq:invcap}), 
 and a ``quantum'' contribution,
 which depends on the Fermi level response of the electrode.
 Note, however, that the labels ``electrostatic'' and ``quantum'' should not
 be overinterpreted. On one hand, the electronic charge density response to
 a bias, which enters the definition of $z_{\rm im}$, is obviously determined
 by quantum effects; on the other hand, the evolution of the Fermi level
 contains residual electrostatic terms that are not accounted 
 for by $z_{\rm im}$, as we shall see shortly.
 Thus, in the following we shall regard both as ``quantum'' 
 contributions, and recover the ``geometric'' term of Eq.~(\ref{eq:invcap})
 by simply referring $z_{\rm im}$ to some well-defined feature of the 
 quantum electrode (e.g., the center of the well, as in the figure).
 We stress that the choice of such reference is somewhat arbitrary -- for example, 
 there are equally good arguments to set $d$ either at the surface or at the center 
 of the well; obviously, the concept of  ``geometric'' distance between electrodes 
 becomes ambiguous when the thickness of the device becomes comparable to the 
 typical interatomic distances.

\subsection{Energy decomposition in a mean-field context}
\label{sec:energydecom}

 In the remainder of this Section we shall assume a singly occupied band,
 whose wavefunctions are given by the product of a plane wave in the plane
 of the quantum electrode (indexed by the in-plane wavenumber 
 ${\bf k_\parallel}$) times an envelope function along $z$,

 \begin{equation}
    \psi_{\bf k_\parallel}({\bf r}) = e^{i \bf k_\parallel \cdot r} \psi(z).
    \label{eq:psikparallel}
 \end{equation}
 
\noindent The one-dimensional envelope function $\psi(z)$ 
 is normalized to unity,
 and corresponds to the self-consistent solution of the following eigenvalue 
 problem,

 \begin{equation}
    \hat{H} |\psi \rangle = \epsilon |\psi \rangle .
    \label{eq:schro}
 \end{equation}

 \noindent Here $\hat{H}= \hat{T}_z + \hat{V}_{\rm H} + \hat{V}_{\rm xc} + 
                 \hat{V}_{\rm ext}$ is 
 the Kohn-Sham Hamiltonian. The first term on the right hand side, 

 \begin{equation}
    \hat{T}_z = -\frac{1}{2 m_\perp} \frac{d^2}{dz^2}, 
    \label{eq:tz}
 \end{equation}

 \noindent is the normal component of the kinetic energy operator,
 where $m_\perp$ is the effective mass along $z$.  
 The other terms are the Hartree (H), exchange-correlation (xc) and 
 external (ext) potentials.
 The Hartree and exchange-correlation potentials, in turn,
 depend on the total and electronic charge densities, where the latter
 is defined in term of the wavefunction $\psi(z)$ as
 
 \begin{eqnarray}
    \rho_{\rm el}(z)  & = & n |\psi(z)|^2,
    \label{eq:rhoel1}
 \end{eqnarray}

 The \emph{total} energy of the capacitor can then be written as a variational 
 functional of $\psi(z)$ that depends parametrically on the areal density of
 particles, $n$, as

 \begin{equation}
    E_{\rm tot}(\psi,n) = E_{\rm K} + E_{\rm H} + E_{\rm xc} + E_{\rm ext} - 
        \lambda (\langle \psi |\psi \rangle -1).
    \label{eq:energy-decom}
 \end{equation}
 
 \noindent Here 

 \begin{equation}
    E_{\rm K} = \frac{\pi n^2}{2 m_\parallel}  + 
        n \langle \psi |\hat{T}_z |\psi \rangle,
 \end{equation}

 \noindent is the kinetic energy density of the noninteracting electrons, 
 which contains the trivial in-plane contribution (depending quadratically 
 on $n$ and inversely on the in-plane effective mass, $m_\parallel$).

 \begin{equation}
    E_{\rm H}[\rho_{\rm tot}] = \frac{1}{2} \int dz \, V_{\rm H} (z) 
                                \rho_{\rm tot}(z)
    \label{eq:electrostaticE}
 \end{equation}

 \noindent is the electrostatic energy density, where the Hartree potential is 
 defined by Eq.~(\ref{eq:poisson}).

 \begin{equation}
    E_{\rm xc}[\rho_{\rm el}] = \int dz \, \epsilon_{\rm xc}[\rho_{\rm el}(z)] 
               \rho_{\rm el}(z)
 \end{equation}

 \noindent is the exchange
 and correlation energy density (describing the electron-electron
 interactions beyond the Hartree approximation, plus the remainder of the 
 many-body kinetic energy). Finally, 

 \begin{equation}
    E_{\rm ext} [\rho_{\rm el}] =  \int dz \, V_{\rm ext} (z) \rho_{\rm el}(z)
 \end{equation} 

 \noindent accounts for a possible external potential. 
 The last term in Eq.~(\ref{eq:energy-decom})
 is a Lagrange multiplier, and serves to guaranteeing the correct 
 normalization of the wavefunction; 
 at the variational minimum one has $\lambda = n \epsilon$.

 We shall now assume that we are at the variational minimum with respect to
 $\psi$, i.e. define

 \begin{equation}
    E_{\rm tot}(n) = \min_\psi E_{\rm tot}(\psi,n).
    \label{eq:energyfunctional}
 \end{equation}

 \noindent To connect the present treatment with the quantities that we 
 introduced in the earlier Sec.~\ref{sec:invcapden}, 
 we shall then take the total derivative of $E_{\rm tot}(n)$ 
 with respect to $n$.
 The latter, in virtue of the Hellmann-Feynman theorem, 
 reduces to a partial derivative and can be easily calculated,

 \begin{equation}
    \frac{d E_{\rm tot}}{d n} = \frac{\partial E_{\rm tot}}{\partial n} = 
        \frac{\pi n}{m_\parallel} + \epsilon - V_{\rm left}.
    \label{eq:detot}
 \end{equation}

 \noindent [Note that the eigenvalue $\epsilon$ and $V_{\rm left}$ are 
 both defined modulo an arbitrary constant, which stems from the 
 electrostatic potential, 
 $V_{\rm H}(z)$ as given by Poisson's equation~(\ref{eq:poisson}); such 
 arbitrariness cancels out when taking their difference.]
 To arrive at this result, we have used the following relationship between 
 self-consistent potentials and energies

 \begin{equation}
    \frac{\delta E_{\rm H,xc}[\rho]}{\delta \rho(z)} = V_{\rm H,xc}(z),
    \label{eq:potHxc}
 \end{equation}

 \noindent and we have subsequently used Eq.~(\ref{eq:schro}) 
 to replace $\langle \psi | \hat{H} | \psi \rangle = \epsilon$.
 Consistent with Eq.~(\ref{eq:defcminusone}), it is easy to get convinced 
 that Eq.~(\ref{eq:detot})
 describes the chemical potential of the quantum electrode, 
 with $V_{\rm left}$ used as a reference, i.e., $\mu_{\rm tot}$.

 To make progress towards an expression for $\mathcal{C}^{-1}$, 
 one can again use the Hellmann-Feynman theorem to write  
 \begin{align}
    \mathcal{C}^{-1} = \frac{d \mu_{\rm tot}}{d n} & = 
       \frac{\pi}{m_\parallel} + 
       \frac{d \left( \epsilon - V_{\rm left} \right) }{dn}
    \nonumber \\
       & = 
       \frac{\pi}{m_\parallel} 
       + \langle \psi | \frac{ d \hat{H} }{dn} | \psi \rangle
       - \frac{ d V_{\rm left} }{dn}.
 \end{align}

 \noindent Note the total derivative signs, which imply that the variation 
 of the self-consistent potentials with $n$, due to the relaxation of the 
 wavefunctions
 $d| \psi \rangle/dn$, must be included.
 In fact, since the Hamiltonian only depends on $n$ via the Hartree and 
 exchange-correlation potentials, we have
 \begin{equation}
    \mathcal{C}^{-1} = \frac{\pi}{m_\parallel} +
                      \Delta_{\rm H}^{\rm left} + \Delta_{\rm xc},
    \label{cm1left}
 \end{equation}
 where the first term on the rhs is the constant kinetic contribution, 
 related to
 the in-plane dispersion, and the other two terms refer to the Hartree (H)
 and exchange-correlation (xc) contributions to the 
 eigenvalue variation, 
 \begin{align}
    \Delta_{\rm H}^{\rm left} & = 
        \langle  \psi \vert \frac{d (\hat{V}_{\rm H} - V_{\rm left})}{dn}  \vert  
                 \psi \rangle,
     \label{eq:deltaHleft} \\
    \Delta_{\rm xc} & = \langle  \psi \vert \frac{d \hat{V}_{\rm xc}}{dn}  \vert  
                 \psi \rangle.
     \label{eq:deltaxc} 
 \end{align}

 To complete the link to the results of the previous Section, 
 note that in Eq.~(\ref{cm1left}) we have incorporated
 the variation of $V_{\rm left}$ into the Hartree contribution.
 This has been done on purpose: Indeed, subtracting $V_{\rm left}$ from
 $V_{\rm H}(z)$
 boils down, from the physical point of view, to fixing the reference of the 
 electrostatic potential.
 The drawback of such a choice is that $\Delta_{\rm H}^{\rm left}$
 depends on the geometry of the capacitor, i.e., on the distance between the 
 plates.
 To circumvent this issue, we can proceed in pretty much the same way as before,
 by setting the convention to $V_{\rm right}=0$ instead, 
 \begin{equation}
 \Delta_{\rm H}^{\rm left} = \Delta_{\rm H}^{\rm right}   + \frac{4\pi z_{\rm im}}{\kappa},
 \end{equation}
 where 
 \begin{equation}
 \Delta_{\rm H}^{\rm right} = \langle  \psi \vert  \frac{d (\hat{V}_{\rm H} - V_{\rm right})}{dn}  \vert  \psi \rangle.
 \label{delright}
 \end{equation}
 Then, $\Delta_{\rm H}^{\rm right}$ reflects the variation of the 
 internal electrostatic energy
 of the ``quantum'' plate (we shall see that this term can be identified
 with the Hartree band bending effect), while the second one is the 
 familiar ``image plane'' contribution.
 Summarizing all the previous results in a single expression, in the 
 spirit of Eq.~(\ref{eq:invc}), 

 \begin{align}
    \mathcal{C}^{-1} & = \frac{4 \pi z_{\rm im} }{\kappa} + \frac{d \mu}{d n}
    \nonumber \\
    & = \frac{4 \pi z_{\rm im} }{\kappa} + 
         \frac{\pi}{m_{\parallel}} +  \Delta_{\rm H}^{\rm right} + \Delta_{\rm xc}.
    \label{eq:invc1band}
 \end{align}
 Similarly, the Fermi level referred to $V_{\rm right}$ is
 \begin{equation}
 \mu = \frac{\pi n}{m_\parallel} + \epsilon - V_{\rm right}.
 \end{equation}

\subsection{Relationship to perturbation theory}
 \label{sec:perturbation}

 With the above derivations, we have achieved a further insight into the
 physics of the inverse capacitance density, by separating it into
 a trivial (constant) kinetic term
 and two contributions (Hartree and xc) that
 originate from the variation of eigenvalue, $\epsilon$, with $n$.
 Before moving on, it is worth spending a few words on the specific role
 played by the relaxation of the electronic wavefunction, since its
 effect is not immediately clear from the above derivations.
 To see this, it is useful to reformulate the inverse capacitance problem
 in the language of linear-response theory.~\cite{Baroni-01} 
 In such a framework,
 the energy functional of Eqs.~(\ref{eq:energy-decom}) and~(\ref{eq:energyfunctional}) can be expanded 
 to second order in the perturbation parameter $n$ around some reference value $n_0$
 [we shall indicate the $l$-th total derivative with respect to $n$ with a ($l$) superscript henceforth],

 \begin{eqnarray}
    E_{\rm tot}(n) &=& E_{\rm tot}(n_0) + 
                        (n-n_0) E^{(1)}(n_0) \nonumber \\ 
                   &&     + \frac{(n-n_0)^2}{2} E^{(2)}(n_0) + \ldots,
    \label{eq:expansionE}
 \end{eqnarray}

 \noindent where the first and second-order terms are, respectively, 
 the chemical potential and the inverse capacitance density,

 \begin{equation}
    E^{(1)} = \mu_{\rm tot}, \qquad  E^{(2)} = \mathcal{C}^{-1}.
 \end{equation}
 An analogous expansion can be operated on the wavefunctions,
 \begin{equation}
   \psi(n) = \psi(n_0) + (n-n_0) \psi^{(1)}(n_0) + \ldots
 \end{equation}

 \noindent One can then write explicit expressions for the
 $2j+1$-th term in the expansion of the energy functionals
 by using the expansion terms of $\psi$ up to order $j$.
 Crucially, the even-order ($2j$-th) energy expansion terms can be
 constructed in such a way that they are \emph{variational}
 in the $j$-th order wavefunctions;~\cite{Gonze-95.2} 
 this implies that the inverse capacitance density can be written as a variational
 functional of the first-order wavefunctions, $\psi^{(1)}$,

 \begin{align}
    E^{(2)}(\psi^{(1)},n) = & \frac{\pi}{m_\parallel} + 
       2n \langle \psi^{(1)} |(H - \epsilon) |\psi^{(1)} \rangle +
    \nonumber \\
      & \rho^{(1)}_{\rm tot} \cdot K_{\rm H} \cdot \rho^{(1)}_{\rm tot} \, 
      + \, \rho^{(1)}_{\rm el} \cdot K_{\rm xc} \cdot \rho^{(1)}_{\rm el}.
    \label{eq:e2}
 \end{align}

 Here $K_{\rm H,xc}$ indicates the Hartree or XC kernels,
 \begin{equation}
    K_{\rm H,xc}(z,z') = 
      \frac{\delta^2 E_{\rm H,xc}}{\delta \rho(z) \delta \rho(z')},
 \end{equation}
 
 \noindent and the two first-order densities refer to the first-order variation
 of either the total or electronic charge density with $n$, 

 \begin{eqnarray}
    \rho^{(1)}_{\rm el} (z) &=& |\psi(z)|^2 + 2 n  \psi(z) \psi^{(1)}(z), 
    \label{eq:rho1el} \\
    \rho^{(1)}_{\rm tot} (z) &=& \rho^{(1)}_{\rm el} (z) - \delta(z).
 \end{eqnarray}
 Such densities, in turn, define the first-order potentials via
 \begin{eqnarray}
   \frac{ d^2 V^{(1)}_{\rm H}(z) }{d z^2} &=& -\frac{4 \pi}{\kappa} \rho^{(1)}_{\rm tot}(z), \label{vh1}\\
   V^{(1)}_{\rm xc}(z) &=&  K_{\rm xc} \cdot \rho^{(1)}_{\rm el} (z),
  \end{eqnarray} 
 It is straightforward to verify that Eq.~(\ref{eq:e2}) reduces to 
 Eq.~(\ref{cm1left})
 at the variational minimum; in the language of Ref.~\onlinecite{Gonze-97.2} 
 these two formulas can be regarded, respectively, as the \emph{stationary} and
 \emph{nonstationary} expressions for the inverse capacitance density.

 In our context, the variational character of Eq.~(\ref{eq:e2})  
 implies that \emph{the relaxation
 of the wavefunctions always lowers the inverse capacitance density}, 
 whatever is the relative contribution of the individual effects.
 (Note that this statement is true for the \emph{total} inverse capacitance
 density, while it may break down once we separate it into a ``quantum'' 
 and ``image plane'' contribution.)
 We shall illustrate this important point with our 
 numerical experiments shortly.

\subsection{Multiband case}
\label{sec:multiband}

 So far we have discussed the inverse capacitance density in the special case
 of a single occupied band. We shall now extend the theory to the case of
 multiple bands.
 Assume that we have $N$ partially occupied bands of the form
 \begin{equation}
 \psi_{{\bf k_\parallel},l}({\bf r}) = e^{i \bf k_\parallel \cdot r} \psi_l(z).
 \label{eq:psi}
 \end{equation}

 \noindent Note that the individual $\psi_l$ need not be eigenstates 
 of the same one-dimensional Schr\"odinger equation; 
 in the most general case (as it happens, for instance,
 at the LaAlO$_{3}$/SrTiO$_{3}$ interface),~\cite{Stengel-11.2}
 different types of subbands may be populated, where each subset 
 feels a different external potential,
 and has a different out-of-plane dispersion, $m_\perp(l)$.
 We shall further assume, for the sake of generality, that the 
 in-plane effective mass of each state is $m_\parallel(l)$ 
 (i.e. they need not be the same for all bands).
 Each band then contributes to the electronic charge density as

 \begin{equation}
    \rho_{\rm el} (z) = \sum_{l} n_{l} \vert \psi_{l} (z) \vert^{2},
    \label{eq:rhomb}
 \end{equation}

 \noindent with

 \begin{equation}
    n_l = \frac{k_{\rm F}^2(l)}{2 \pi}, \qquad \sum_l n_l = n,
    \label{nl}
 \end{equation}

 \noindent where $k_{\rm F}^2(l)$ is the Fermi momentum, 
 and to the total kinetic energy as

 \begin{equation}
    E_{\rm K}(l) = \frac{\pi n_l^2}{2 m_\parallel(l)}, 
       \qquad \sum_l E_{\rm K}(l) = E_{\rm K}.
 \end{equation}
 Finally, since there is a unique Fermi level, the quantity
 \begin{equation}
    \mu_{\rm tot} = \frac{\pi n_l}{m_\parallel(l)} + \epsilon_l - V_{\rm left}
    \label{mul}
  \end{equation}

 \noindent must be the same for all bands.
 Solving for $n_l$ in Eq.~(\ref{mul}) and replacing in Eq.~(\ref{nl}), we obtain

 \begin{equation}
   \frac{\sum_l m_\parallel(l)}{\pi} \mu_{\rm tot} = 
   n + \sum_l \frac{m_\parallel(l)}{\pi} (\epsilon_l - V_{\rm left}),
 \end{equation}

 \noindent  and finally

 \begin{equation}
    \mu_{\rm tot} = \frac{\pi n}{ \sum_l m_\parallel(l)} + 
    \frac{\sum_l m_\parallel(l) \epsilon_l}{\sum_l m_\parallel(l)} - 
    V_{\rm left},
    \label{eq:multiband}
 \end{equation}
 
 \noindent which yields the chemical potential as a function of 
 the \emph{total} charge density,
 $n$, and of a weighted sum of the single-particle eigenvalues.

 As it was done in Sec.~\ref{sec:energydecom} for the single band case,
 the inverse capacitance density can be computed as

 \begin{equation}
    \mathcal{C}^{-1} = \frac{d \mu_{\rm tot}}{d n} = 
       \frac{\pi}{\sum_l m_\parallel(l)}  
       + \frac{\sum_l m_\parallel(l) \frac{d \epsilon_l}{dn}}
              {\sum_l m_\parallel(l)}
       - \frac{d V_{\rm left} }{dn}
    \label{eq:C-1multiband}
 \end{equation}

 The most remarkable consequence of Eq.~(\ref{eq:C-1multiband}) is a 
 drastic lowering of the
 positive contribution to the compressibility that is due to the 
 in-plane kinetic energy.
 Indeed, if we assume that the masses are all equal, the kinetic contribution 
 to $\mathcal{C}^{-1}$
 is $\pi / (N m_\parallel)$, i.e., it reduces to a fraction of the 
 single-band value $\pi / m_\parallel$.
 Regarding the eigenvalue contribution, it is straightforward to show that

 \begin{equation}
    \frac{d \epsilon_l}{dn} = \langle \psi_l | 
         \left( \hat{V}_{\rm H}^{(1)} + \hat{V}_{\rm xc}^{(1)} \right) | \psi_l \rangle,
    \label{eq:multiband-eigen}
 \end{equation}
 i.e., even if the states $\psi_l$ are eigenfunctions of different Hamiltonian
 operators, only the variation of the global self-consistent potential 
 is relevant for calculating their contribution to $\mathcal{C}^{-1}$.

\subsection{Computational method}
 \label{sec:computationalmethod}

 We have implemented the model presented in the previous Sections
 by self-consistently solving the one-particle Kohn-Sham Hamiltonian described 
 in Eq.~(\ref{eq:schro}).
 To perform the integrals, we use Numerov's algorithm on a real-space grid,
 whose fineness can be controled with a single energy cutoff
 (the kinetic energy of the plane wave that can be represented in the
 grid without aliasing). 
 The exchange and correlation interactions are treated at the level 
 of the local density
 approximation (LDA)~\cite{Kohn-65,Ceperley-80}
 to density functional theory (DFT).~\cite{Hohenberg-64}
 The energy eigenvalues of Eq.~(\ref{eq:schro}) corresponding to
 eigenfunctions with the correct asymptotic behavior 
 (smooth decay outside the slab) are searched using the 
 ``shooting method'' and the 
 ``double integration technique''.~\cite{Numerov-Gianozzi}
 Within this method, the eigenvalues of the bound states are 
 always bounded between the minimum value of the total potential inside
 the supercell and the energy at left of the classical plate, taken as zero. 
 At odds with other techniques that rely on the solution of the Poisson
 equation [Eq.~(\ref{eq:poisson})] with fast-Fourier transform techniques,
 where the eigenvalues are computed with respect to the average of the
 electrostatic potential in the unit cell (not always a well defined 
 quantity~\cite{Kleinman-81})
 our reference energy is well defined and the eigenvalues between different
 calculations perfectly comparable.

 \psfrag{zbohr}[cc][cc]{$z$ (Bohr)}
 \psfrag{Psi}[cc][cc]{$\psi_{i}(z)$}
 \psfrag{rhoel}[cc][cc]{$\rho(z)$}
 \psfrag{VHar}[cc][cc]{$V_{\rm H} (z)$}

 \section{Electron compressibility in confined metallic gases
          with frozen wave functions}
 \label{sec:confinedgasesfrozen}

 It is clear from Eq.~(\ref{cm1left}) for the single band case, or its 
 generalization to the multiple band case in
 Eqs.~(\ref{eq:C-1multiband})-(\ref{eq:multiband-eigen}), 
 that there are three main ingredients to the inverse capacitance density
 of an electron gas: 
 The first one is a positive constant, independent of the electron density,
 that comes from the contribution of the in-plane kinetic energy.
 The second and the third are, respectively, 
 the Hartree and the exchange-correlation contributions to the 
 eigenvalue variations.
 These three contributions have already been discussed at length in the
 context of confined electron gases. To prepare for the discussion of 
 the new effects, in this Section we shall briefly review the established
 results, and link to the relevant literature whenever possible.
 
 \subsection{Ideal two-dimensional case}
 \label{sec:ideal2D}

 Pioneer works on negative capacitance
 relied on an ideal two-dimensional electron gas model
 (the electrons are exactly confined into an idealized 2D plane,
 with a homogeneous charge density) to rationalize the experimental
 results.
 In such a limit, the subbands are separated by an infinite energy so only the
 lowest one is occupied, and
 the electronic charge density has a shape of a Dirac delta
 located at $z=d$, i.e. at the distance between the classical electrode
 and the ideal plane.

 In the limit of a two-dimensional electrode,
 the exchange energy per unit area $E_{\rm x}$ 
 is negative and depends on the charge density as 

 \begin{equation}
    E_{\rm x} = \epsilon_{\rm x}^{\rm 2D}(n) n,
    \label{eq:eexchenerden}
 \end{equation}
 \noindent where $\epsilon_{\rm x}^{\rm 2D}(n)$ is the exchange energy
 per electron,

 \begin{equation}
    \epsilon_{\rm x}^{\rm 2D}(n) = - \frac{4}{3 \kappa} 
                 \sqrt{\frac{2}{\pi}} n^{1/2}.
    \label{eq:exchangeenergypere}
 \end{equation}

 \noindent Specializing the definition of the potential,
 Eq.~(\ref{eq:potHxc}), to the exchange energy
 then Eqs.~(\ref{eq:deltaHleft})-(\ref{eq:deltaxc}) take the form

 \begin{eqnarray}
    \Delta^{\rm left}_{\rm H} &=& \int dz \, V^{(1)}_{\rm H}(z)
           \left[\delta(z-d) - \delta(z) \right] \nonumber \\
            &=& \frac{dV_{\rm right}}{dn} - \frac{dV_{\rm left}}{dn} =
           \frac{ 4 \pi d}{\kappa},
    \label{eq:deltah1km} \\
    \Delta_{\rm xc} &=& \int dz \, V^{(1)}_{\rm xc}(z) \delta(z-d) 
            =  - \frac{1}{\kappa} \sqrt{\frac{2}{\pi}} n^{-1/2},
    \label{eq:xc1km}
 \end{eqnarray}

 \noindent where we have used that in this ideal two-dimensional case
 $z_{\rm im} = d$. Thus, following Eq.~(\ref{cm1left}), the inverse
 of the capacitance density amounts to 

 \begin{equation}
    \mathcal{C}^{-1} = \frac{ 4 \pi d}{\kappa} + 
    \left( \frac{\pi}{m_{\parallel}} - \frac{1}{\kappa}
    \sqrt{\frac{2}{\pi}}
    n^{-1/2} \right).
    \label{eq:dmudndrivenex}
 \end{equation}

\begin{figure}
    \begin{center}
       \includegraphics[width=\columnwidth]{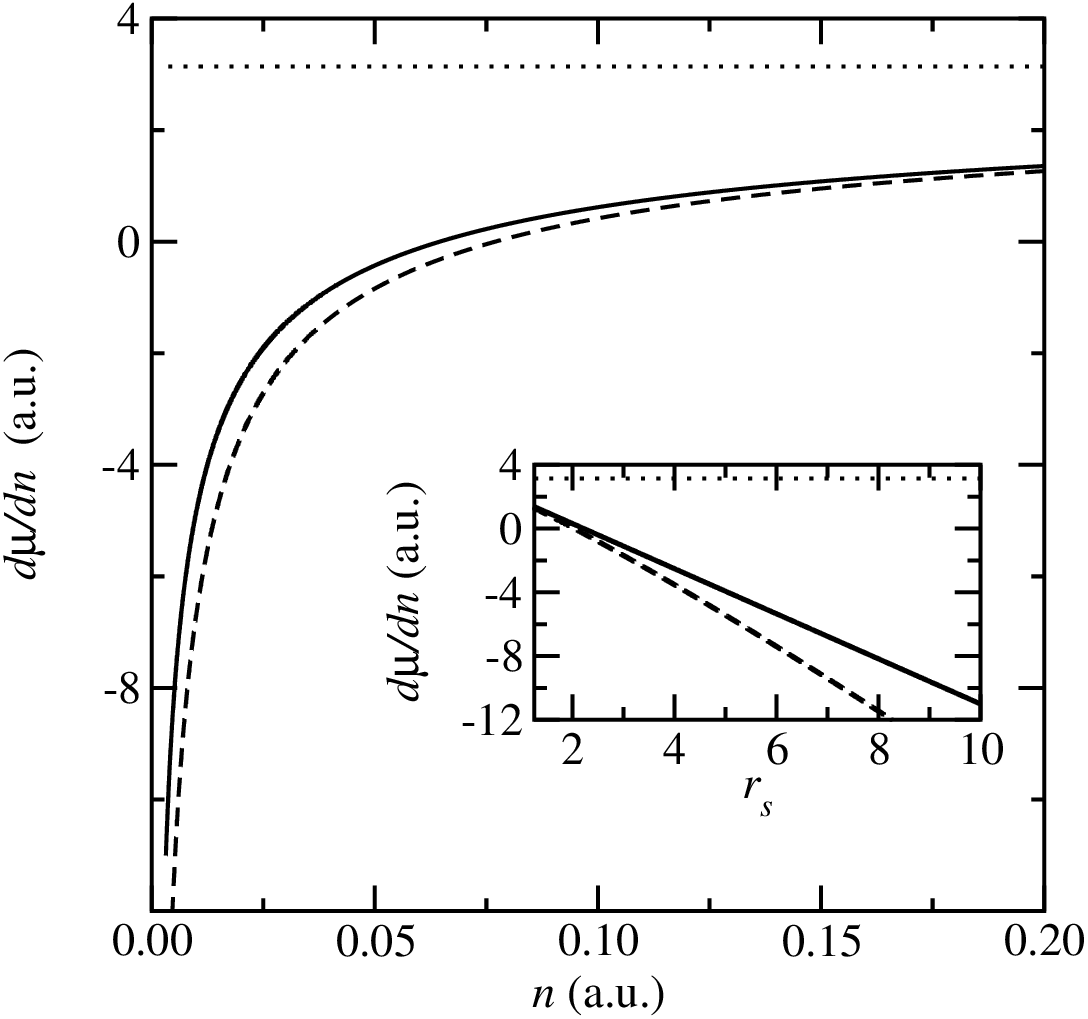}
       \caption{ Contributions to the electron compressibility  
                 as a function of the charge density of the sheet 
                 in the ideal 2D limit.
                 Black dotted line shows the constant kinetic contribution.
                 Solid black line is the electron compressibility
                 (sum of the kinetic and exchange contribution).
                 Dashed black curve considers, in addition, the correlation
                 contribution.
                 Inset: same results with respect the parameter $r_{\rm s}$.
                 Results have been obtained for 
                 $m_{\parallel} = \kappa = 1$.
               }
       \label{fig:KM}
    \end{center}
 \end{figure}

 \noindent We can identify the 
 quantity in brackets with the electron compressibility, $d\mu/dn$.
 Its behavior as a function of $n$ is illustrated
 in Fig.~\ref{fig:KM}. We can see 
 that in the high-density limit 
 (small $r_{\rm s}$) the electron compressibility is positive
 and approaches the non-interacting regime, where the
 kinetic energy dominates.
 As the density is reduced, $d\mu/dn$
 decreases, and within the present Hartree-Fock approximation, 
 changes its sign at a critical density $n_{c} = {2 m_{\parallel}^{2}}/
 (\kappa^{2} \pi^{3})$.
 For lower densities, the electron compressibility 
 is negative and therefore, as highligthed in the Introduction,
 the total capacitance can become larger than the classical 
 geometrical value. 

 So far, we have neglected the correlation energy.
 Indeed, its effects~\cite{Tanatar-89} 
 are expected to be minor
 within the density range of interest for typical
 semiconductors.~\cite{Eisenstein-94} 
 To assess the importance of correlation in the present context, 
 we have recalculated 
 $\mathcal{C}^{-1}$ by incorporating it explicitly 
 (see dashed line in Fig.~\ref{fig:KM});
 its impact is marginal, as expected.

 \subsection{Confined electron gas of a finite thickness}
 \label{sec:finitethickness}
 
 In most practical cases, a confined electron gas significantly
 deviates from the ideal 2D model of Sec.~\ref{sec:ideal2D} because 
 of finite-thickness effects.~\cite{Eisenstein-94}
 In the following we shall recap the impact of 
 {\em Coulomb softening}, affecting the mutual repulsion of the electrons,
 and of the {\em Hartree band-bending}, i.e. a Stark-like shift of the quantum
 well states that is due to the external potential.

 \subsubsection{Coulomb softening}
 \label{sec:coulomb-softening}

 Coulomb softening has to do  
 with the modification of the exchange interactions 
 within the ``thickened'' gas. 
 To gauge its importance, we shall assume that the gas is
 confined by an infinite square well of thickness $w$.
 This is justified in semiconductor quantum wells,
 whenever the conduction band offsets are much deeper
 than the mean electron energy.
 The electronic ground-state wavefunction
 can be then determined analytically, 

 \begin{equation}
    \psi(z) = \sqrt{\frac{2}{w}} \sin \left( \frac{\pi z}{w} \right),
    \label{eq:gsinfinitewell}
 \end{equation}
 leading to a three-dimensional charge density of the form
 \begin{equation}
    \rho_{\rm el}(n,z) = n |\psi(z)|^2 = 
              \frac{2n}{w} \sin^{2} \left( \frac{\pi z}{w} \right).
    \label{eq:chargedensitysw}
 \end{equation}

\begin{figure}
    \begin{center}
       \includegraphics[width=\columnwidth]{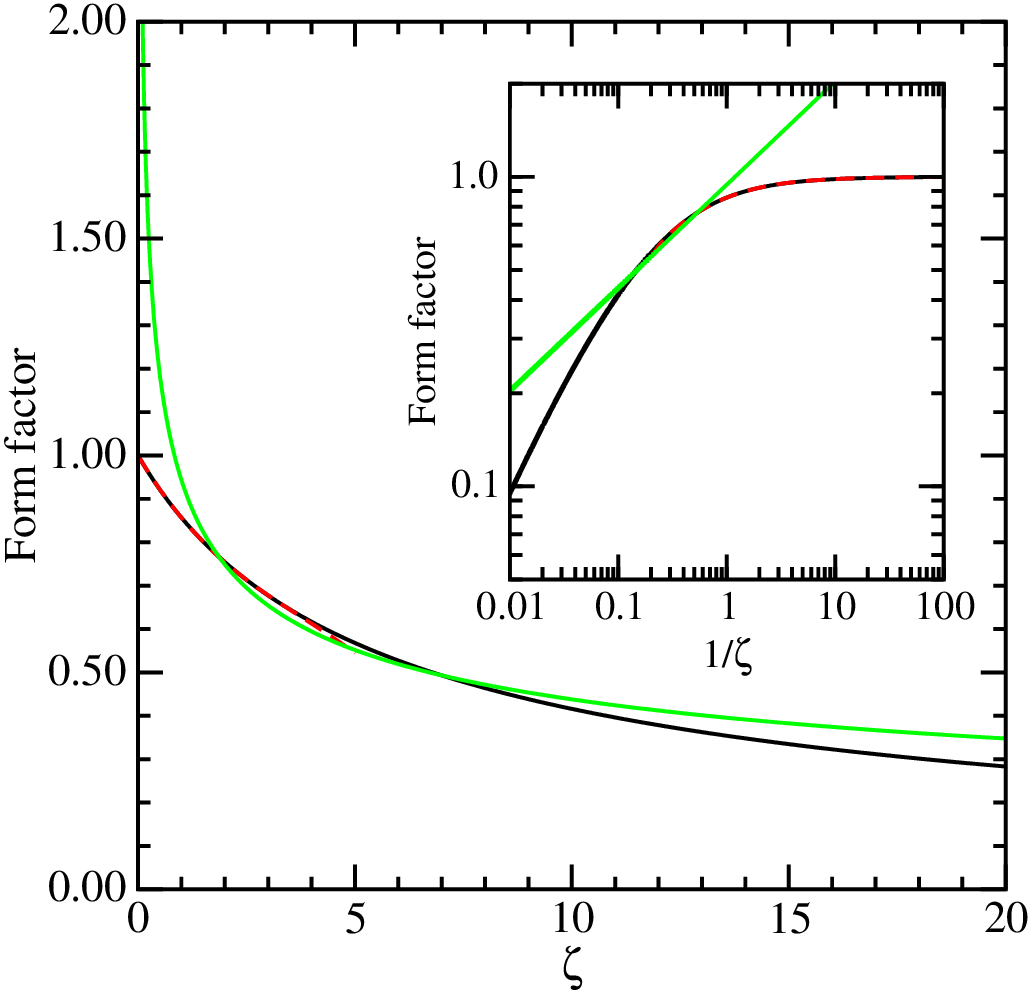}
       \caption{ (Color online) 
                 Form factor as given by the exact exchange functional
                 (black solid line), by a polynomial fit to 
                 low values of $\zeta$
                 as fitted by Eisenstein {\it et al.} 
                 in Ref.~\onlinecite{Eisenstein-94}
                 (red dashed lines), and by the LDA approximation 
                 (solid green line).
                 Inset: same data, but in a log-log scale and with 
                 the abscissas inverted.
               }
       \label{fig:form-factor}
    \end{center}
 \end{figure}

 Assuming that only the lowest subband is occupied within the
 relevant range of $n$, the necessary corrections to the exchange energy 
 per electron, Eq.~(\ref{eq:exchangeenergypere}), can be 
 summarized~\cite{Stern-74} in a form factor, $F(\zeta)$,

 \begin{equation}
    \epsilon_{\rm x} (n, w) = \epsilon_{\rm x}^{\rm 2D}(n) F(\zeta),
    \label{eq:sternx}
 \end{equation}

 \noindent where $\zeta = w /r_{\rm s} = w \sqrt{\pi n}$ 
 is the dimensionless
 ratio between the width of the potential well and the mean spacing 
 (in units of Bohr radius $a_0$) between electrons in the plane,
 as defined in Eq.~(\ref{eq:defnrs}).
 A simple polynomial fit, valid for an infinite square well
 ground-state wavefunction and for $0<\zeta<3.5$, is reported
 in the Appendix of Ref.~\onlinecite{Eisenstein-94}.
 Here, in order to obtain
 a more complete picture, we have calculated numerically $F(\zeta)$ 
 over a broader range of densities/thicknesses.
 The results are shown in Fig.~\ref{fig:form-factor}.
 Clearly, the exchange energy  
 is always smaller than in the ideal 2D case 
 (the form factor, black line in Fig.~\ref{fig:form-factor} is
 smaller than 1 at any $n$).
 
 Many practical implementations, including the numerical model that
 we shall use in this work, rely on the use of approximate
 functionals for the exchange and correlation, e.g. the 
 LDA.~\cite{Kohn-65,Ceperley-80}
 It is therefore important, at this stage, to gauge the accuracy 
 of the LDA exchange in describing the compressibility of a 
 confined electron gas.
 In fact, just like in the case of the exact (Hartree-Fock) treatment, 
 we can write $\epsilon_{\rm x}^{\rm LDA}(n,w)$ as the
 ideal 2D exchange energy times a form factor,
 \begin{equation}
    \epsilon^{\rm LDA}_{\rm x}(n,w) = \epsilon_{\rm x}^{\rm 2D}(n) 
    F^{\rm LDA}(\zeta),
    \label{eq:exchangelda}
 \end{equation}
whose explicit form is
 \begin{equation}
    F^{\rm LDA}(\zeta) = A \zeta^{-\frac{1}{3}}.
    \label{eq:form-factor-lda}
 \end{equation}
 ($A=0.9436555$ is a dimensionless constant.)
 The derivation of Eq.~(\ref{eq:form-factor-lda}) is given in 
 Appendix~\ref{app:form-factor-lda}.

 The approximate LDA form factor is compared with the exact Hartree-Fock limit
 in Fig.~\ref{fig:form-factor}.
 While the two functions roughly agree for $1 <\zeta<10$, 
 the LDA result is clearly wrong both in the low-density and high-density limits 
 for a given $w$ (or, equivalently, in the large and small thickness limit for a given $n$.) 
 In either case, LDA overestimates the exchange energy, 
 and such an overestimation becomes severe at small $\zeta$
 where $F^{\rm LDA}$ erroneously diverges as $\zeta^{-\frac{1}{3}}$.
 (This is the limit of small $w$ for a fixed $n$, 
 where the exact result tends to 1).
 These results indicate that outside the range $1 <\zeta<10$, LDA 
 may give a very inaccurate description of exchange effects, 
 and one should be extremely careful when drawing physical conclusions therein.

\begin{figure*}
  \begin{tabular}{cc}
  \includegraphics[width=0.45\textwidth]{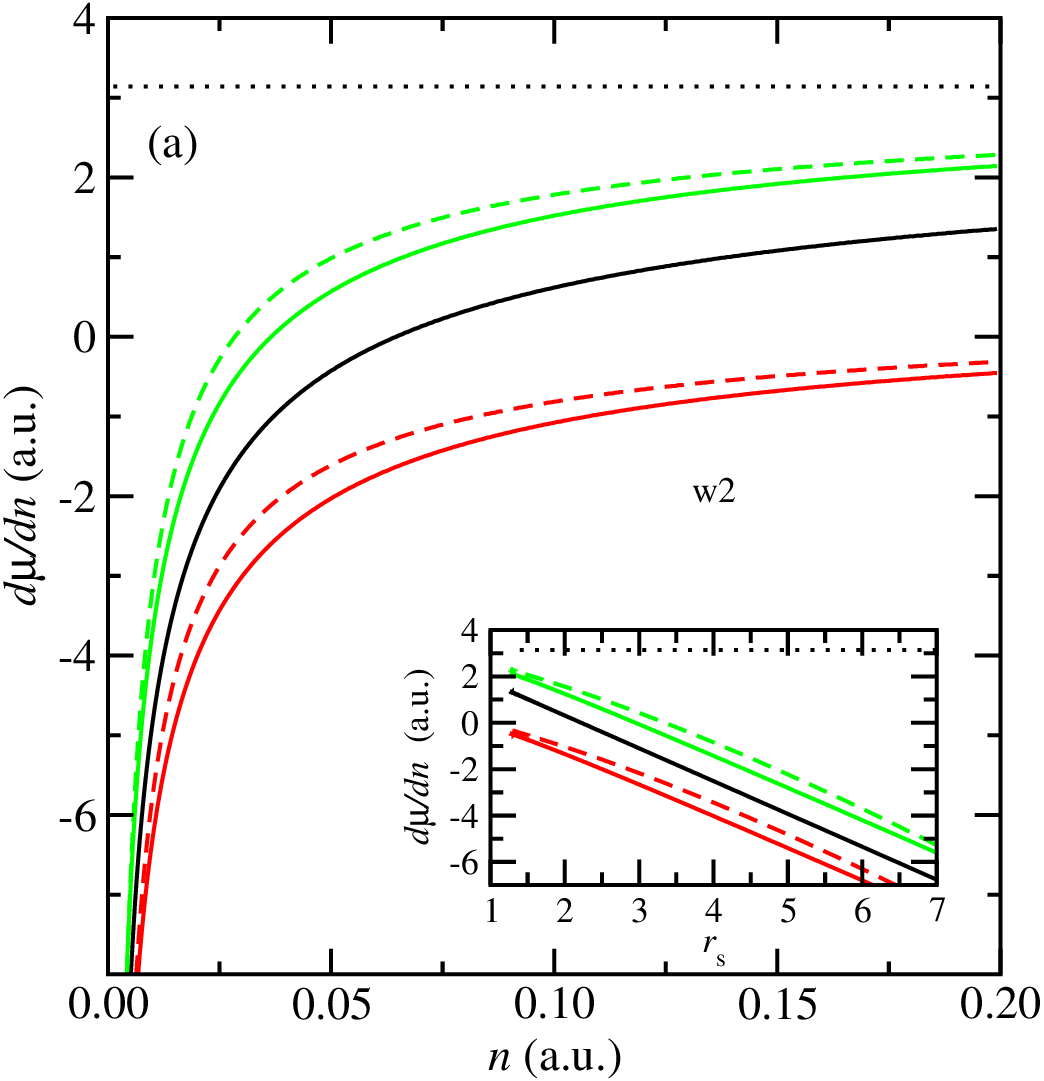} &
  \includegraphics[width=0.45\textwidth]{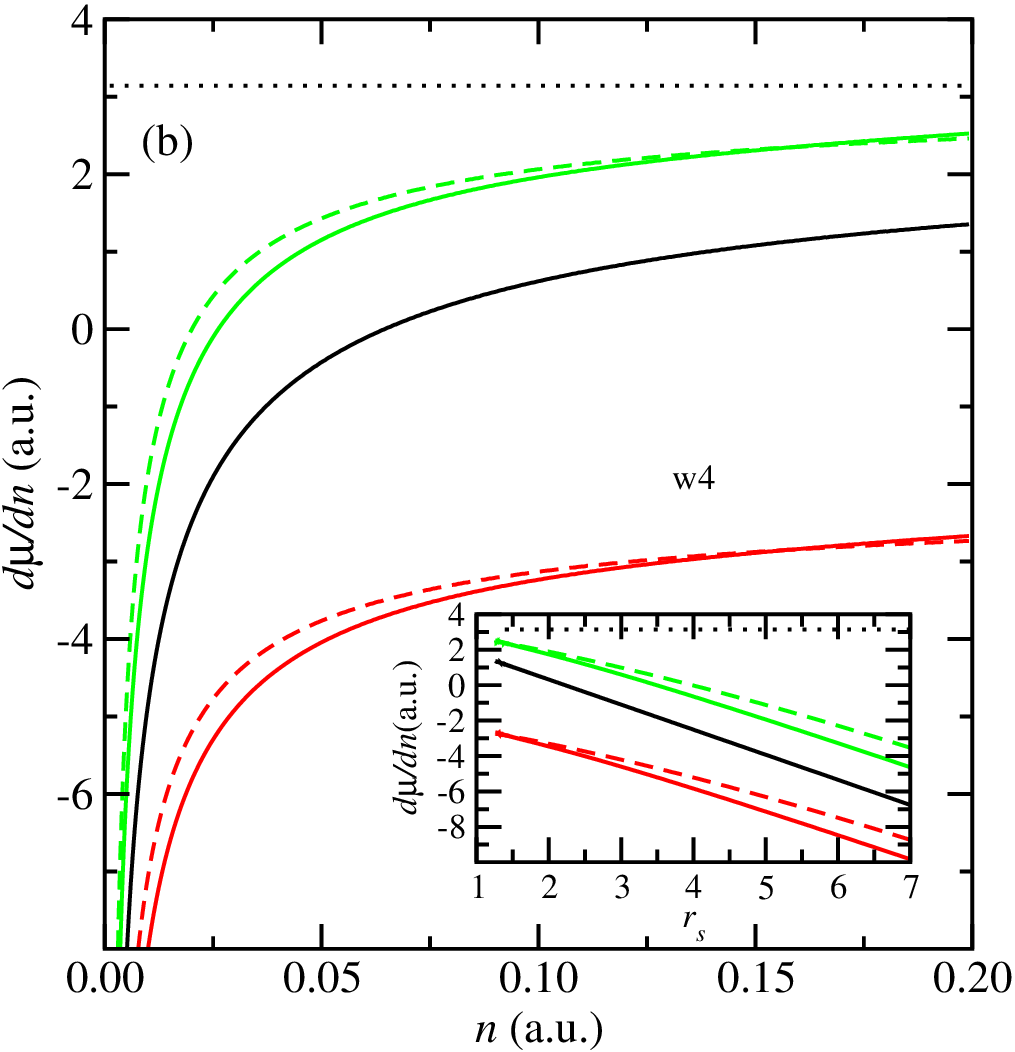} \\
  \end{tabular}
    \caption{(Color online) Various contributions to the 
             electron compressibility as a function of the 
             electron density, as given by
             the LDA exchange (dashed curves) or by the 
             ``thickened'' (exact) exchange functional (solid curves).
             Results in panel (a) and (b) have been obtained for a thickness of 
             $w = 2.0$ Bohr and $w = 4.0$ Bohr respectively.
             The analytical result for the ideal 2D
             limit ($w \rightarrow 0$ limit) is also shown as a 
             thick black curve for comparison.
             The dotted line at $d\mu/dn=\pi$ shows the contribution of the
             in-plane kinetic energy.
             Green curves represent the results considering only the
             thickened exchange,
             as discussed in Sec.~\ref{sec:coulomb-softening}, 
             in addition to the 
             noninteracting constant coming from the kinetic contribution.
             Red curves contain also the 
             ``Hartree band-bending'' as explained in 
             Sec.~\ref{sec:hartree-band-bending}.
             Insets: same results with respect the parameter $r_{\rm s}$.
             Results have been obtained for 
             $m_{\parallel} = \kappa = 1$.
             a.u. stands for atomic units.
            }
    \label{fig:finitethickness}
 \end{figure*}

 Results including the contribution of this
 ``Coulomb-softening'' effect
 on the electronic compressibility are shown 
 with green lines in Fig.~\ref{fig:finitethickness}.
 To quantify this softening, we use both the form
 factor computed numerically under the assumption of 
 a frozen wave function corresponding to the ground state of an infinite 
 square well potential (shown in Fig.~\ref{fig:form-factor}), 
 and the LDA form factor [Eq.~(\ref{eq:form-factor-lda})].
 Then the exchange energy per unit area is computed as in 
 Eq.~(\ref{eq:eexchenerden}),
 together with its second derivative with respect the charge density
 to evaluate numerically the contribution of the thickened exchange 
 [Eq.~(\ref{eq:deltaxc})] to the electron compressibility.
 After adding the contant contribution coming from the kinetic energy term,
 the most important conclusions that can be drawn are: 
 (i) due to the softening
 of the negative contributions coming from the exchange energy,
 this effect reduces the tendency of the electronic compressibility 
 towards negative values (green curves in Fig.~\ref{fig:finitethickness}
 are always above the analytical results for the ideal 2D limit, represented
 by the black curves); and
 (ii) in the ranges of densities and thicknesses tried in 
 Fig.~\ref{fig:finitethickness}, the LDA approximation provides a
 qualitative (even semiquantitatively) correct picture when 
 compared with the exact functional.  

 \psfrag{w2}[cc][cc][1.00]{$w$ = 2.0 Bohr}
 \psfrag{w4}[cc][cc][1.00]{$w$ = 4.0 Bohr}

 \subsubsection{Hartree band-bending}
 \label{sec:hartree-band-bending}

 The Hartree band-bending effect is embodied in the $\Delta^{\rm right}_{\rm H}$
 contribution to the inverse capacitance density of Eq.~(\ref{eq:invc1band}),
 and is related to the first-order variation of the Hartree potential with $n$.
 At a given $n$ the Hartree potential is given by the 
 Poisson equation of Eq.~(\ref{eq:poisson}),
 which results in the following double integral,

 \begin{align}
    V_{\rm H} (z)
      & = - \frac{4 \pi}{\kappa} 
            \int_{-\infty}^z dt  \int_{-\infty}^t dt' \rho_{\rm tot}(t') 
    \nonumber \\
      & = \frac{4 \pi n z}{\kappa} - \frac{4 \pi}{\kappa} 
            \int_{-\infty}^z dt  \int_{-\infty}^t dt' n |\psi(t')|^2.
 \end{align}

 \noindent Assuming that the wavefunction is frozen, then
 $V_{\rm H} (z)$ is linear in $n$, 
 so its variation can be trivially computed,

 \begin{equation}
    \frac{\delta V_{\rm H}(z)}{\delta n} = 
        \frac{4 \pi z}{\kappa}
      - \frac{4 \pi}{\kappa} 
        \int_{-\infty}^z dt  \int_{-\infty}^t dt' |\psi(t')|^2.
    \label{eq:dervhdn}
 \end{equation}

 \noindent The integrals of Eq.~(\ref{eq:deltaHleft}) and Eq.~(\ref{eq:dervhdn}) 
 can be calculated numerically, resulting in

 \begin{equation}
     \Delta^{\rm left}_{\rm H}
     = \frac{4 \pi \bar{z}}{\kappa} -1.29862 w,
     \label{eq:dere1dnint}
 \end{equation}

 \noindent i.e. this is a constant contribution that simply scales 
 linearly with the thickness of the quantum well.
 Since we have assumed that the electronic wave functions are frozen,
 $z_{\rm im} = \bar{z}=d$ as explained in Sec.~\ref{sec:invcapden}.
 Therefore, the first term of the right hand side of Eq.~(\ref{eq:dere1dnint})
 is the geometric 
 contribution to the inverse of the
 capacitance density, while the remainder is the 
 desired Hartree band-bending term, $\Delta^{\rm right}_{\rm H}=-1.29862 w$.

 Numerical results including the contribution of $\Delta^{\rm right}_{\rm H}$
 are shown in Fig.~\ref{fig:finitethickness}.
 Consistent with earlier studies, the (negative)
 Hartree band-bending clearly overcompensates the Coulomb softening,
 resulting in an overall enhancement of the negative compressibility
 effect (red curves in Fig.~\ref{fig:finitethickness} are always
 lower than the black ones).
 The thicker the film, the larger the reduction in the electron
 compressibility.
 
 \psfrag{dmudnlabel}[ll][ll][1.20]{$d\mu/dn$ (a.u.)}
 \psfrag{njellium}[cc][cc][1.20]{$n$ (a.u.)}
 \begin{figure}
    \includegraphics[width=1.00\columnwidth]{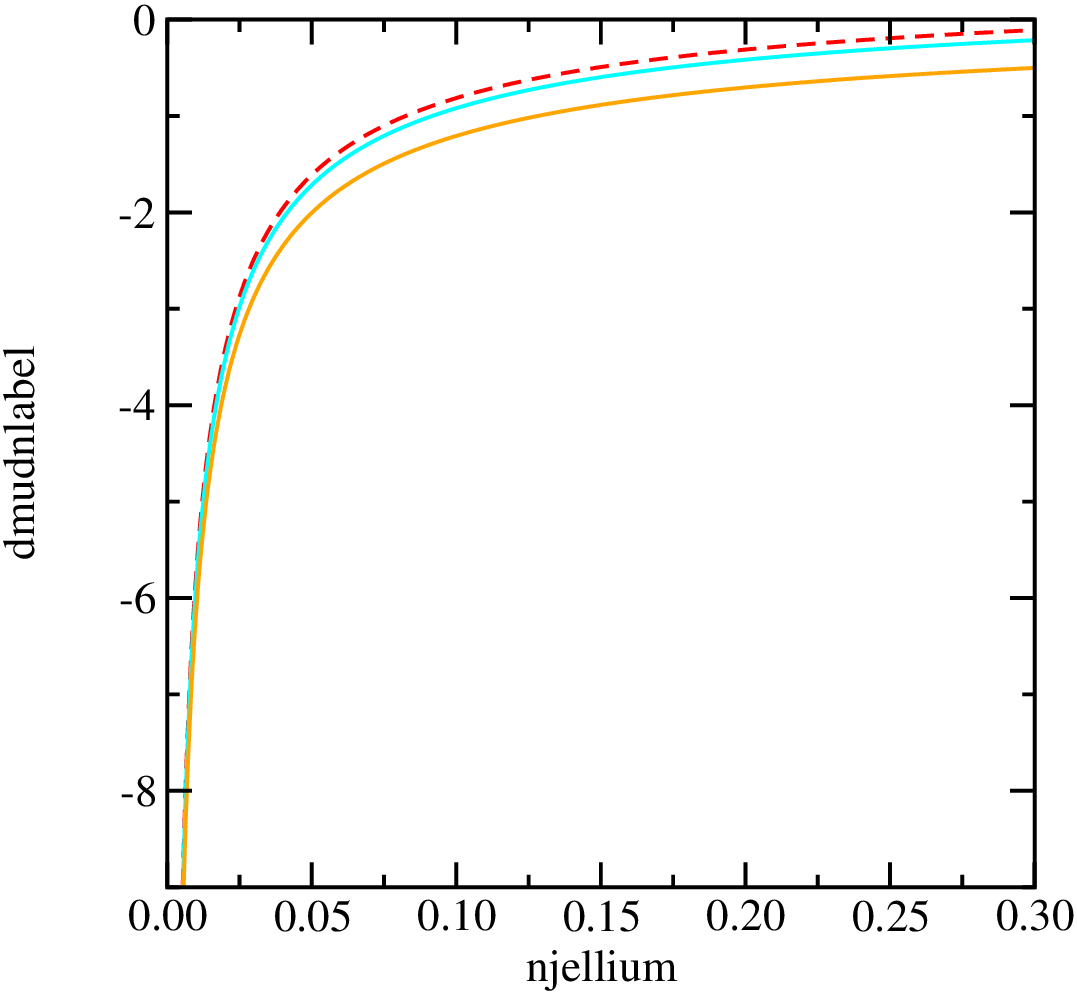}
    \caption{ (Color online) 
              Electronic compressibility as a function of the charge density
              including the Coulomb softening and the Hartree band bending
              effects for quantum wells of different depths.
              The exchange is treated at the LDA level. 
              The depth of the quantum well is infinite for the 
              red-dashed line [same as in Fig.~\ref{fig:finitethickness}(a)], 
              250 Ha for the cyan line,
              and 22 Ha for the orange line.
              The width of the well is $w$ = 2.0 Bohr.
              All magnitudes in atomic units.
            }
    \label{fig:depths}
 \end{figure}

\subsection{Quantum wells}
\label{sec:quantumwells}

 The above arguments are valid for an electron gas confined
 in a potential well of infinite depth. To prepare for the
 discussion of our numerical results in the next Section, we shall briefly
 study here the more realistic case of a quantum well (QW) with finite depth.
 This will allow us to gain a first insight on the impact of confinement
 effects.

 To do so, we first calculate the finite-QW ground-state wavefunction
 of the bare well ($n=0$) for three different depths
 ($V_{\rm ext}$= $\infty$, 250 Ha, and 22 Ha).
 Next, we insert this wavefunction into Eq.~(\ref{eq:dervhdn}) 
 to compute the Hartree band-bending at different values of $n$, 
 neglecting relaxation as before.
 The results are plotted in Fig.~\ref{fig:depths}.
 Almost no difference is found between the infinite 
 quantum well and the 250 Ha one.
 Nevertheless, the shallower the quantum well, the more
 extended the wavefunctions, which results in a lowering of the 
 electron compressibility
 as if the QW were still infinite but slightly wider
 [the prefactor in front of the second term of the rhs of 
 Eq.~(\ref{eq:dere1dnint}) changes to -1.351 for a depth of 250 Ha,
 and to -1.495 for a depth of 22Ha]
 
 Note that the Coulomb softening effect appears to be much 
 less sensitive to the well depth compared to the Hartree band-bending
 term shown in Fig.~\ref{fig:depths}. This observation is general to 
 all our work. Indeed, we shall see in the following 
 Section that it is the Hartree band bending effect that undergoes the most
 dramatic enhancements when the confinement is reduced or
 lifted altogether.

 \section{Electron compressibility in confined metallic gases
          with self-consistent wave functions}
 \label{sec:confinedgasesscf}

 After reviewing the basic phenomenology of confined 2D electrodes,
 we shall now move to presenting our main results, obtained
 by using the self-consistent numerical 
 solver described in Sec.~\ref{sec:computationalmethod}.
 In particular, we shall focus on two distinct aspects of the problem:
 (i) the self-consistent relaxation of the wavefunctions, which 
 leads to a displacement of the electronic charge density from the
 geometric center of the quantum wells (Sec.~\ref{sec:res-traditional-qw}), 
 and
 (ii) the effect of the population of higher subbands 
 (Sec.~\ref{sec:jellium_unconstrained}).

 \subsection{Traditional quantum well}
 \label{sec:res-traditional-qw}

 To validate our numerical implementation against the
 analytical results of the previous Section
 we consider, first of all, 
 the ``traditional quantum well'' capacitor
 of Fig.~\ref{fig:cartoonmodel} (referred to as t-QW henceforth).
 Here the electrons in the quantum electrode 
 are confined in a narrow layer by an external 
 (usually large) potential, $V_{\rm ext}$, mimicking
 the conduction-band offsets at the QW boundaries.
 Within the setup discussed in this subsection we shall 
 assume that $\rho_{\rm jell} = 0$ in Eq.~(\ref{eq:rhotot}),
 so the QW is charged with a charge density $n$, that is
 compensated by the charge density $-n$ located in the classical
 electrode.
 This configuration  
 essentially corresponds to the model of 
 Eisenstein and coworkers.~\cite{Eisenstein-94}

\begin{figure}
    \includegraphics[width=1.00\columnwidth]{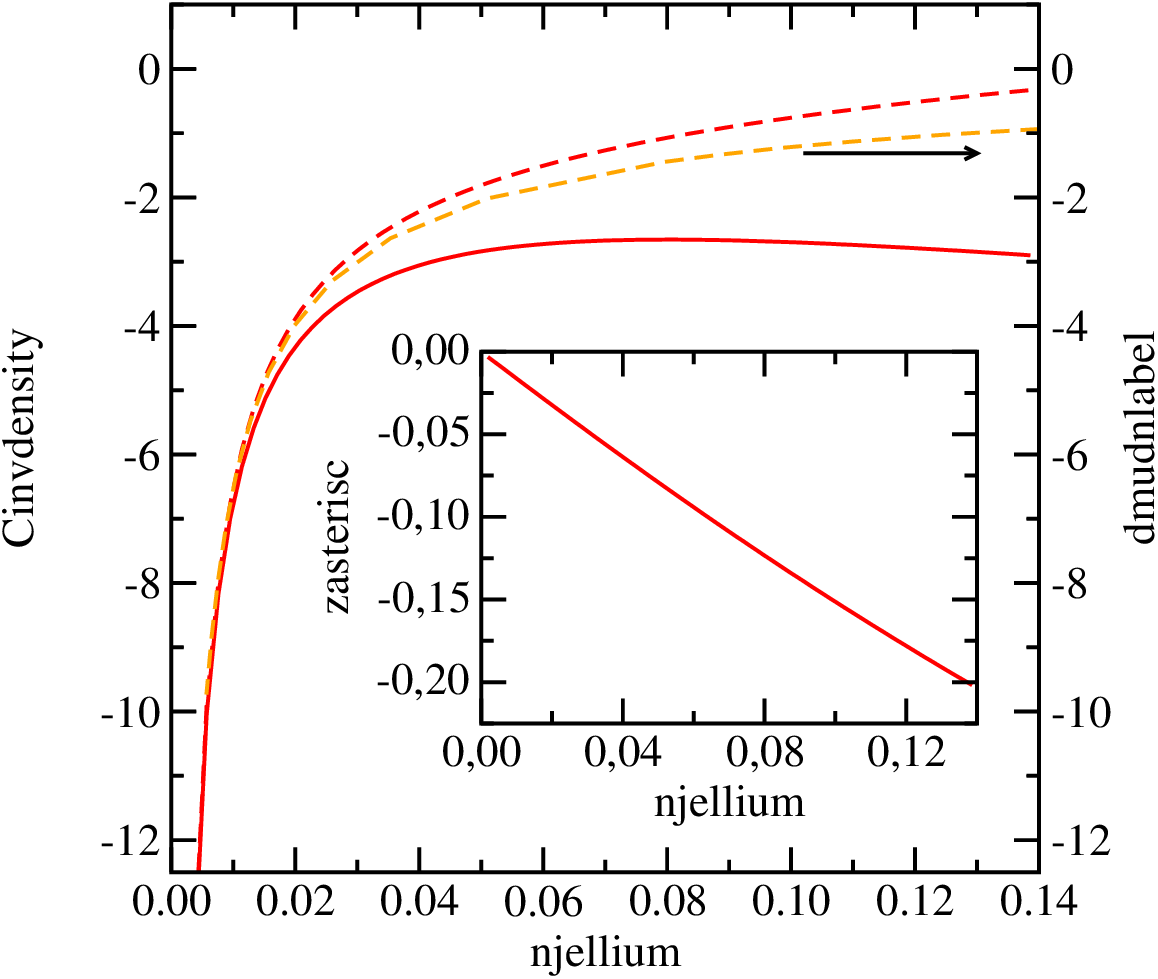}
    \caption{ (Color online) 
              Quantum contributions to $\mathcal{C}^{-1}$ in the
              traditional quantum well (t-QW) model ($w$ = 2.0 Bohr) 
              as a function of the electronic density $n$.
              The solid red curve is the full relaxed-wavefunction result.
              The Fermi level contribution [Eq.~(\ref{eq:invc1band})]
              is represented by a dashed red curve.
              The dashed orange curve represents 
              the electron compressibility for the frozen wave-function 
              results for a well of the same depth,
              $V_{\rm ext} =-22$ Ha.
              Inset: evolution of  $z_{\rm im}^{\ast}$ (i.e. the 
              image plane location with respect to the geometric 
              center of the QW) as a function of $n$.
            }
    \label{fig:eisenstein}
 \end{figure}

 Figure~\ref{fig:eisenstein} shows the evolution of the
 inverse capacitance density, $\mathcal{C}^{-1}$, computed as in 
 Eq.~(\ref{eq:defcminusone}), and of the electron compressibility 
 (``Fermi level'' contribution), $d\mu/dn$. 
 [To remove any dependency on the geometry of the capacitor, 
 we have substracted from $\mathcal{C}^{-1}$ a ``geometrical capacitance'' 
 equal to $-4 \pi d/\kappa$, where $d$ is the distance between the 
 classical plate and the center of the quantum well.
 Consistently, $z_{\rm im}$ will be referred to this specific feature
 of the quantum electrode, so we define $z_{\rm im}^{\ast} = z_{\rm im} -d$.]
 To make contact with the formalism of the previous Section,  
 we also show the analytical results for $d\mu/dn$, calculated
 by using the ground-state wavefunction of the \emph{finite} well
 at $n=0$ as described in Sec.~\ref{sec:quantumwells}.
 As the effects of wavefunction (WF) relaxation are neglected within 
 the analytic model of Sec.~\ref{sec:quantumwells} we shall refer to the dashed
 orange curve as to the ``frozen-WF'' results.
 Recall that these values are always negative in the 
 range of $n$ that we consider here, due to the combination of
 the thickened exchange (Sec.~\ref{sec:coulomb-softening}) and 
 Hartree band bending (Sec.~\ref{sec:hartree-band-bending}) effects.

 But on top of these effects, the self-consistency on a QW of finite 
 thickness introduces additional ingredients, whose impact on 
 the physics was not clearly identified in earlier works.~\cite{Eisenstein-94}
 Indeed, respect to the frozen-WF result, the electron compressibility 
 as defined in Eq.~(\ref{eq:invc1band}) is slightly more positive, while the
 total inverse capacitance is significantly more negative.
 The difference between the latter two quantities consists in
 the ``image charge'' contribution, 
 reflected in the $4 \pi z_{\rm im}/\kappa$ term in
 Eq.~(\ref{eq:invc1band}), which is due to
 the displacement of the electronic charge density from the geometric 
 center of the QW.
 Because of the electrostatic deformation of the electronic
 cloud, $z_{\rm im}$ is always smaller than the distance between the
 classical plate and the center of the QW, as shown in the inset of 
 Fig.~\ref{fig:eisenstein}.
 This is, therefore, an extra source of negative capacitance that largely overcomes
 the small upward shift of $d\mu/dn$ with respect to the frozen-WF 
 values. (The physical reasons behind both the overcompensation and the upward shift
 will be clarified in the next paragraphs.)

 \psfrag{inversecap}[ll][ll][0.60]{$\mathcal{C}^{-1}- \frac{4\pi d}{\kappa}$}
 \psfrag{zasterisc}[ll][ll][1.00]{$z_{\rm im}^{\ast}$ (a.u.)}
 \psfrag{Cinvdensity}[ll][ll][1.20]{$\mathcal{C}^{-1}$ (a.u.)}
 \psfrag{zasterisc2}[cc][cc][1.00]{$z_{\rm im}^{\ast}$}
 \psfrag{sigmaplusunit}[cc][cc][1.20]{$n$ (a.u.)}

 \subsection{Jellium slab with an external confinement potential}
 \label{sec:jellium-confined}

 For realistic simulations of many physical systems, it
 is convenient to treat the compensating charge as a uniform 
 background within the quantum well itself, in the spirit of the 
 jellium models.~\cite{Kopp-09}
 In order to separate the effects introduced by charge compensation 
 and confinement, we first carried out a set of simulations
 where we kept the same QW confining potential 
 as in Sec.~\ref{sec:res-traditional-qw}, but added a 
 jellium background,
 $\rho_{\rm jell}(z) \ne 0$ in Eq.~(\ref{eq:rhotot}), on top.
 The jellium density is defined as

 \begin{equation}
   \rho_{\rm jell}(z) = -n_{\rm jell} W(z),
    \label{eq:rhojell}
 \end{equation}

 \noindent where $W(z)$ is a ``window function'' equal to
 $1/w$ inside the well and zero outside. 
 From a practical point of view, at every value of 
 $n$ we shall first calculate the ground state of the neutral capacitor, 
 by setting $n_{\rm jell}=n$; the inverse capacitance density
 is then defined as the second derivative of the total energy
 with respect to $n$ \emph{at fixed} $n_{\rm jell}$.

 \begin{figure}
    \begin{tabular}{cc}
       \includegraphics[width=1.00\columnwidth]{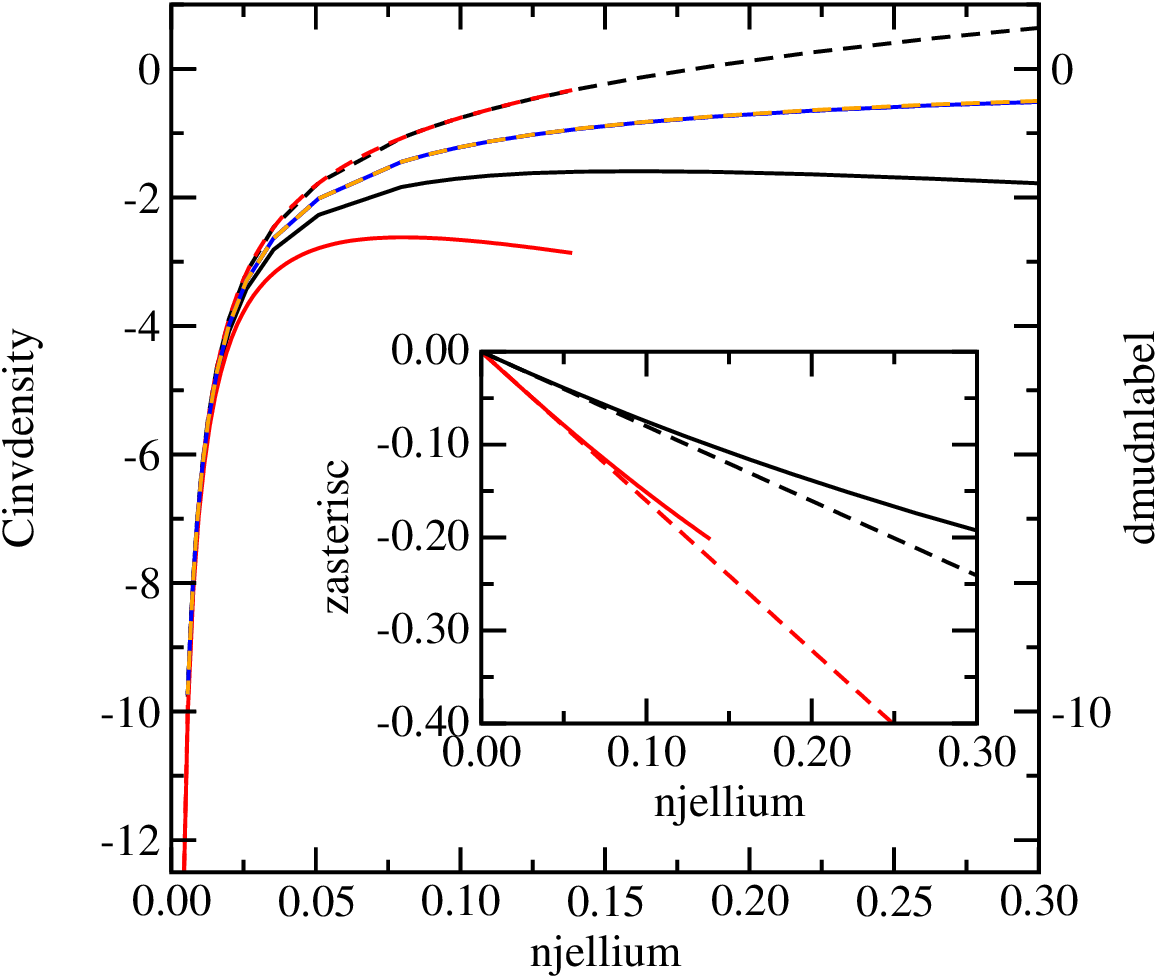} &
    \end{tabular}
    \caption{ (Color online) 
              Evolution of the inverse capacitance densities
              (solid lines; left y-axis) as a function of the
              charge density. 
              The electron compressibility 
              contributions [Eq.~(\ref{eq:invc1band})]
              are represented by dashed lines (right y-axis).
              Black lines represents the results of the 
              confined jellium slabs
              where the electronic wave functions are allowed to relax. 
              Blue lines are the results obtained when the electronic
              wave functions are frozen to the ones of a neutral jellium slab.
              Red lines represent the results of the traditional quantum well,
              already shown in Fig.~\ref{fig:eisenstein}, 
              while the dashed orange
              line represents the analytical results for the
              electron compressibility for a quantum well 
              of the same thickness and depth and under the same LDA approach, 
              already shown in Fig.~\ref{fig:depths}.
              Inset: evolution of the position of the image charge plane
              (referred to $d$ as explained in the text)
              as a function of the charge density. 
              Numerical results have been obtained for 
              $m_{\parallel} = \kappa = 1$, 
              $w$ = 2.0 Bohr, $d$ = 12.5 Bohr,
              and $V_{\rm ext}$ = -22 Ha.
            }
    \label{fig:results-jellium-well}
 \end{figure}

 The evolution of $\mathcal{C}^{-1}$, 
 together with its decomposition into the electron compressibility ($d\mu/dn$) 
 and image-plane ($z_{\rm im}$) contributions
 [Eq.~(\ref{eq:invc1band})], as a function of the density in the jellium
 background are shown as black curves in Fig.~\ref{fig:results-jellium-well};
 for comparison, we also report the results of Fig.~\ref{fig:eisenstein} 
 for the bare quantum well as red curves.
 Remarkably, 
 the evolution of the $d\mu/dn$ values with $n$ accurately
 match, regardless of
 whether we compensate the electronic charge with a jellium density (present case, j-QW henceforth)
 or not (traditional quantum well, t-QW).
 Conversely, the total inverse capacitance density (inclusive of
 the image-plane contribution) is significantly more negative in
 the t-QW case.

 To help explain these results, we follow a similar strategy as in the
 previous Section and compute, in addition to the self-consistent
 results, the frozen-wavefunction (WF) compressibilities. 
  [Here, at difference with the previous case where the $n=0$ 
  wavefunction was used throughout, we calculate the ``frozen WF''
  result by using, at each $n$, the ground-state wavefunction of 
  the neutral capacitor ($n=n_{\rm jell}$).]
  The results are plotted as a solid blue curve, and compared 
  with the frozen-WF data of Fig.~\ref{fig:eisenstein} (dashed orange);
  again, the two curves show an essentially perfect overlap.
  Obviously, at $n=0$ the two curves must coincide, 
  as the j-QW model reduces to 
  the t-QW case in the limit of small densities. 
  Deviations at finite $n$ can only stem from the
  deformation of the (symmetric) ground-state wavefunction due to the
  self-consistent potentials [for example, the superposition of 
  $\rho_{\rm el}(z)$ and $\rho_{\rm jell}(z)$ produces a 
  nonvanishing electrostatic potential whenever $n=n_{\rm jell} \neq 0$]. 
  Based on our result, we conclude that such 
  deviations are negligible within our choice of geometry and
  computational parameters.

 \psfrag{zimjel}[ll][ll][0.75]{$z_{\rm im}^{\ast}$(jellium + QW)}
 \psfrag{zimeis}[ll][ll][0.75]{$z_{\rm im}^{\ast}$(traditional QW)}
 \psfrag{zbarjel}[ll][ll][0.75]{$\bar{z}$(jellium + QW)}
 \psfrag{zbareis}[ll][ll][0.75]{$\bar{z}$(traditional QW)}
 \psfrag{rhoelec}[cc][cc][1.20]{$\rho_{\rm el}(z)$}
 \psfrag{rho1z}[cc][cc]{$\rho^{(1)} (z)$}
 \begin{figure}
    \begin{center}
       \includegraphics[width=\columnwidth]{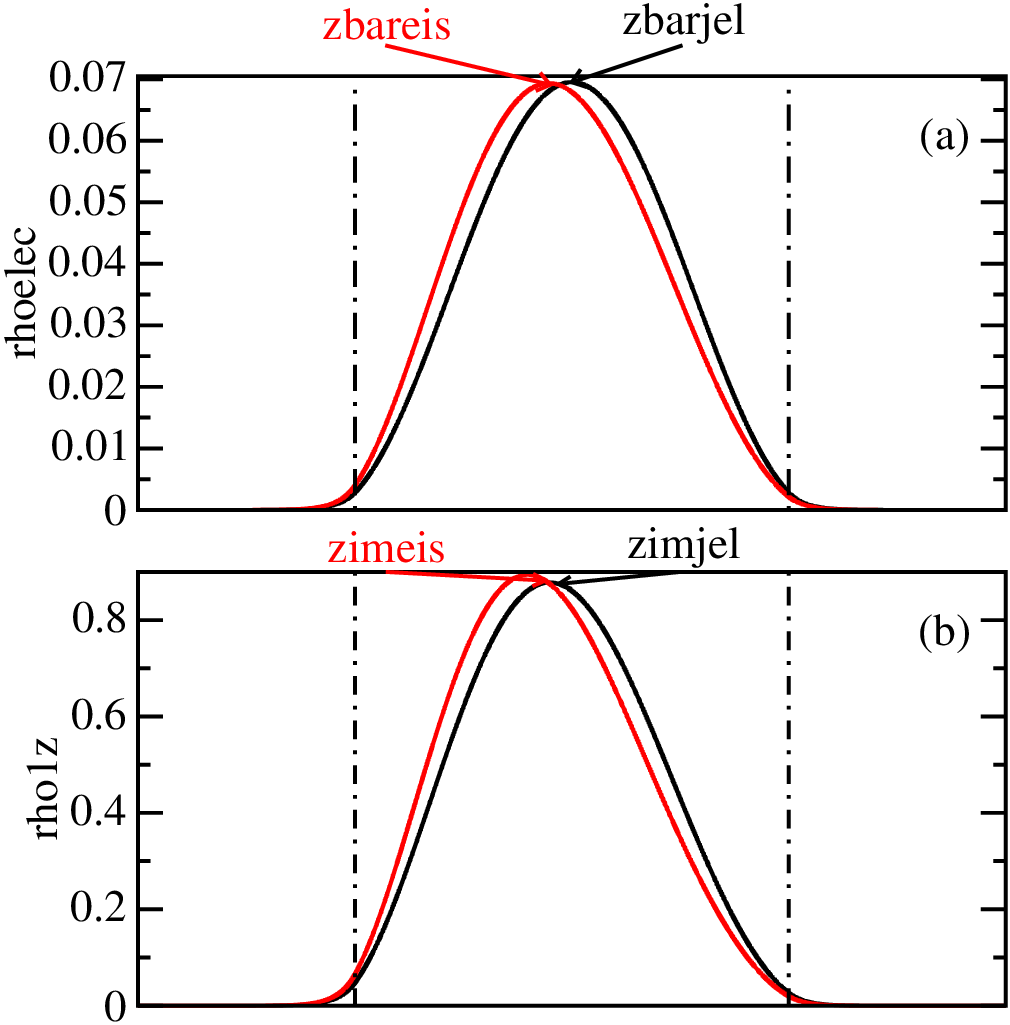}
       \caption{ (Color online) 
                 (a) Electronic charge density profile of the 
                 confined jellium model in a neutral configuration
                 (black solid line), and the traditional
                 quantum well model (red solid line).
                 (b) First order variation of the electronic charge density
                 in the two models.
                 Vertical dot-dashed lines represent the geometrical limits
                 of the slab.
                 The results have been obtained for a slab
                 of $w$ = 2.0 Bohrs, $m_{\parallel}$ = 
                 $\kappa=1$, $n = 0.079$ Bohr$^{-2}$,
                 and $V_{\rm ext}$ = -22 Ha. 
                 All magnitudes in atomic units
               }
       \label{fig:charge-density-jellium-qw}
    \end{center}
 \end{figure}

 This latter observation allows us to interpret the combined 
 effects of jellium compensation and wavefunction relaxation 
 (or lack thereof) in terms of an approximate model. In particular,
 for sufficiently small $\delta n$ we can assume, for the charged 
 capacitor at $n=n_{\rm jell} + \delta n$, that the relaxed 
 wavefunction is
 \begin{equation}
   \psi \sim \psi^{(0)}(n=0) + \delta n \psi^{(1)}(n=0). 
 \end{equation}
 Following a few derivation steps (see Appendix~\ref{app:relax-wavefun}),
 we could account for all the results presented so far.
 In particular, for both the t-QW and the j-QW models we find
 the same Hartree band-bending contribution,
 \begin{equation}
   \Delta^{\rm relax}_{\rm H} \sim \Delta^{\rm frozen}_{\rm H} + n z_{\rm B},
 \end{equation} 
 where $z_{\rm B}$ is defined as the dipole moment of 
 $\psi^{0}(z) \psi^{1}(z)$,
 \begin{equation}
    z_{\rm B} = \int dz \, z \psi^{0}(z) \psi^{1}(z).
    \label{eq:defzB}
 \end{equation} 
 Moreover, the value of $z_{\rm im}^*$ is twice as large in the t-QW case,
 \begin{equation}
    z_{\rm im}^{\rm t-QW} \sim 4n z_{\rm B}, \qquad 
    z_{\rm im}^{\rm j-QW} \sim 2n z_{\rm B},
 \end{equation}
 consistent with our self-consistent results 
 (see inset of Fig.~\ref{fig:results-jellium-well}).
 (To corroborate this point we report, in the same inset,
 the predictions of the approximate model for $z_{\rm im}$.) 
 Note that, within the
 approximations we used, the same quantity $z_{\rm B}$ 
 is responsible for both the upshift of
 $d\mu/dn$ and the off-centering of the electronic charge (image-plane effect).

 The different behavior of $z_{\rm im}^{\rm t-QW}$ and 
 $z_{\rm im}^{\rm j-QW}$, which constitutes the main result of this subsection,
 can be intuitively rationalized by observing the 
 respective evolution of the ground-state and first-order electronic
 states with $n$.
 In the t-QW model, the ground-state electronic wavefunction is 
 increasingly distorted for increasing $n$, since it
 is feeling a progressively stronger attraction by the classical
 electrode;
 conversely, in the j-QW case, the ground-state electronic 
 wavefunction is always symmetric,
 and with a shape that is roughly unsensitive to $n_{\rm jell}$ [see 
 Fig.~\ref{fig:charge-density-jellium-qw}(a)]. 
 This effect propagates to the first-order densities 
 [Fig.~\ref{fig:charge-density-jellium-qw}(b)], which in turn
 define $z_{\rm im}$ following Eq.~(\ref{eq:averagez}) and
 Eq.(\ref{eq:defzim}).
 In particular, in the t-QW case, the distortion of $\psi(n)$ adds up to the 
 contribution of the first-order wavefunctions, resulting in a 
 $\rho^{(1)}_{\rm el} (z)$
 whose off-centering is twice as large compared to the j-QW case.

 Before concluding this part, we shall take the opportunity to
 corroborate, in light of the numerical results presented so far, 
 an important point that we have already mentioned in 
 Sec.~\ref{sec:perturbation}: 
 Due to the variational character of the problem, 
 the relaxation of the wave functions always
 lowers the inverse capacitance density.
 This fact is immediately clear when we compare the  
 relaxed (solid black line) and unrelaxed (solid blue line) results
 for $\mathcal{C}^{-1}$ in Fig.~\ref{fig:results-jellium-well}. 
 As we have already emphasized, the electron compressibility
 $d\mu/dn$ shows an opposite trend:
 the relaxed (dashed black curve) results are systematically 
 more positive than the frozen-wavefunction values (blue curve). 
 This, however, is not in contradiction with the above 
 argument: 
 $d\mu/dn$, unlike $\mathcal{C}^{-1}$, cannot be written as
 a variational functional of $\psi^{(1)}$, and therefore is not 
 bound to decrease upon relaxation.

 \subsection{Jellium slab in the absence of strong confinement potential}
 \label{sec:jellium_unconstrained}
 
\begin{figure}
    \includegraphics[width=1.00\columnwidth]{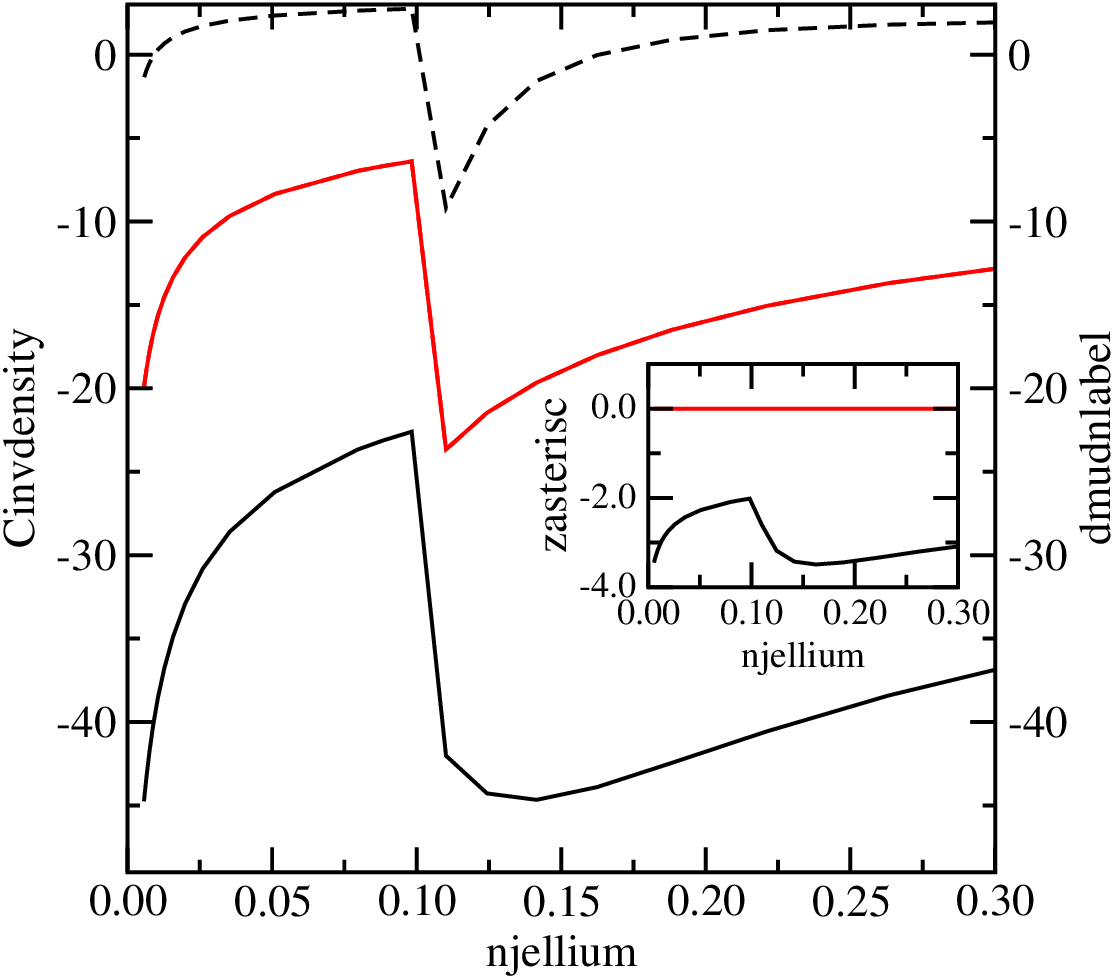} 
    \caption{ (Color online) 
              Quantum contributions to $\mathcal{C}^{-1}$ in the
              jellium slab in the absence of a strong confinement potential
              (solid lines; left y-axis) as a function of the 
              electronic density $n$.
              The solid black line is the fully relaxed wavefunction result.
              The solid red line represents the results at the
              frozen wavefunction level.
              The black dashed line represents the electron compressibility 
              [Eq.~(\ref{eq:invc1band})] (right y-axis).
              Inset: evolution of $z_{\rm im}^{\ast}$ as a function of $n$.
              Numerical results have been obtained for
              $m_{\parallel} = \kappa = 1$,
              $w$ = 2.0 Bohr.
            }
    \label{fig:results-jellium-nowell}
 \end{figure}

 In Fig.~\ref{fig:results-jellium-nowell} we plot
 the behavior of the inverse of the capacitance density
 as a function of $n$ for a 2.0 Bohr-thick jellium slab, this
 time \emph{without} the confinement potential; this means that
 here the electron gas is only kept in place by the positive 
 jellium background.
 As before, we also show the Fermi-level contribution, 
 the frozen-WF results and the image-plane location (inset).
 Remarkably, $\mathcal{C}^{-1}$ is much more negative than in the 
 confined case discussed in the previous Section [note the difference in
 vertical scale compared, e.g., with Fig.~\ref{fig:results-jellium-well}].
 This fact is already clear at the frozen-WF level, and is further
 enhanced by an unusually large negative contribution from $z_{\rm im}$.
 Moreover, at a density of about $n_{\rm crit}\sim 0.10$ a.u. 
 we observe a dramatic dip 
 in both $\mathcal{C}^{-1}$ and $d\mu/dn$; such discontinuity, as 
 we shall see shortly, is due to the transition from the one-band to
 the two-band regime. 
 (Because of the weaker confinement, the energy separation
 between the different subbands is much smaller than in the previous examples, 
 and multiple subbands may become occupied even at moderate values of the 
 in-plane charge density.)
 We shall discuss these two regimes separately in the following.

\begin{figure}
    \begin{center}
 \psfrag{phiz}[cc][cc]{$\psi(z)$}
 \psfrag{Deltarho}[cc][cc]{$\rho_{\rm el}^{(1)} (z)$}
 \psfrag{DeltaVH}[cc][cc]{$\left( V_{\rm H}-V_{\rm right}\right)^{(1)} (z)$}
       \includegraphics[width=\columnwidth]{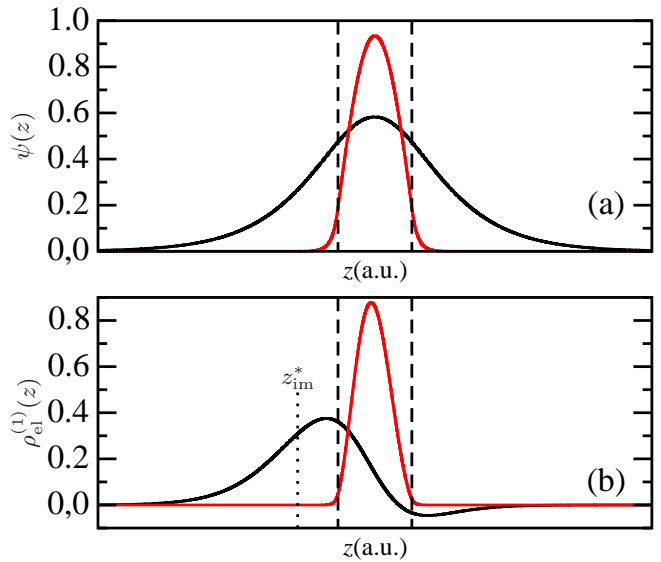}
       \caption{ (Color online) 
                 (a) Real space wave function of
                 the lowest subband in a jellium slab without (black
                 solid line) or with (red solid line) an extra confinement 
                 potential in the quantum well.
                 (b) Profile of the total derivative of the self-consistent
                 electronic charge density with and without the confinement 
                 potential. 
                 $z_{\rm im}^{\ast}$ represents the location of the image charge
                 plane, defined as in Eq.~(\ref{eq:defzim}).
                 The results have been obtained for a slab
                 of $w$ = 2.0 Bohrs, $m_{\parallel}$ = $\kappa =1$, 
                 $V_{\rm ext}$ = -22 Ha,
                 $r_{\rm s}$ = 2.0. All magnitudes in atomic units.
               }
       \label{fig:comparpozo}
    \end{center}
 \end{figure}

 \begin{figure}
    \begin{center}
       \includegraphics[width=\columnwidth]{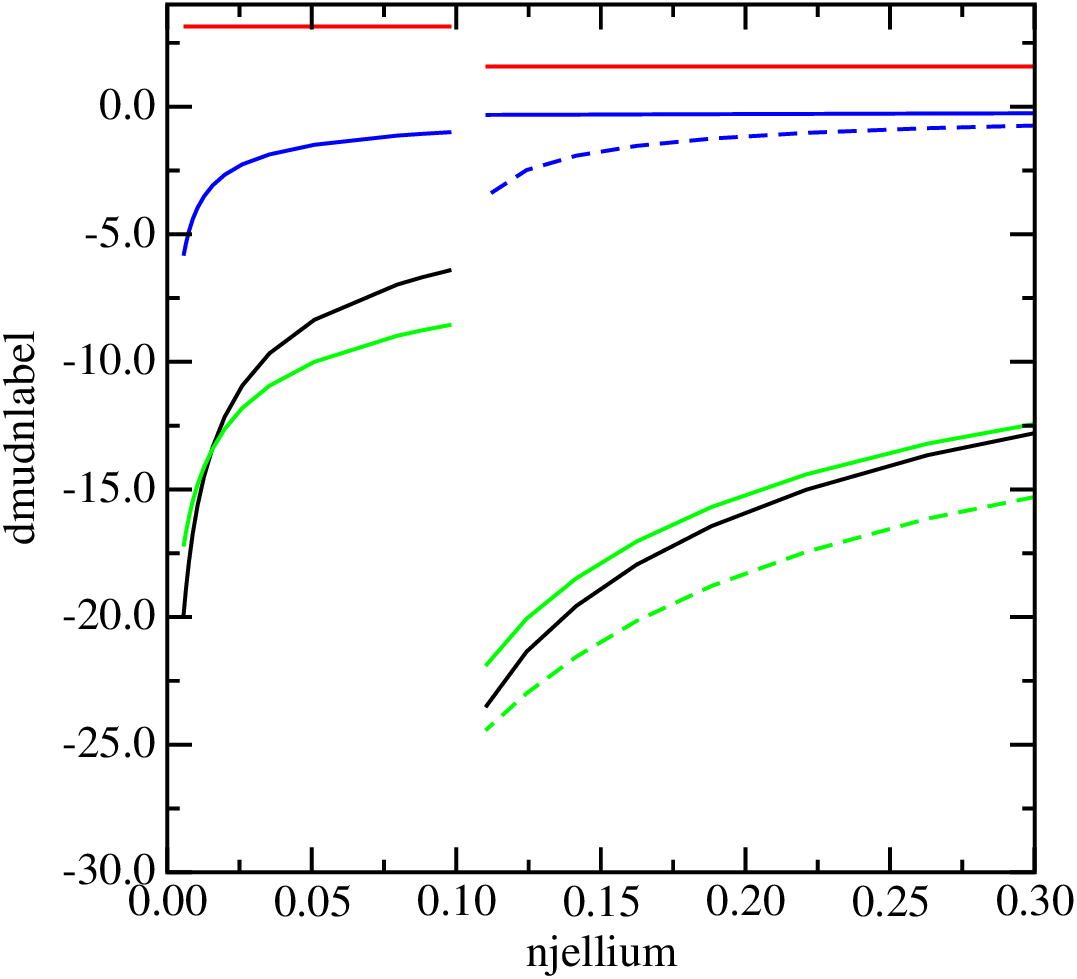}
       \caption{ (Color online) 
                 Evolution of the electron compressibility $d\mu/dn$ as a
                 function of the charge density in the jellium slab 
                 in the absence of an external confinement potential
                 (black line), 
                 and its decomposition into 
                 Kinetic ($\pi/m_{\parallel}$; red), 
                 Hartree 
                 [$\Delta^{\rm right}_{\rm H}(l)$; green], and
                 exchange-correlation [$\Delta_{\rm xc}(l)$; blue]
                 contributions, where $l$ is an index for the occupied subband.
                 The Hartree and exchange-correlation contributions 
                 coming from the first band are plotted in solid lines,
                 while the contributions from the second band are represented
                 by dashed lines.
                 The electron wave functions are frozen 
                 to the ones obtaied in a neutral slab $n = n_{\rm jell}$.
                 Results have bee obtained for a slab
                 of $w$ = 2.0 Bohr, $m_{\parallel}$ = $\kappa =1$.
               }
       \label{fig:evoleigen-nowl-frozen-relax}
    \end{center}
 \end{figure}

 \subsubsection{Low-density regime}
 \label{sec:low-density}
 
 In the dilute regime (for low enough charge densities),
 only the lowest subband is occupied, as it happened in all the models
 analyzed up to now.
 The ground-state wavefunction, however, is now
 significantly more extended [Fig.~\ref{fig:comparpozo}(a)], and also more 
 prone to be deformed when the capacitor is charged 
 [Fig.~\ref{fig:comparpozo}(b)]; 
 this implies that all the effects discussed in the previous sections are 
 dramatically amplified.
 
 At the frozen-WF level, as we pointed out in Sec.~\ref{sec:finitethickness}, 
 the increase in the (negative) Hartree band-bending term 
 (green line in Fig.~\ref{fig:evoleigen-nowl-frozen-relax})
 largely overcompensates for the
 reduction in the exchange contribution due to Coulomb softening
 (blue line in Fig.~\ref{fig:evoleigen-nowl-frozen-relax}).
 This results in a drastic decrease in the electron compressibility 
 compared with a j-QW model of the same thickness. 

 Upon relaxation, while the electron compressibility becomes positive,
 the overall inverse capacitance further decreases 
 (solid black line in Fig.~\ref{fig:results-jellium-nowell})
 compared to the frozen-WF case, in agreement with the behavior of j-QW model.
 Interestingly, however, here $z_{\rm im}$ is substantially 
 more shifted towards the classical electrode, and follows a 
 qualitatively different trend in the limit of small $n$.
 ($z_{\rm im}$ is a roughly linear function of $n$ in the j-QW 
 case, while here it appears to diverge towards $-\infty$.)
 Both features, as we said, are due to the absence of a
 confining potential; this greatly enhances the charge-density
 response of the system to an external bias, and the more so
 in the dilute limit (the electrostatic attraction
 due to the jellium background vanishes
 for $n\rightarrow 0$).
 
 \subsubsection{High-density regime}
 \label{sec:band-crossing}

 \psfrag{eps1}[cc][cc]{$\epsilon_{1}$}
 \psfrag{eps2}[cc][cc]{$\epsilon_{2}$}
 \psfrag{musymbol}[cc][cc]{$\mu$}
 \psfrag{wavefun}[cc][cc]{$\psi_{i}$}
 \psfrag{eps2rs0}[ll][ll][0.60]{$\epsilon_{2}(r_{s}, \sigma_{\rm ext}= 0)$}
 \psfrag{kparallel}[cc][cc]{${\bf k_\parallel}$}
 \psfrag{geDOS}[cc][cc]{$g(E)$}
 \begin{figure}
    \begin{center}
       \includegraphics[width=\columnwidth]{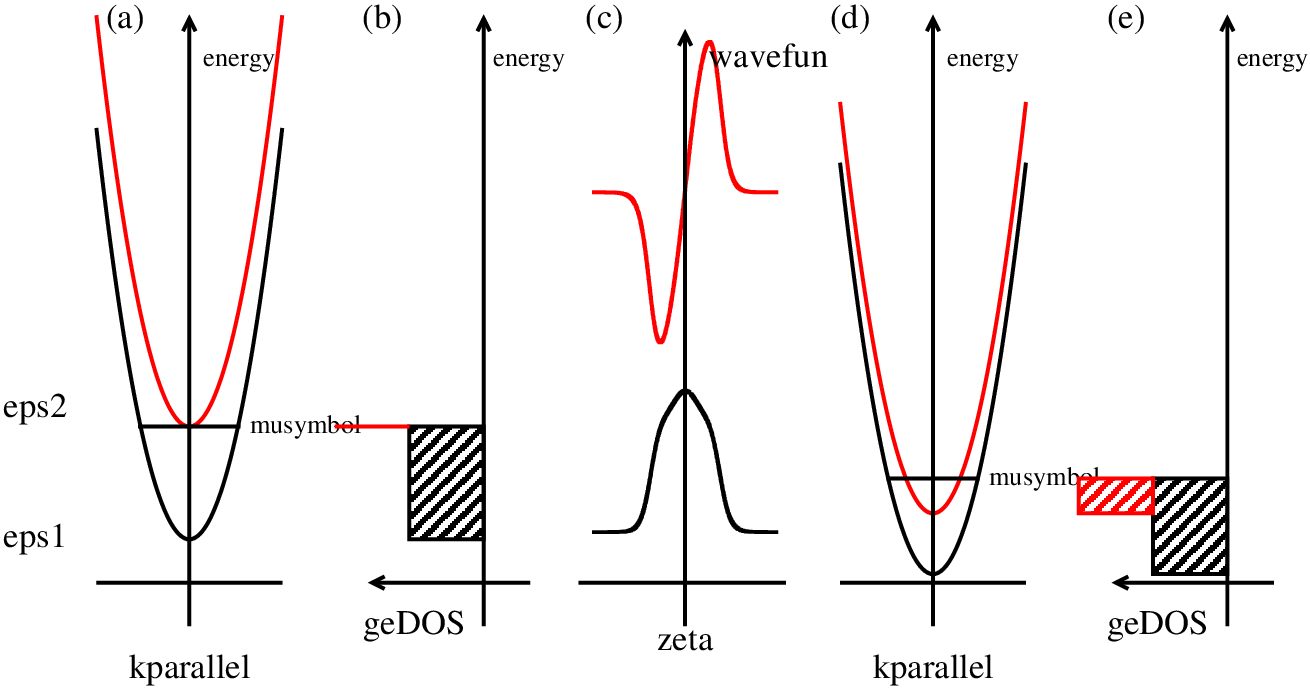}
       \caption{ (Color online) 
                 (a) Energies of the first (black solid line) 
                 and second (red solid line) subbands 
                 as a function of the parallel $k$-point.
                 (b) Contribution of the two subbands to the 
                 density of states, $g(E)$, at the critical density 
                 when the energy 
                 of the highest occupied state in the first subband 
                 equals the eigenvalue of the second subband.
                 (c) Radial shape of the eigenfunction of the one-dimensional
                 Schr\"odinger equation along $z$ for the two subbands.
                 Panels (d) and (e) are the same as (a) and (b), but 
                 for an increased charge density that produces the lowering of
                 the chemical potential.
                 Note how, despite the fact that the charge density is 
                 larger in (d)-(e) (larger occupied areas of the shaded
                 rectangles), the chemical potential $\mu$ is lower than
                 in (a)-(b).
               }
       \label{fig:twobandmodel}
    \end{center}
 \end{figure}

 Due to the absence of a confinement potential,  
 there is a critical density [see the schematic illustration 
 in Figs.~\ref{fig:twobandmodel}(a)-(b)], 
 where the Fermi level crosses the eigenvalue of the 
 first excited subband.
 For larger densities, the two lowest subbands are partially occupied.
 As we have already demonstrated in Sec.~\ref{sec:multiband}, 
 all contributions to the inverse 
 capacitance are discontinuous at the transition. We shall illustrate them
 in detail hereafter, by framing our discussion around the frozen-WF results
 (Fig.~\ref{fig:evoleigen-nowl-frozen-relax}).
 These provide a clearer insight on the underlying physics;
 on the other hand, wavefunction relaxation does not introduce anything new that
 hasn't been discussed in the earlier paragraphs.

 \psfrag{dn1dn2}[cc][cc][1.20]{$dn_{l}/dn$}
 \begin{figure}
    \begin{center}
       \includegraphics[width=\columnwidth]{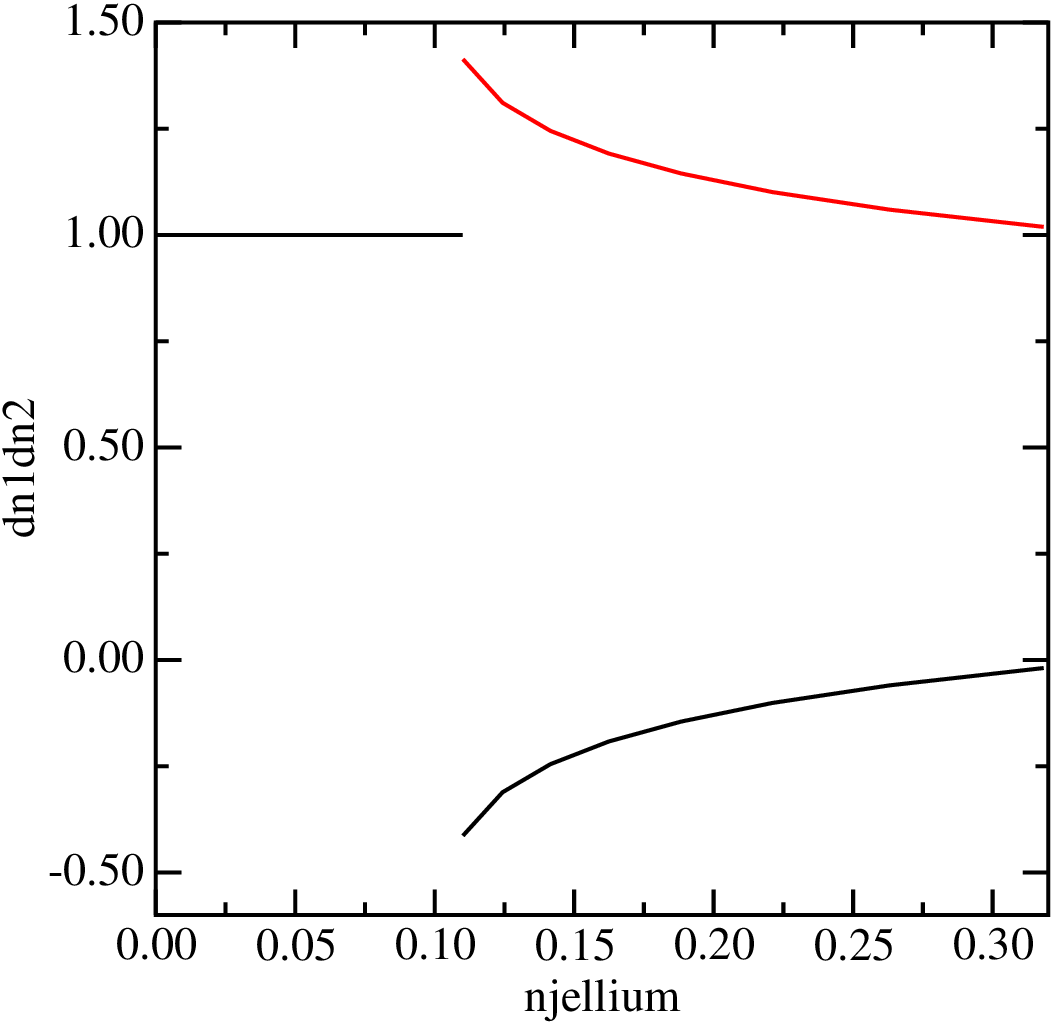}
       \caption{ (Color online) Filling rate of the first (black line)
                 and second (red line) subbands as a function of $n$
                 for a jellium slab of $w$ = 2.0 Bohrs
                 with frozen wave functions.
                 (The individual curves are $dn_{l}/dn$, and must sum to unity
                 as $\sum_{l} n_{l} = n$).
               }
       \label{fig:dn1dn2}
    \end{center}
 \end{figure}

 First, since we are considering that the in-plane
 effective masses are the same for all the bands,
 the in-plane kinetic energy term decreases by
 a factor of two at the transition (red segments in 
 Fig.~\ref{fig:evoleigen-nowl-frozen-relax}), and is
 otherwise constant within either regime. 
 The remainder of $d\mu/dn$ originates from the
 variation with $n$ of the Hamiltonian eigenvalues, and
 can be further split into Hartree and exchange-correlation
 contributions of each occupied band.
 [In particular, one takes the weighted average of
 the contributions of all occupied bands,
 according to Eq.~(\ref{eq:C-1multiband}).]

 \psfrag{Psil}[cc][cc]{$\psi_{l} (z)$}
 \psfrag{Psilsquare}[cc][cc]{$\left| \psi_{l} \right|^{2} (z)$}
 \psfrag{Potentials1}[cc][cc]{$\left( V_{\rm H}-V_{\rm right}\right)^{(1)}$}
 \psfrag{zeta}[cc][cc]{$z$}
 \begin{figure}
    \begin{center}
       \includegraphics[width=\columnwidth]{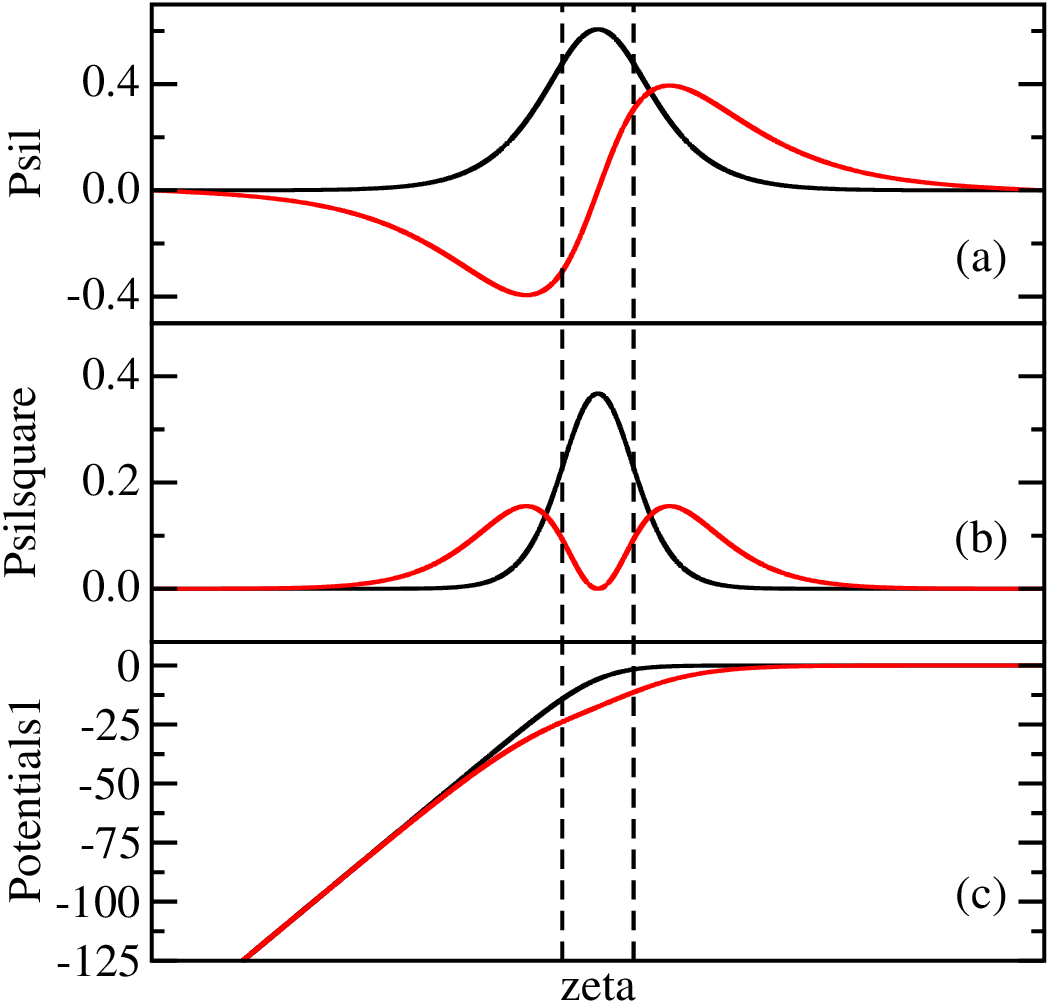}
       \caption{ (Color online) 
                 (a) Real-space wave functions of the lowest-energy subband
                 (black line)
                 and the first excited state (red line).
                 The corresponding contributions to the electronic charge
                 density is shown in panel (b).
                 (c) Profile of the first-order change in the 
                 electrostatic potential obtained by a double integral
                 of $|\psi_1(z)|^2$ (black line) and $|\psi_2(z)|^2$
                 (red line). 
                 The electrostatic potential is measured with respect to the
                 right, as this is the quantity that enters in the 
                 determination of the electron compressibility.
                 Dashed vertical lines mark the geometrical limits of the
                 slab.
                 The results have been obtained for a slab
                 of $w$ = 2.0 Bohrs, $m_{\parallel}$ = $\kappa =1$, 
                 and $r_{\rm s}$ = 1.7. All magnitudes in atomic units.
       }
       \label{fig:perturb-twobands}
    \end{center}
 \end{figure}

 As illustrated in Fig.~\ref{fig:evoleigen-nowl-frozen-relax}, 
 the drop of the electron compressibility right after the critical 
 density $n_{\rm crit}$ is clearly dominated by the electrostatic contribution,
 which undergoes an abrupt drop at the transition.
 Interestingly, above $n_{\rm crit}$ the Hartree contributions of 
 both bands, $\Delta^{\rm right}_{\rm H}(1)$ and 
 $\Delta^{\rm right}_{\rm H}(2)$, are equally 
 large and negative. 
 To see why, we need to look at the behavior of the first-order 
 Hartree potential $\hat{V}_{\rm H}^{(1)}$ [recall Eq.(\ref{delright})], 
 which is in turn related  to the first-order electronic
 density, $\rho_{\rm el}^{(1)}(z)$ via a Poisson equation [Eq.(\ref{vh1})]. 
 Below the transition
 $\rho_{\rm el}^{(1)}(z)$ corresponds the square 
 modulus of the only occupied band,
 \begin{equation}
 \rho^{(1)}_{n<n_{\rm crit}}(z) = |\psi_1(z)|^2.
 \end{equation}
 [Recall that we are working within the frozen-WF regime, which 
 implies discarding the contribution of the first-order $\psi$ 
 in Eq.~(\ref{eq:rho1el}).]
 Conversely, above the transition we have
 \begin{equation}
 \rho^{(1)}_{n>n_{\rm crit}}(z) = \frac{dn_1}{dn}|\psi_1(z)|^2 + \frac{dn_2}{dn}|\psi_2(z)|^2, 
 \label{eq:rho1nlarge}
 \end{equation}
 i.e., the first-order density (and hence, the first-order potential) is now
 a linear combination of the squared moduli of the individual wavefunctions.
 [Note that the coefficients $dn_l/dn$ must sum up to 1, since $n = \sum_l n_l$.]
 Remarkably, while $n_l$ are continuous functions of $n$ everywhere, their derivative
 is discontinuous at the transition. 
 As shown in Fig.~\ref{fig:dn1dn2},
 right above $n_{\rm crit}$ the filling rate of the first band becomes \emph{negative},
 while the second band hosts the entirety of the excess charge \emph{plus} the
 amount that is being transferred from the first band.

 This abrupt change in the filling rate would have no effect whatsoever on the 
 Hartree band bending if $|\psi_2(z)|^2$ were equal to $|\psi_1(z)|^2$.
 The key point is that $|\psi_2(z)|^2$ is a much broader 
 charge distribution than $|\psi_1(z)|^2$, as one can clearly see
 from Fig.~\ref{fig:perturb-twobands}(b). 
 This means that $V^{(1)}(z)$, which is related to
 $\rho^{(1)}(z)$ via a double integration, will be significantly 
 more negative for $n>n_{\rm crit}$ than for $n<n_{\rm crit}$
 in a neighborhood of the jellium slab [see Fig.~\ref{fig:perturb-twobands}(c)].
 As a consequence, both $\Delta_{\rm H}(1)$ and $\Delta_{\rm H}(2)$, defined
 as the mean value of $V^{(1)}(z)$ on the first and second eigenstate of
 the ground-state Hamiltonian, undergo an abrupt decrease at $n=n_{\rm crit}$.
 In other words, one can say that right at the transition the
 quantum electrode starts to accummulate charge into the second
 quantum state; this is spatially much broader than the lowest band,
 resulting in a greatly enhanced Hartree band-bending effect, 
 consistent with what we have seen in all examples discussed so far
 (the thicker the electron gas, the more negative its compressibility).

 This effect, of course, can be maintained only in a limited range of $n$
 values.
 As the jellium charge density progressively increases, 
 the electrostatic potential that keeps the electrons in place
 becomes deeper, the eigenfunctions shrink accordingly and the
 Hartree band-bending effect becomes weaker.
 In fact, the same mechanism occurs in the dilute regime discussed 
 above, where $\Delta_{\rm H}(1)$ shows an analogous monotonic increase with $n$.
 This is yet another consequence of the absence of a confining potential: 
 Recall that in the j-QW model $\Delta_{\rm H}(1)$ is a \emph{constant} at the frozen-WF level,
 as the ground-state wavefunctions undergo negligible changes with $n$.
 Note that the same arguments are equally valid whenever a new band starts to
 be populated; however, we expect the enhancement of $d\mu/dn$ to become progressively 
 smaller as the number of degrees of freedom increases (recall that the overall $d\mu/dn$ 
 is written as a weighted average over all bands).

 \subsection{Asymmetric confinement}
 \label{sec:asymmetric}

 \begin{figure}
    \begin{center}
       \includegraphics[width=\columnwidth]{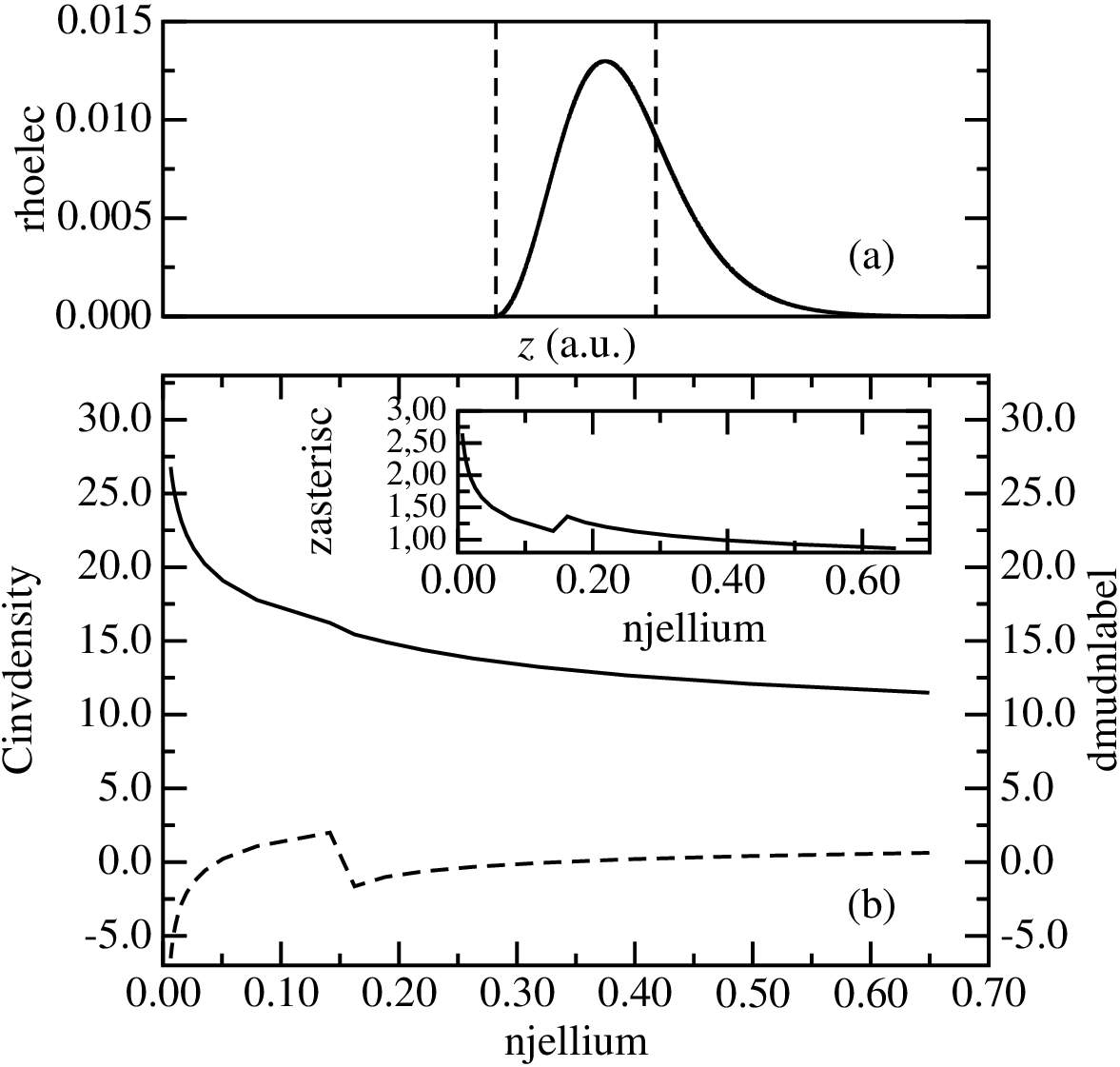}
       \caption{ (a) Electronic charge density in a jellium slab with 
                 an Heaviside step potential at the surface closer to the
                 classical electrode.
                 Vertical dashed lines represent the geometrical surfaces
                 of the slab.
                 The inverse capacitance density (solid line, left y-axis),
                 and the electron compressibility (dashed line, right y-axis)
                 as a function of the charge density 
                 is shown in panel (b). 
                 Inset: Position of the image charge plane with respect the
                 geometrical center of the quantum electrode.
                 The results have been obtained for a slab of 
                 $w = 3.1$ Bohr of thickness, 
                 $r_{\rm s}$ = 3.0 [panel (a)], and
                 $m_{\parallel}$ = $\kappa$ = 1.
               }
       \label{fig:asymmetry}
    \end{center}
 \end{figure}

 Since part of the renewed interest on negative compressibility is due to 
 some recent experiments carried out in polar interfaces between
 two insulators (SrTiO$_{3}$/LaAlO$_{3}$),~\cite{Li-11}
 it is important to discuss, at least qualitatively, how
 the effects described in the previous Sections manifest themselves 
 in the latter context.
 As suggested in Ref.~\onlinecite{Stengel-11.2}, the role 
 of LaAlO$_{3}$ is primarily to confine the conduction 
 electrons to the SrTiO$_{3}$ side, and define the electrostatic boundary 
 conditions via an external surface charge density.
 This can be effectively modeled, within the methodology developed in 
 this work, by using an \emph{asymmetric} confining potential 
 of Heaviside type, while controlling the electrical boundary conditions 
 via the parameter $n$.
 In addition, we shall include a thin jellium slab on the ``SrTiO$_3$'' side
 of the Heaviside potential, to guarantee a stable numerical solution.
 (As we shall see, the presence of the compensating jellium has little 
 impact on the results at the qualitative level; it can be physically thought as 
 a finite density of dopants that diffuse through the interface because of
 ``intermixing''.)
 This way, the electronic charge is confined only on one side of the 
 slab, while it is free to relax on the other side.
 This asymmetry is clear in the spatial distribution of
 the ground-state electronic density, Fig.~\ref{fig:asymmetry}(a). 
 
\begin{figure}
    \begin{center}
       \includegraphics[width=\columnwidth]{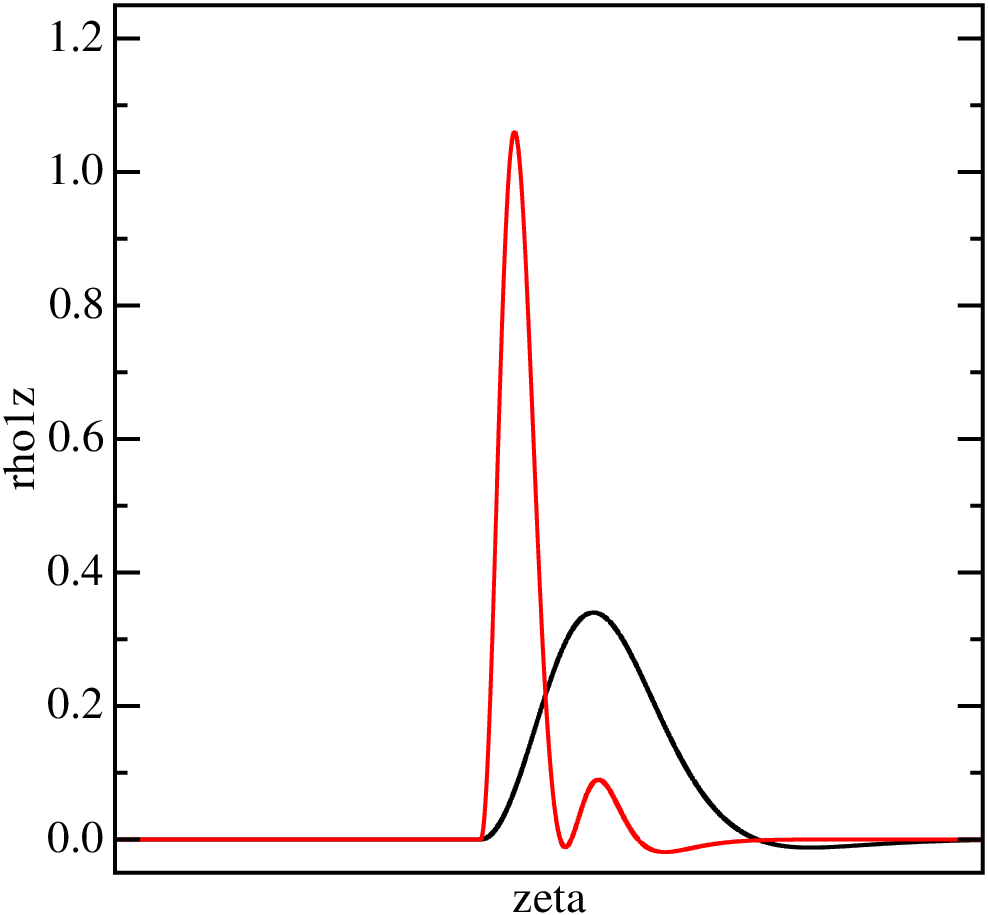}
       \caption{ (Color online) Profile of the total derivative 
                 of the self-consistent electronic charge density in a 
                 jellium with an asymmetric confinement potential.
                 Black line obtained for $r_{\rm s}$ = 7.0, and 
                 red line for $r_{\rm s}$ = 0.7.
                 The rest of the parameters as in
                 Fig.~\ref{fig:asymmetry}.
                 All magnitudes in atomic units.
               }
       \label{fig:rho1-asym}
    \end{center}
 \end{figure}

 The calculated behaviour of the inverse capacitance density 
 is shown in Fig.~\ref{fig:asymmetry}(b), together with
 the usual decomposition into Fermi-level and image-charge contributions.
 Remarkably, the \emph{total} $\mathcal{C}^{-1}$
 undergoes a monotonic decrease with the charge density $n$,
 i.e. it shows an opposite trend
 than in all the situations analyzed up to now.
 This is clearly due to to the evolution of the image charge plane
 contribution, shown in the inset.
 (The electron compressibility displays
 roughly the same behaviour as the one described in 
 Sec.~\ref{sec:band-crossing}.)
 In the low-density regime the presence of the
 Heaviside potential prevents the wave function from occupying
 the region between the two plates, while the weaker and 
 weaker confining potential of the jellium lets the electronic
 density spread arbitrarily far to the other side (see 
 Fig.~\ref{fig:rho1-asym}). This translates into a divergence of 
 $z^{\ast}_{\rm im}$ in the dilute limit, in stark contrast with the QW 
 cases discussed earlier. 
 At higher densities, higher subbands become occupied; their 
 nodes lead then to oscillations 
 in the first-order charge density (see red line in Fig.~\ref{fig:rho1-asym}),
 effectively pushing its center of mass closer and closer to the
 electrode surface.
 This effect, however, cannot overcome the surface confining barrier, 
 which means that, in the limit of large $n$, $z^{\ast}_{\rm im}$
 will slowly approach its asymptotic value of zero.
 Note that $z^{\ast}_{\rm im}$ (and hence the total inverse capacitance)
 is always \emph{positive} in Fig.~\ref{fig:asymmetry}(b).
 This is mostly a matter of convention, though:  
 Due to the geometry of the problem we find it more appropriate here to 
 use the potential step as a reference to define the classical capacitance,
 unlike in the symmetrical QW cases of the previous sections.

\begin{figure}
    \begin{center}
       \includegraphics[width=\columnwidth]{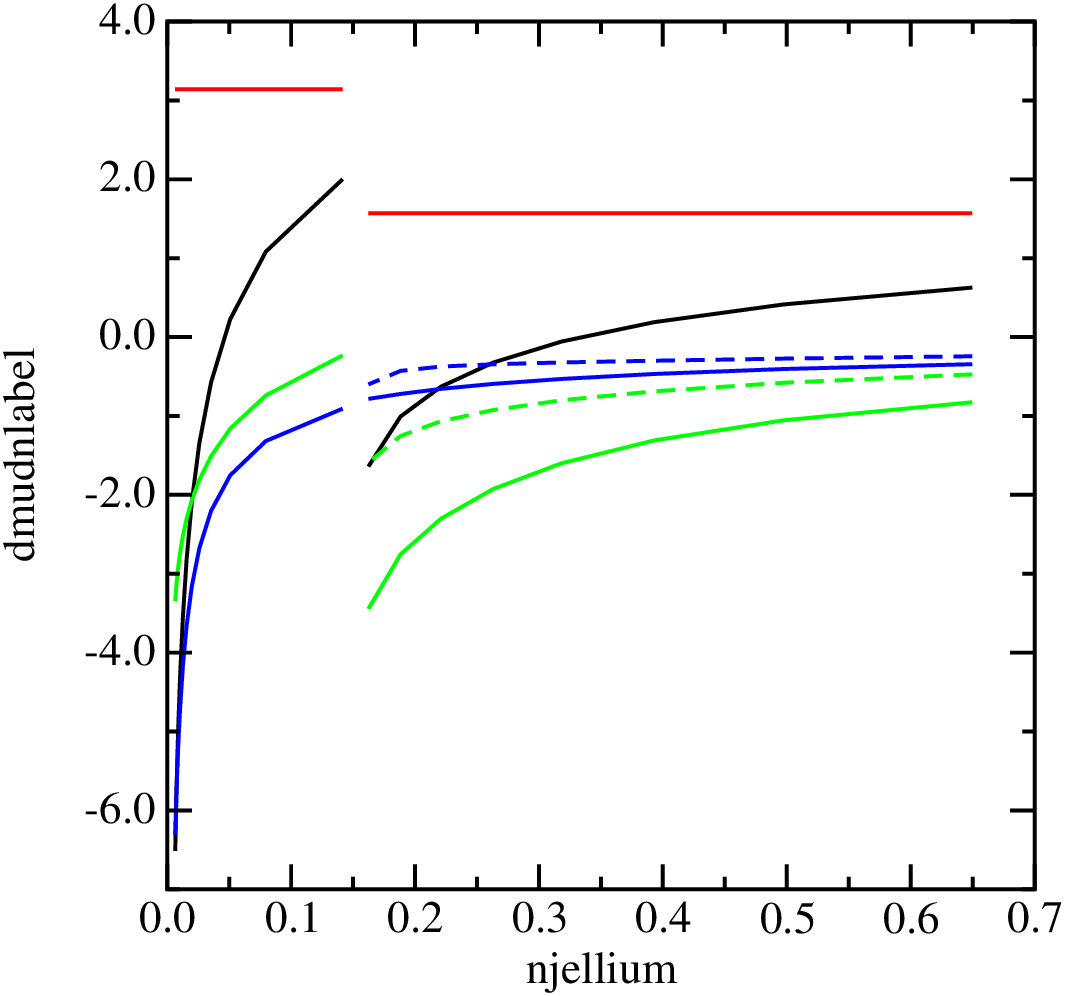}
       \caption{ (Color online) Evolution of the electron compressibility
                 $d\mu/dn$ as a function of the charge density in the jellium
                 slab with an asymmetric potential. Meaning of the lines as
                 in Fig.~\ref{fig:evoleigen-nowl-frozen-relax}.
                 Results have bee obtained for a slab
                 of $w$ = 3.1 Bohr, and $m_{\parallel}$ = $\kappa =1$.
                 All magnitudes in atomic units.
               }
       \label{fig:decom-asym}
    \end{center}
 \end{figure}

 As we said above, the Fermi-level contribution qualitatively follows the
 same trend as in the cases discussed earlier. To substantiate this point, we
 show the breakdown of $d\mu/dn$ into the individual contributions in
  Fig.~\ref{fig:decom-asym}.
 (This analysis is in all respects analogous to that of 
 Fig.~\ref{fig:evoleigen-nowl-frozen-relax},
 with the sole difference that here we discuss the \emph{relaxed-wavefunction}
 values, rather than the frozen ones.)
 As in Fig.~\ref{fig:evoleigen-nowl-frozen-relax},
 transition from the single- to multiple-band regime produces a 
 clear discontinuity,
 which is driven by a simultaneous reduction of the in-plane kinetic energy and 
 a dip in the electrostatic Hatree band-bending terms.
 Interestingly, Fig.~\ref{fig:decom-asym} shows an apparent reversal of the
 role of the bands above the transition. 
 (In Fig.~\ref{fig:evoleigen-nowl-frozen-relax}
 $\Delta_{\rm H}(1)$ is always less negative than $\Delta_{\rm H}(2)$, 
 while here the opposite is true.)
 This should not be misinterpreted, though: Indeed, the upper band is still 
 the primary responsible for the discontinuous change in $V_{\rm H}^{(1)}$ 
 across the transition. Yet, $\Delta_{\rm H}(l)$ corresponds to the 
 mean value of $V_{\rm H}^{(1)}$ on the $l$-th eigenfunction;
 the first band is spatially located closer to the
 surface, and hence can probe a region where $V_{\rm H}^{(1)}$ is deeper.
 In any case, the discontinuity almost disappears when the contributions of 
 $z^{\ast}_{\rm im}$ and $d\mu/dn$ are summed up, 
 leading to a simple monotonic behavior 
 of $\mathcal{C}^{-1}$.

 \section{Discussion}
 \label{sec:discussion}

 The fact that the inverse capacitance density is always positive in our 
 asymmetric confinement model appears, at first sight, problematic in 
 light of the experimental results of Refs.~\onlinecite{Li-11,Tinkl-12}. 
 Our results also starkly disagree with the conclusions
 of Kopp and Mannhart~\cite{Kopp-09} regarding the purported universality 
 of the quantum capacitance effect. 
 On the contrary, here we find a remarkable variability in the physical behavior
 of each individual system depending on geometry, confinement and other factors. 
 Interestingly, most of this variability is carried by two contributions 
 that are electrostatic in nature: the Hartree band-bending and the 
 image-charge effects. 
 This result disproves earlier assumptions that exchange and kinetic effects 
 wouls dominate~\cite{Kopp-09},
 and prompts to a profound rethinking of the established interpretations.

 Regarding the LaAlO$_{3}$/SrTiO$_{3}$ experiments,~\cite{Li-11,Tinkl-12} 
 we regard it as highly unlikely that models based on the free-electron gas 
 such as those presented in this work (or in KM) will be able to explain the 
 observed effects.
 Here, there are several indications that a more sophisticated description of
 electron correlations might be needed.
 For example, Ti 3$d$ orbitals in oxides are known~\cite{Mannhart-10} to form a 
 two dimensional electron liquid rather than an electron gas;
 strong electronic correlations may then localize the electrons and form 
 polaronic quasiparticles.
 In this regime, the localized charges could couple with image charges in the
 other metal electrode and produce a dramatic increase of the 
 capacitance as suggested by Skinner and Shklovskii.~\cite{Skinner-10}
 (This model has been also recalled to explain the larger capacitances
 with respect the classical values in gated carbon 
 nanotubes,~\cite{Fu-15}
 or in a black phosphorus thin film sandwiched between two layers of hexagonal 
 boron nitride with a few-layer graphene as terminal electrodes.~\cite{Wu-16})
 As we said, such scenarios are far beyond the range of 
 applicability of our model, though.
 More accurate simulations, e.g. explicitly including  the underlying 
 atomic structure 
 and the strong electron-electron interactions, would be desirable in the future
 to settle these important points.

 Our new predictions for a negative electron compressibility 
 that is associated with the population of higher subbands 
 could be experimentally verified
 even in traditional semiconductor quantum wells (QWs).
 In fact some of our results might be behind the first-order phase transition
 observed in GaAs/Al$_x$Ga$_{1-x}$ QWs, when the first excited subband
 is occupied with electrons, as the Fermi level is tuned into
 resonance with the excited subband by applying a dc voltage.~\cite{Goni-02}
 In Ref.~\onlinecite{Goni-02}, the transition was attributed to a
 discontinuous jump of the exact 2D exchange potential every time a
 subband is occupied.
 A discontinuity in the exchange-correlation potential upon adding electrons
 to the ground-state is, in principle, not expected within LDA;  
 and indeed, in our calculations we observe a discontinuity in
 the \emph{derivative} of the band occupations  
 (and hence of the band eigenvalues) with respect to $n$, 
 while the band eigenenergies are always continuous functions of $n$.
 At first sight, this seems to imply that exact exchange is crucial 
 to obtaining a qualitative agreement
 with the experiments. However, we believe that this need not be the case,
 and that our results may even 
 provide an alternative explanation to the experimental observations of 
 Ref.~\onlinecite{Goni-02}. 
 The key point is that the voltage, rather than the electron density,
 is controlled in the experiments of Ref.~\onlinecite{Goni-02}.
 We stress that, based on our results of  
 Sec.~\ref{sec:jellium_unconstrained},
 the exchange energy and potentials are largely irrelevant in 
 determining the quantum capacitance in the weak confinement limit.
 In such a regime, the physics is dominated by Hartree band-bending effects, 
 which are electrostatic in nature. Based on the above arguments, 
 the latter are therefore sufficient to produce a strong
 first-order transition as a function of the applied voltage when the 
 Fermi level approaches the second subband.

 Another remarkable feature of our simulations is a very
 large enhancement of the susceptibility in the negative capacitance regime.
 In a neutral jellium slab,
 the lowest and the first excited bands display opposite parities,
 and the total density is symmetric with respect the center of the slab.
 If an external electric field is applied, like the one produced by
 the classical plate after charging the slab,
 then the two bands can be hybridized, leading to an asymmetric
 electronic cloud as reflected by $z_{\rm im}$ in our
 simulations.
 This effect is particularly large when the first excited band starts
 filling and the two bands are essentially degenerate:
 the hybridization is almost costless and yields a large change in the
 dipole, that could be measured as a huge enhancement of the
 susceptibility of the system.
 Recent experiments in ferroelectric superlattices
 support the idea that the presence of regions with very
 large susceptibilities, like interfaces
 and domain walls~\cite{Zubko-16,Yadav-19},
 are fundamental to understand the boost of the capacitance
 in these systems. 

 Also, the aforementioned systems might be a perfect playground
 to check the role played by the effective mass or the dielectric
 constant of the medium in the description of real physical systems.
 It seems unlikely to us that the contribution of the valence electrons to the
 many-body interactions within the gas of carriers can be summarized
 by a single number (the effective dielectric constant).
 Even more, assuming that this approximation is valid, 
 one can wonder what dielectric constant should be used:
 the static (including the response of the underlying lattice),
 or the high-frequency (taking into account only the electronic response).
 While the ions shouldn't mediate  
 the exchange and correlation effects between the electrons, 
 the use of the static dielectric constant has been the common approach 
 in the semiconductor community.
 Such an approach appears problematic, however, in 
 oxides, where the static and the high-frequency dielectric
 constants may differ by orders of magnitude.
 As the dielectric constant appears at the denominator in 
 the Hartree and exchange-correlation energies, then all the 
 useful (old and new) mechanisms that we discussed in this work
 would be suppressed and the system would behave like a free-electron gas.
 An insight on these issues would require the use of sophisticated
 many-body techniques; we regard it as an interesting topic for future studies. 

 \section{Conclusions}
 \label{sec:conclusions}

 Since the milestone works in semiconductor quantum wells aimed to 
 understand the quantum Hall effect,~\cite{Eisenstein-94}
 the model based on the competition of the kinetic energy
 with the quantum exchange-energy in the 
 electron-electron interactions has been pointed out as the root
 of the negative electron compressibility in two-dimensional metals.
 These have been applied even in systems as different as two-dimensional
 electron gases at oxide interfaces.~\cite{Li-11}

 In this work we have proven how they must be taken with care when other
 systems are studied, such as delta-doping layers or two-dimensional metals
 where the confinement of the electrons to the well might be much smaller than
 in the previous systems.
 In such a situation, the kinetic and the exchange-correlation effects 
 are not the only pieces that are important, but other actors enter into play
 such as the Hartree interactions between more extended electron systems,
 the population of subbands of increasing energy, or the displacement
 of the center of charge of the electronic clouds that tend to decrease the
 effective distance between the plates of the capacitor.
 We have quantified all of them in an step by step basis using a jellium slab
 as a toy model.
 Exploiting the quantum nature of the metallic electrodes to
 overcome the classical limits on capacitor performance
 appears as a promising research avenue. The results reported here
 open the door to the rational design of devices based on
 negative electron compressibility and related effects.

 \section{Acknowledgments}

 We acknowledge P. de Castro-Manzano, J. Mannhart and Th. Kopp
 for useful discussions.
 This work was supported by the Spanish Ministery
 of Economy and Competitiveness through the
 MINECO Grant No. FIS2012-37549-C05-04 and
 No. FIS2015-64886-394-C5-2-P, and P.G.F. acknowledges support from
 Ram\'on y Cajal grant No. RyC-2013-12515.
 M. S. acknowledges the support of MINECO Grants No. MAT2016-77100-C2-2-P
 and No. SEV-2015-0496, and of Generalitat de Catalunya 
 (Grant No. 2017 SGR1506).
 This project has received funding from the European Research Council (ERC) 
 under the European Union’s Horizon 2020 research and innovation program 
 (Grant Agreement No. 724529).

 \appendix

 \section{Form factor within the local density approximation}
 \label{app:form-factor-lda}

 Considering Slater's expression for the exchange energy 
 of the homogeneous electron gas,~\cite{Dirac-30,Slater-51}
 then it can be written in a local form, which in atomic units reads as

 \begin{equation}
    \epsilon^{\rm LDA}_x (\rho) = -\frac{3}{4 \kappa} 
        \left( \frac{3}{\pi} \right)^{\frac{1}{3}} \rho^{\frac{1}{3}},
 \end{equation}

 \noindent where $\kappa$ is the relative permittivity of the medium.
 The exchange energy is then given by the three-dimensional integral
 \begin{equation}
    E^{\rm LDA}_x = \int d^3 r \, \epsilon_x 
         \left[ \rho_{\rm el} ({\vec r})\right] 
         \rho_{\rm el} ({\vec r}) = 
        -\frac{3}{4 \kappa} \left( \frac{3}{\pi} \right)^{\frac{1}{3}}
         \int d^3 r \, \rho_{\rm el}^{\frac{4}{3}}({\vec r}).
\end{equation}
 The exchange energy per unit area for the particular case under study
 (square well wavefunction) can be written as

 \begin{equation}
    E^{\rm LDA}_x / S = -\frac{3}{4 \kappa} 
                         \left( \frac{3}{\pi} \right)^{\frac{1}{3}} 
                         \left( \frac{2n}{w} \right)^{\frac{4}{3}} 
                         \int_0^w dz \, \sin^\frac{8}{3} (\pi z / w).
 \end{equation}
 This can be, in turn, further simplified to

 \begin{equation}
    E^{\rm LDA}_x / S = -\frac{3}{4 \kappa} 
                         \left( \frac{3}{\pi} \right)^{\frac{1}{3}} 
                         \left( \frac{2n}{w} \right)^{\frac{4}{3}} 
                         \frac{w}{\pi} C,
 \end{equation}

 \noindent where the constant $C$ can be calculated numerically as

 \begin{equation}
    C = \int_0^\pi dz \, \sin^\frac{8}{3} (z) = 1.4003141.
 \end{equation}
 Finally, we obtain the exchange energy per electron in the confined 2D gas,
 \begin{equation}
    \epsilon_x^{\rm LDA}(n,w) = -\frac{3 C}{2 \pi \kappa} 
        \left( \frac{6}{\pi} \right)^{\frac{1}{3}}  
        \left( \frac{n}{w} \right)^{\frac{1}{3}}.
 \end{equation}

 Remarkably, just like in the case of the exact treatment, 
 we can write $\epsilon_x^{\rm LDA}(n,w)$ as the
 ``ideal'' 2D exchange energy times a form factor,

 \begin{equation}
    \epsilon^{\rm LDA}_x(n,w) = \epsilon_x^{\rm 2D}(n) F^{\rm LDA}(\zeta).
    \label{eq:exchangelda2}
 \end{equation}

 \noindent The explicit form factor is

 \begin{eqnarray}
    F^{\rm LDA}(\zeta) &=& \frac{3 C}{2 \pi \kappa} 
                           \left( \frac{6}{\pi} \right)^{\frac{1}{3}} 
                           \frac{3 \kappa}{4} 
                           \sqrt{\frac{\pi}{2}} n^{-\frac{1}{2}} 
                           \left( \frac{n}{w} \right)^{\frac{1}{3}} 
    \nonumber \\
                       &=& \frac{9 C}{8 \pi} 
                           \left( \frac{6}{\pi} \right)^{\frac{1}{3}} 
                           \sqrt{\frac{\pi}{2}} \pi^{\frac{1}{6}} 
                           \zeta^{-\frac{1}{3}} 
    \nonumber \\
                       &=& A \zeta^{-\frac{1}{3}},
    \label{eq:form-factor-lda-app}
\end{eqnarray}

 \noindent where $A=0.9436555$ is a dimensionless constant.

 \section{Effect of the relaxation of the wave function}
 \label{app:relax-wavefun}

 In this Appendix we shall analyze the influence of the jellium background
 charge density in the position of the image charge and in the 
 Hartree band-bending contribution to the electron compressibility.
 For the remainder of this Appendix, we shall refer to all the
 physical quantities related with the traditional quantum well
 described in Sec.~\ref{sec:res-traditional-qw} with the subscript ``TQW'',
 while all the quantites related with the jellium model,
 Sec.~\ref{sec:jellium-confined}, will be labelled
 as ``jell''.

 For the traditional quantum well, the ground state
 for any value of the charge density $n$ can be approached as

 \begin{equation}
    \psi^{(0)}_{\rm TQW} (z; n) \approx \psi^{(0)} (z; n = 0) + 
           n \psi^{(1)}(z;n = 0).
    \label{eq:psi0tqw}
 \end{equation}

 \noindent Thus we can write the electron density, upto second order in 
 $n$, in terms of the wave function and its derivative computed at $n=0$,

 \begin{align}
    \rho^{(0)}_{\rm TQW, el} (z; n) = & 
    n \vert \psi^{(0)}_{\rm TQW} (z; n)\vert^{2}
    \nonumber \\
     = & n \vert \psi^{(0)} (z; n = 0) \vert^{2} 
    \nonumber \\
       & + 2n^{2} \vert \psi^{(0)} (z; n = 0) \psi^{(1)} (z; n = 0) \vert,
 \end{align}

 \noindent and, straightforwardly, the first-order variation of the
 electronic charge density
 
 \begin{align}
    \rho^{(1)}_{\rm TQW, el} (z; n) 
       = & \vert \psi^{(0)} (z; n = 0) \vert^{2} 
    \nonumber \\
         & + 4n \vert \psi^{(0)} (z; n = 0) \psi^{(1)} (z; n = 0) \vert.
    \label{eq:rho1tqw}
 \end{align}

 Now, we switch our attention to the jellium case.
 The most important difference with respect the previous situation is that
 we have a dependency with respect two parameters: $n$ and $n_{\rm jell}$.
 Now it can be assumed that the ground state and its first-order derivative
 are not significantly modified when the charge density in the
 jellium changes,

 \begin{align}
    \psi^{(0)}_{\rm jell} (z; n_{\rm jell}, n) &  \approx 
    \psi^{(0)}_{\rm jell} (z; n_{\rm jell}=0, n=0) 
    \nonumber \\
    &  = \psi^{(0)} (z; n = 0),
    \label{eq:psi0jel}
 \end{align}
 
 \begin{align}
    \psi^{(1)}_{\rm jell} (z; n_{\rm jell}, n) &  \approx 
    \psi^{(1)}_{\rm jell} (z; n_{\rm jell}=0, n=0) 
    \nonumber \\
    &  = \psi^{(1)} (z; n = 0).
 \end{align}

 \noindent We stress at this point that the ground state and
 first-order wave functions are exactly the same for the traditional
 quantum well and the jellium when $n_{\rm jell}=0$ and $n=0$.
 The approach comes when we assume that they remain the same for any 
 density.

 Starting from the previous wave functions, the electron charge density
 can be computed in the jellium setup as,

 \begin{align}
    \rho^{(0)}_{\rm jell, el} (z; n_{\rm jell}, n) = & 
        n 
        \vert \psi^{(0)}_{\rm jell} (z; n_{\rm jell}, n)\vert^{2}
    \nonumber \\
     \approx & n \vert \psi^{(0)} (z; n = 0) \vert^{2}. 
 \end{align}

 \noindent For the first-order change in the charge density, 
 we realize that in the process of charging/discharging the capacitor
 the variable that enters into play is $n$, while the background charge
 density $n_{\rm jell}$ remains constant.
 Therefore, we have to take the partial derivative with respect to $n$, 
 keeping $n_{\rm jell}$ frozen

 \begin{align}
    \rho^{(1)}_{\rm jell, el} (z; n_{\rm jell}, n) 
       = & \frac{\partial \rho^{(0)}_{\rm jell,el}(z;n_{\rm jell},n)}
                {\partial n} \Big|_{n_{\rm jell}}
    \nonumber \\
       = & \vert \psi^{(0)} (z; n = 0) \vert^{2} 
    \nonumber \\
         & + 2 n 
             \vert \psi^{(0)} (z; n = 0) \psi^{(1)} (z; n = 0) \vert.
    \label{eq:rho1jell}
 \end{align}

 \psfrag{psi0psi1}[ll][ll]{$\psi^{(0)}(z), \psi^{(1)}(z)$}
 \psfrag{nAnB}[ll][ll]{$n_{\rm A}(z), n_{\rm B}(z)$}
 \psfrag{vAvB}[ll][ll]{$V_{\rm A}(z), V_{\rm B}(z)$}
 \psfrag{zaxis}[ll][ll]{$z$}
 \begin{figure}
    \begin{center}
       \includegraphics[width=\columnwidth]{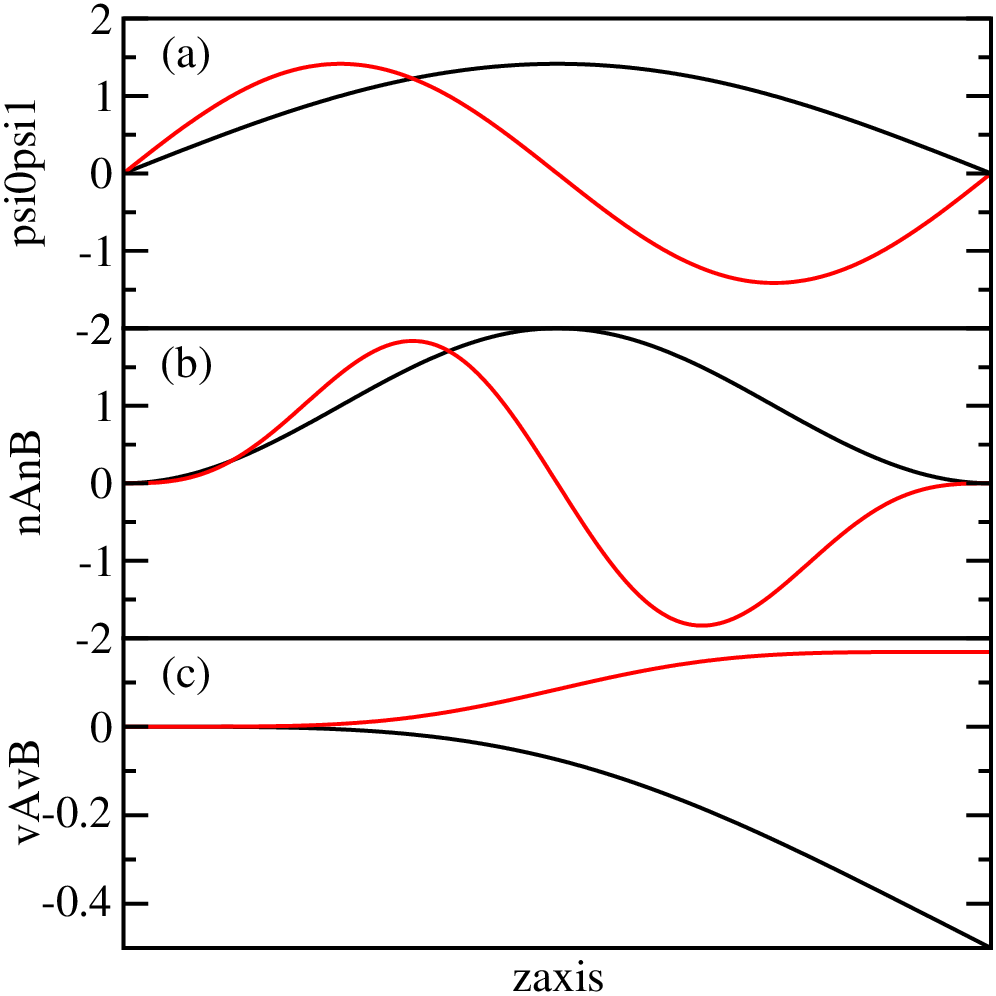}
       \caption{ (Color online) 
                 (a) Ground state (black line) and first-excited (red line) 
                 electronic eigenfunctions of a particle in an 
                 infinite quantum well potential in one dimension.
                 (b) Associated $n_{\rm A}$ (black) and $n_{\rm B}$ (red)
                 potentials as defined in the text.
                 (c) Electrostratic potentials computed after a double
                 integral of $n_{\rm A}$ ($V_{\rm A}$; black line) 
                 and $n_{\rm B}$ ($V_{\rm B}$; red line).
               }
       \label{fig:modelinfwell}
    \end{center}
 \end{figure}

 Now, and for the sake of simplicity, we shall revolve around a simplified
 version of the problem, based on the infinite quantum well, that captures
 the most important physical ingredients.
 From now on, we assume that $\psi^{(0)}$ and $\psi^{(1)}$ are
 the ground and first excited states of an infinite square quantum well.
 As shown in Fig.~\ref{fig:modelinfwell}(a), 
 these functions are orthonormal and, respectively, 
 symmetric and antisymmetric with respect the center of the quantum well.
 From these wave functions, we can compute the related densities
 $n_{\rm A}$ and $n_{\rm B}$, defined as 

 \begin{subequations}
    \begin{align}
       n_{\rm A} (z) & = \vert \psi^{(0)}(z) \vert^{2},
       \label{eq:na}
       \\
       n_{\rm B} (z) & = \psi^{(0)} (z) \psi^{(1)} (z),
       \label{eq:nb}
    \end{align}
 \end{subequations}

 \noindent and plotted in Fig.~\ref{fig:modelinfwell}(b). 
 We can use these two densities to approximate the first-order changes in the
 charge densities for the traditional quantum well [Eq.~(\ref{eq:rho1tqw})],
 and the jellium model [Eq.~(\ref{eq:rho1jell})]

 \begin{subequations}
    \begin{align}
       \rho^{(1)}_{\rm TQW,el} (z) & = n_{\rm A} (z) + 4 n n_{\rm B}(z),
       \label{eq:rho1tqw-2}
       \\
       \rho^{(1)}_{\rm jell,el} (z) & = n_{\rm A} (z) + 2 n n_{\rm B}(z).
       \label{eq:rho1jell-2}
    \end{align}
 \end{subequations}

 From the previous two equations we can immediately compute the  
 position of the image charge of the two models, following
 Eq.~(\ref{eq:averagez}) and Eq.~(\ref{eq:defzim}), as

 \begin{equation}
    z_{\rm im} = \int dz \, z \rho^{(1)}(z),
 \end{equation}

 \noindent so, making use of the symmetry of the charge densities,

 \begin{subequations}
    \begin{align}
       z_{\rm im}^{\rm TQW} = 4 n z_{B},
       \label{eq:zimtqw}
       \\
       z_{\rm im}^{\rm jell} = 2 n z_{B},
       \label{eq:zimjell}
    \end{align}
 \end{subequations}

 \noindent where $z_{B}$ is the first moment of the charge distribution
 given by $n_{\rm B}$,

 \begin{equation}
    z_{B} = \int z \: n_{\rm B}(z) \: dz.
 \end{equation}

 Integrating the charge densities $n_{\rm A}$ and $n_{\rm B}$
 we obtain two different potentials, 
 coined in Fig.~\ref{fig:modelinfwell}(c) as $V_{\rm A}$ and $V_{\rm B}$,
 respectively.
 For the sake of simplicity in this simplified model we have
 assumed that the classical electrode is located at the right of the
 quantum well, i. e. the opposite convention as used in
 the rest of the work. This local change in the convention
 does not affect the conclusions that can be drawn from the model.
 The offset in the potential for $V_{\rm B}$ corresponds
 to the dipole moment associated with $n_{\rm B}$,
 as can be proven by an integration by parts,

 \begin{equation}
    \Delta V_{\rm B} = V_{\rm B} (+\infty) - 
                       V_{\rm B} (-\infty) = 4 \pi z_{B}.
 \end{equation}

 To compute the Hartree contributions to the 
 electron compressibility $d\mu/dn$, Eq.~(\ref{delright}),

 \begin{subequations}
    \begin{align}
       \Delta \epsilon^{(1)}_{\rm H, TQW} (n)
       & = \int dz \, V^{(1)}_{\rm H,TQW} (z;n) 
           \vert \psi_{\rm TQW}(z;n) \vert^{2},
       \label{eq:eps1tqw}
       \\
       \Delta \epsilon^{(1)}_{\rm H, jell}(n)
       & = \int dz \, V^{(1)}_{\rm H,jell}(z;n)
           \vert \psi_{\rm jell}(z;n) \vert^{2},
       \label{eq:eps1jell}
    \end{align}
 \end{subequations}

 \noindent where we have used, according to 
 Eq.~(\ref{eq:psi0tqw}) and Eq.~(\ref{eq:psi0jel})

 \begin{subequations}
    \begin{align}
       \psi_{\rm TQW} (z;n) & = \psi^{(0)} (z;n=0) + n \psi^{(1)} (z;n=0),
       \label{eq:psitqw}
       \\
       \psi_{\rm jell} (z;n) & = \psi^{(0)} (z;n=0). 
       \label{eq:psijell}
    \end{align}
 \end{subequations}

 \noindent and the potentials are the double integrals
 of $\rho^{(1)}_{\rm TQW,el}$ and $\rho^{(1)}_{\rm jell,el}$,
 respectively [Eq~(\ref{vh1})].
 Then, neglecting terms beyond second order in $n$,

 \begin{subequations}
    \begin{align}
       \Delta \epsilon^{(1)}_{\rm H, TQW}
       & =  V_{\rm A} \cdot n_{\rm A} + 4 n V_{\rm B} \cdot n_{\rm A} + 
         2n V_{\rm A} \cdot n_{\rm B},
       \label{eq:eps1tqw-2}
       \\
       \Delta \epsilon^{(1)}_{\rm H, jell} 
       & = V_{\rm A} \cdot n_{\rm A} + 2 n V_{\rm B} \cdot n_{\rm A},
       \label{eq:eps1jell-2}
    \end{align}
 \end{subequations}

 \noindent where we have used the shorthand notation

 \begin{equation}
    V_{x} \cdot n_{y} = \int dz \, V_{x} (z) n_{y} (z). 
 \end{equation}

 Paying attention to the symmetry of the potentials 
 $V_{\rm A}$ and $V_{\rm B}$ in Fig.~\ref{fig:modelinfwell}(c),
 they can be written as

 \begin{subequations}
    \begin{align}
       V_{\rm A} (z) & = -\frac{z}{2} + f_{\rm S}(z),
       \label{eq:vasym}
       \\
       V_{\rm B} (z) & = \frac{z_{B}}{2} + f_{\rm A}(z),
       \label{eq:vbsym}
    \end{align}
 \end{subequations}

 \noindent where $f_{\rm S}$ and $f_{\rm A}$ are 
 two functions that are, respectively, symmetric and antisymmetric
 with respect to the change $z \rightarrow -z$.
 Since $n_{\rm A}$ is normalized to unity, then 

 \begin{subequations}
    \begin{align}
       V_{\rm A} \cdot n_{\rm B} & = -\frac{z_{B}}{2}, 
       \label{eq:vanbsym}
       \\
       V_{\rm B} \cdot n_{\rm A} & = \frac{z_{B}}{2}. 
       \label{eq:vbnasym}
    \end{align}
 \end{subequations}

 \noindent So we arrive to the final conclusion,

 \begin{equation}
     \Delta \epsilon^{(1)}_{\rm H, TQW} = \Delta \epsilon^{(1)}_{\rm H, jell}
     = V_{\rm A} \cdot n_{\rm A} + z_{B} n.
     \label{eq:equalhartree}
 \end{equation}

 The first term in Eq.~(\ref{eq:equalhartree}) is the Hartree band-bending
 for frozen wave functions, while the second is a linear correction with $n$
 that appears when the wave functions are allowed to relax.
 With a similar derivation it is easy to prove that combining 
 $V^{(1)}_{\rm H,jell}$ with $\psi_{\rm TQW}$ 
 the linear part cancels out.

 \section{Units}
 \label{app:units}

 The results discussed in this work have been 
 calculated assuming a
 relative permittivity $\kappa=1$ and an effective electron mass
 $m^\ast=1$.
 Nevertheless, they can be used to intepret
 experiments where the materials under study present
 a different value of these two parameters.
 For this purpose, some physical quantities must be rescaled 
 by the factors included in
 Table~\ref{table:conversionfactor}.

 \begin{table}
    \caption{Conversion factors to rescale the numerical results 
             obtained in this work to a material with 
             effective mass $m^{\ast}$ and relative 
             dielectric constant $\kappa$.}
    \begin{tabular}{lc}
       \hline
       \hline
       Quantity                          & 
       Factor                            \\
       \hline
       Length                            & 
       $\kappa / m^\ast$                 \\
       Energy                            & 
       $m^\ast / \kappa^2$               \\
       Areal density                     &
       $(m^\ast / \kappa)^2$             \\
       Electronic compressibility        & 
       $1 / m^\ast$                      \\
       \hline
       \hline
    \end{tabular}
    \label{table:conversionfactor}
 \end{table}
 
 Finally, note that the resulting $\partial \mu / \partial n$ is in 
 atomic units of inverse capacitance density. 
 To convert to a length (Bohr), this number needs to be
 rescaled by $\kappa / 4\pi$.

\begin{figure}
    \begin{center}
       \includegraphics[width=0.7\columnwidth]{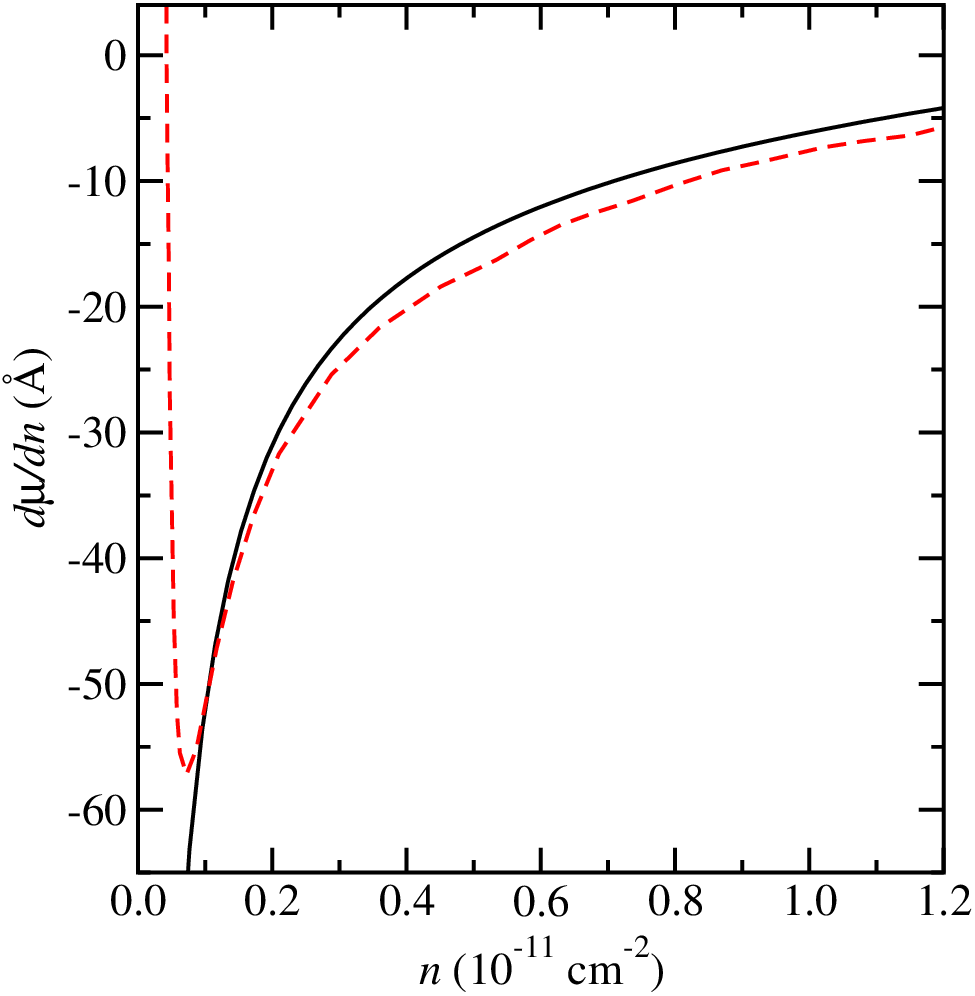}
       \caption{ (Color online) 
                 Electronic compressibility as a function of electron density
                 assuming the parameters of a realistic
                 GaAs/Al$_{0.3}$Ga$_{0.7}$As quantum well,
                 $m^{\ast}/m_{\rm e} = 0.067$, 
                 $\kappa$ = 12.6 $\epsilon_{0}$, and $V_{\rm ext}$ = 250 meV.
                 The results are the same as in Fig.~\ref{fig:eisenstein}(b)
                 but the physical magnitudes have been rescaled
                 according to the rules
                 given in Table~\ref{table:conversionfactor}.
                 Black solid line represents the theoretical results
                 obtained for the traditional quantum well.
                 Red dashed line is the experimental curve,
                 taken from Ref.~\onlinecite{Eisenstein-94}.
               }
       \label{fig:gaas}
    \end{center}
 \end{figure}

 As an example, and as a convincing way to validate our calculations
 against experimental measurements, we have rescaled the 
 results obtained in a simple quantum well potential
 to simulate the behaviour of the electronic compressibility versus 
 the charge density in GaAs/Al$_{x}$Ga$_{1-x}$As quantum wells, as measured in 
 Ref.~\onlinecite{Eisenstein-94}.
 Taking $V_{\rm ext}$ to mimic the conduction-band offset
 of GaAs and Al$_{0.3}$Ga$_{0.7}$As (250 meV), 
 assuming simple parabolic bands with effective mass $m^{\ast}/m_{e} = 0.067$,
 and considering the dielectric constant to be 
 $\kappa = 12.6 \epsilon_{0}$ (where $\epsilon_{0}$ is
 the permittivity of free space), we obtain the results
 of Fig.~\ref{fig:gaas} that compares very well with the 
 reported experimental values.
 The theoretically predicted divergence of $d\mu/dn$ at low temperature 
 is suppressed, presumably, by disorder in the low density 
 regime.~\cite{Eisenstein-94}

%


\begin{thebibliography}{39}%
\makeatletter
\providecommand \@ifxundefined [1]{%
 \@ifx{#1\undefined}
}%
\providecommand \@ifnum [1]{%
 \ifnum #1\expandafter \@firstoftwo
 \else \expandafter \@secondoftwo
 \fi
}%
\providecommand \@ifx [1]{%
 \ifx #1\expandafter \@firstoftwo
 \else \expandafter \@secondoftwo
 \fi
}%
\providecommand \natexlab [1]{#1}%
\providecommand \enquote  [1]{``#1''}%
\providecommand \bibnamefont  [1]{#1}%
\providecommand \bibfnamefont [1]{#1}%
\providecommand \citenamefont [1]{#1}%
\providecommand \href@noop [0]{\@secondoftwo}%
\providecommand \href [0]{\begingroup \@sanitize@url \@href}%
\providecommand \@href[1]{\@@startlink{#1}\@@href}%
\providecommand \@@href[1]{\endgroup#1\@@endlink}%
\providecommand \@sanitize@url [0]{\catcode `\\12\catcode `\$12\catcode
  `\&12\catcode `\#12\catcode `\^12\catcode `\_12\catcode `\%12\relax}%
\providecommand \@@startlink[1]{}%
\providecommand \@@endlink[0]{}%
\providecommand \url  [0]{\begingroup\@sanitize@url \@url }%
\providecommand \@url [1]{\endgroup\@href {#1}{\urlprefix }}%
\providecommand \urlprefix  [0]{URL }%
\providecommand \Eprint [0]{\href }%
\providecommand \doibase [0]{http://dx.doi.org/}%
\providecommand \selectlanguage [0]{\@gobble}%
\providecommand \bibinfo  [0]{\@secondoftwo}%
\providecommand \bibfield  [0]{\@secondoftwo}%
\providecommand \translation [1]{[#1]}%
\providecommand \BibitemOpen [0]{}%
\providecommand \bibitemStop [0]{}%
\providecommand \bibitemNoStop [0]{.\EOS\space}%
\providecommand \EOS [0]{\spacefactor3000\relax}%
\providecommand \BibitemShut  [1]{\csname bibitem#1\endcsname}%
\let\auto@bib@innerbib\@empty
\bibitem [{\citenamefont {Moore}(1965)}]{Moore-65}%
  \BibitemOpen
  \bibfield  {author} {\bibinfo {author} {\bibfnamefont {G.~E.}\ \bibnamefont
  {Moore}},\ }\href@noop {} {\bibfield  {journal} {\bibinfo  {journal}
  {Electronics}\ }\textbf {\bibinfo {volume} {38}},\ \bibinfo {pages} {114}
  (\bibinfo {year} {1965})}\BibitemShut {NoStop}%
\bibitem [{Note1()}]{Note1}%
  \BibitemOpen
  \bibinfo {note} {According to Moore's law, the number of transistors that can
  be placed on an integrated circuit would approximately double every eighteen
  months.}\BibitemShut {Stop}%
\bibitem [{\citenamefont {Waldrop}(2016)}]{Waldrop-16}%
  \BibitemOpen
  \bibfield  {author} {\bibinfo {author} {\bibfnamefont {M.~M.}\ \bibnamefont
  {Waldrop}},\ }\href@noop {} {\bibfield  {journal} {\bibinfo  {journal}
  {Nature (London)}\ }\textbf {\bibinfo {volume} {530}},\ \bibinfo {pages}
  {145} (\bibinfo {year} {2016})}\BibitemShut {NoStop}%
\bibitem [{\citenamefont {Markov}(2014)}]{Markov-14}%
  \BibitemOpen
  \bibfield  {author} {\bibinfo {author} {\bibfnamefont {I.~L.}\ \bibnamefont
  {Markov}},\ }\href@noop {} {\bibfield  {journal} {\bibinfo  {journal} {Nature
  (London)}\ }\textbf {\bibinfo {volume} {512}},\ \bibinfo {pages} {147}
  (\bibinfo {year} {2014})}\BibitemShut {NoStop}%
\bibitem [{\citenamefont {Sapoval}\ and\ \citenamefont
  {Hermann}(1993)}]{Sapoval}%
  \BibitemOpen
  \bibfield  {author} {\bibinfo {author} {\bibfnamefont {B.}~\bibnamefont
  {Sapoval}}\ and\ \bibinfo {author} {\bibfnamefont {C.}~\bibnamefont
  {Hermann}},\ }\href@noop {} {\emph {\bibinfo {title} {Physics of
  semiconductors}}}\ (\bibinfo  {publisher} {Springer-Verlag},\ \bibinfo
  {address} {New York},\ \bibinfo {year} {1993})\BibitemShut {NoStop}%
\bibitem [{\citenamefont {Feynman}\ \emph {et~al.}(1964)\citenamefont
  {Feynman}, \citenamefont {Leighton},\ and\ \citenamefont
  {Sands}}]{Feynman-II}%
  \BibitemOpen
  \bibfield  {author} {\bibinfo {author} {\bibfnamefont {R.}~\bibnamefont
  {Feynman}}, \bibinfo {author} {\bibfnamefont {R.~B.}\ \bibnamefont
  {Leighton}}, \ and\ \bibinfo {author} {\bibfnamefont {M.~L.}\ \bibnamefont
  {Sands}},\ }\href@noop {} {\emph {\bibinfo {title} {The Feynman lectures on
  Physics. Volume II}}}\ (\bibinfo  {publisher} {Addison-Wesley Publishing},\
  \bibinfo {address} {Reading, Massachusetts, USA},\ \bibinfo {year}
  {1964})\BibitemShut {NoStop}%
\bibitem [{\citenamefont {Salahuddin}\ and\ \citenamefont
  {Datta}(2008)}]{Salahuddin-08}%
  \BibitemOpen
  \bibfield  {author} {\bibinfo {author} {\bibfnamefont {S.}~\bibnamefont
  {Salahuddin}}\ and\ \bibinfo {author} {\bibfnamefont {S.}~\bibnamefont
  {Datta}},\ }\href@noop {} {\bibfield  {journal} {\bibinfo  {journal} {Nano
  Lett.}\ }\textbf {\bibinfo {volume} {8}},\ \bibinfo {pages} {405} (\bibinfo
  {year} {2008})}\BibitemShut {NoStop}%
\bibitem [{\citenamefont {Cano}\ and\ \citenamefont
  {Jim\'enez}(2010)}]{Cano-10}%
  \BibitemOpen
  \bibfield  {author} {\bibinfo {author} {\bibfnamefont {A.}~\bibnamefont
  {Cano}}\ and\ \bibinfo {author} {\bibfnamefont {D.}~\bibnamefont
  {Jim\'enez}},\ }\href@noop {} {\bibfield  {journal} {\bibinfo  {journal}
  {Appl. Phys. Lett.}\ }\textbf {\bibinfo {volume} {97}},\ \bibinfo {pages}
  {133509} (\bibinfo {year} {2010})}\BibitemShut {NoStop}%
\bibitem [{\citenamefont {Gao}\ \emph {et~al.}(2014)\citenamefont {Gao},
  \citenamefont {Khan}, \citenamefont {Marti}, \citenamefont {Nelson},
  \citenamefont {Serrao}, \citenamefont {Ravichandran}, \citenamefont
  {Ramesh},\ and\ \citenamefont {Salahuddin}}]{Gao-14}%
  \BibitemOpen
  \bibfield  {author} {\bibinfo {author} {\bibfnamefont {W.}~\bibnamefont
  {Gao}}, \bibinfo {author} {\bibfnamefont {A.}~\bibnamefont {Khan}}, \bibinfo
  {author} {\bibfnamefont {X.}~\bibnamefont {Marti}}, \bibinfo {author}
  {\bibfnamefont {C.}~\bibnamefont {Nelson}}, \bibinfo {author} {\bibfnamefont
  {C.}~\bibnamefont {Serrao}}, \bibinfo {author} {\bibfnamefont
  {J.}~\bibnamefont {Ravichandran}}, \bibinfo {author} {\bibfnamefont
  {R.}~\bibnamefont {Ramesh}}, \ and\ \bibinfo {author} {\bibfnamefont
  {S.}~\bibnamefont {Salahuddin}},\ }\href@noop {} {\bibfield  {journal}
  {\bibinfo  {journal} {Nano Lett.}\ }\textbf {\bibinfo {volume} {14}},\
  \bibinfo {pages} {5814} (\bibinfo {year} {2014})}\BibitemShut {NoStop}%
\bibitem [{\citenamefont {Zubko}\ \emph {et~al.}(2016)\citenamefont {Zubko},
  \citenamefont {Wojde\l{}}, \citenamefont {Hadjimichael}, \citenamefont
  {Fernandez-Pena}, \citenamefont {Sen\'e}, \citenamefont {Luk'yanchuk},
  \citenamefont {Triscone},\ and\ \citenamefont {\'I{\~n}iguez}}]{Zubko-16}%
  \BibitemOpen
  \bibfield  {author} {\bibinfo {author} {\bibfnamefont {P.}~\bibnamefont
  {Zubko}}, \bibinfo {author} {\bibfnamefont {J.~C.}\ \bibnamefont
  {Wojde\l{}}}, \bibinfo {author} {\bibfnamefont {M.}~\bibnamefont
  {Hadjimichael}}, \bibinfo {author} {\bibfnamefont {S.}~\bibnamefont
  {Fernandez-Pena}}, \bibinfo {author} {\bibfnamefont {A.}~\bibnamefont
  {Sen\'e}}, \bibinfo {author} {\bibfnamefont {I.}~\bibnamefont {Luk'yanchuk}},
  \bibinfo {author} {\bibfnamefont {J.-M.}\ \bibnamefont {Triscone}}, \ and\
  \bibinfo {author} {\bibfnamefont {J.}~\bibnamefont {\'I{\~n}iguez}},\
  }\href@noop {} {\bibfield  {journal} {\bibinfo  {journal} {Nature (London)}\
  }\textbf {\bibinfo {volume} {534}},\ \bibinfo {pages} {524} (\bibinfo {year}
  {2016})}\BibitemShut {NoStop}%
\bibitem [{Note2()}]{Note2}%
  \BibitemOpen
  \bibinfo {note} {In a capacitor, the charge density stored on the plates is
  proportional to the normal component of the electric displacement field.
  Since, in atomic units, $Q/S = D/ (4\pi ) = (E + 4 \pi P)/ (4 \pi )$, where
  $E$ is the normal component of the electric field, and in a typical
  ferroelectric material $4 \pi P \gg E$, then $Q/S \approx P$.}\BibitemShut
  {Stop}%
\bibitem [{\citenamefont {Kopp}\ and\ \citenamefont
  {Mannhart}(2009)}]{Kopp-09}%
  \BibitemOpen
  \bibfield  {author} {\bibinfo {author} {\bibfnamefont {T.}~\bibnamefont
  {Kopp}}\ and\ \bibinfo {author} {\bibfnamefont {J.}~\bibnamefont
  {Mannhart}},\ }\href@noop {} {\bibfield  {journal} {\bibinfo  {journal} {J.
  Appl. Phys.}\ }\textbf {\bibinfo {volume} {106}},\ \bibinfo {pages} {064504}
  (\bibinfo {year} {2009})}\BibitemShut {NoStop}%
\bibitem [{\citenamefont {Mannhart}\ and\ \citenamefont
  {Schlom}(2010)}]{Mannhart-10}%
  \BibitemOpen
  \bibfield  {author} {\bibinfo {author} {\bibfnamefont {J.}~\bibnamefont
  {Mannhart}}\ and\ \bibinfo {author} {\bibfnamefont {D.~G.}\ \bibnamefont
  {Schlom}},\ }\href@noop {} {\bibfield  {journal} {\bibinfo  {journal}
  {Science}\ }\textbf {\bibinfo {volume} {327}},\ \bibinfo {pages} {1607}
  (\bibinfo {year} {2010})}\BibitemShut {NoStop}%
\bibitem [{Note3()}]{Note3}%
  \BibitemOpen
  \bibinfo {note} {In the standard textbook picture, when an electron is added
  to a metallic system it fills the lowest unoccupied energy state; as a
  consequence, the chemical potential increases.}\BibitemShut {Stop}%
\bibitem [{\citenamefont {Ashoori}\ and\ \citenamefont
  {Silsbee}(1992)}]{Ashoori-92}%
  \BibitemOpen
  \bibfield  {author} {\bibinfo {author} {\bibfnamefont {R.~C.}\ \bibnamefont
  {Ashoori}}\ and\ \bibinfo {author} {\bibfnamefont {R.~H.}\ \bibnamefont
  {Silsbee}},\ }\href@noop {} {\bibfield  {journal} {\bibinfo  {journal} {Solid
  State Commun.}\ }\textbf {\bibinfo {volume} {81}},\ \bibinfo {pages} {821}
  (\bibinfo {year} {1992})}\BibitemShut {NoStop}%
\bibitem [{\citenamefont {Eisenstein}\ \emph {et~al.}(1994)\citenamefont
  {Eisenstein}, \citenamefont {Pfeiffer},\ and\ \citenamefont
  {West}}]{Eisenstein-94}%
  \BibitemOpen
  \bibfield  {author} {\bibinfo {author} {\bibfnamefont {J.~P.}\ \bibnamefont
  {Eisenstein}}, \bibinfo {author} {\bibfnamefont {L.~N.}\ \bibnamefont
  {Pfeiffer}}, \ and\ \bibinfo {author} {\bibfnamefont {K.~W.}\ \bibnamefont
  {West}},\ }\href@noop {} {\bibfield  {journal} {\bibinfo  {journal} {Phys.
  Rev. B}\ }\textbf {\bibinfo {volume} {50}},\ \bibinfo {pages} {1760}
  (\bibinfo {year} {1994})}\BibitemShut {NoStop}%
\bibitem [{\citenamefont {Li}\ \emph {et~al.}(2011)\citenamefont {Li},
  \citenamefont {Richter}, \citenamefont {Paetel}, \citenamefont {Mannhart},\
  and\ \citenamefont {Ashoori}}]{Li-11}%
  \BibitemOpen
  \bibfield  {author} {\bibinfo {author} {\bibfnamefont {L.}~\bibnamefont
  {Li}}, \bibinfo {author} {\bibfnamefont {C.}~\bibnamefont {Richter}},
  \bibinfo {author} {\bibfnamefont {S.}~\bibnamefont {Paetel}}, \bibinfo
  {author} {\bibfnamefont {J.}~\bibnamefont {Mannhart}}, \ and\ \bibinfo
  {author} {\bibfnamefont {R.~C.}\ \bibnamefont {Ashoori}},\ }\href@noop {}
  {\bibfield  {journal} {\bibinfo  {journal} {Science}\ }\textbf {\bibinfo
  {volume} {332}},\ \bibinfo {pages} {825} (\bibinfo {year}
  {2011})}\BibitemShut {NoStop}%
\bibitem [{\citenamefont {Tinkl}\ \emph {et~al.}(2012)\citenamefont {Tinkl},
  \citenamefont {Breutschaft}, \citenamefont {Richter},\ and\ \citenamefont
  {Mannhart}}]{Tinkl-12}%
  \BibitemOpen
  \bibfield  {author} {\bibinfo {author} {\bibfnamefont {V.}~\bibnamefont
  {Tinkl}}, \bibinfo {author} {\bibfnamefont {M.}~\bibnamefont {Breutschaft}},
  \bibinfo {author} {\bibfnamefont {C.}~\bibnamefont {Richter}}, \ and\
  \bibinfo {author} {\bibfnamefont {J.}~\bibnamefont {Mannhart}},\ }\href@noop
  {} {\bibfield  {journal} {\bibinfo  {journal} {Phys. Rev. B}\ }\textbf
  {\bibinfo {volume} {86}},\ \bibinfo {pages} {075116} (\bibinfo {year}
  {2012})}\BibitemShut {NoStop}%
\bibitem [{sam()}]{same-kappa}%
  \BibitemOpen
  \href@noop {} {}\bibinfo {note} {From now on, and on the sake of simplicity,
  we assume that there is no discontinuity in this parameter at the
  electrode/dielectric interface, so the effective dielectric constant of the
  electrode is also $\kappa$.}\BibitemShut {Stop}%
\bibitem [{Note4()}]{Note4}%
  \BibitemOpen
  \bibinfo {note} {Note that in Ref.~\protect \rev@citealpnum {Eisenstein-94} a
  slightly more complicated double-well setup was used, in order to facilitate
  the comparison with experimental measurements; the precise relationship
  between our model in Fig.~\ref {fig:cartoonmodel} and that of Ref.~\protect
  \rev@citealpnum {Eisenstein-94} will be discussed in Sec.~\ref
  {sec:confinedgasesscf}.}\BibitemShut {Stop}%
\bibitem [{\citenamefont {Baroni}\ \emph {et~al.}(2001)\citenamefont {Baroni},
  \citenamefont {de~Gironcoli}, \citenamefont {Corso},\ and\ \citenamefont
  {Giannozzi}}]{Baroni-01}%
  \BibitemOpen
  \bibfield  {author} {\bibinfo {author} {\bibfnamefont {S.}~\bibnamefont
  {Baroni}}, \bibinfo {author} {\bibfnamefont {S.}~\bibnamefont
  {de~Gironcoli}}, \bibinfo {author} {\bibfnamefont {A.~D.}\ \bibnamefont
  {Corso}}, \ and\ \bibinfo {author} {\bibfnamefont {P.}~\bibnamefont
  {Giannozzi}},\ }\href@noop {} {\bibfield  {journal} {\bibinfo  {journal}
  {Rev. Mod. Phys.}\ }\textbf {\bibinfo {volume} {73}},\ \bibinfo {pages} {515}
  (\bibinfo {year} {2001})}\BibitemShut {NoStop}%
\bibitem [{\citenamefont {Gonze}(1995)}]{Gonze-95.2}%
  \BibitemOpen
  \bibfield  {author} {\bibinfo {author} {\bibfnamefont {X.}~\bibnamefont
  {Gonze}},\ }\href@noop {} {\bibfield  {journal} {\bibinfo  {journal} {Phys.
  Rev. A}\ }\textbf {\bibinfo {volume} {52}},\ \bibinfo {pages} {1086}
  (\bibinfo {year} {1995})}\BibitemShut {NoStop}%
\bibitem [{\citenamefont {Gonze}\ and\ \citenamefont {Lee}(1997)}]{Gonze-97.2}%
  \BibitemOpen
  \bibfield  {author} {\bibinfo {author} {\bibfnamefont {X.}~\bibnamefont
  {Gonze}}\ and\ \bibinfo {author} {\bibfnamefont {C.}~\bibnamefont {Lee}},\
  }\href@noop {} {\bibfield  {journal} {\bibinfo  {journal} {Phys. Rev. B}\
  }\textbf {\bibinfo {volume} {55}},\ \bibinfo {pages} {10355} (\bibinfo {year}
  {1997})}\BibitemShut {NoStop}%
\bibitem [{\citenamefont {Stengel}(2011)}]{Stengel-11.2}%
  \BibitemOpen
  \bibfield  {author} {\bibinfo {author} {\bibfnamefont {M.}~\bibnamefont
  {Stengel}},\ }\href@noop {} {\bibfield  {journal} {\bibinfo  {journal} {Phys.
  Rev. Lett.}\ }\textbf {\bibinfo {volume} {106}},\ \bibinfo {pages} {136803}
  (\bibinfo {year} {2011})}\BibitemShut {NoStop}%
\bibitem [{\citenamefont {Kohn}\ and\ \citenamefont {Sham}(1965)}]{Kohn-65}%
  \BibitemOpen
  \bibfield  {author} {\bibinfo {author} {\bibfnamefont {W.}~\bibnamefont
  {Kohn}}\ and\ \bibinfo {author} {\bibfnamefont {L.~J.}\ \bibnamefont
  {Sham}},\ }\href@noop {} {\bibfield  {journal} {\bibinfo  {journal} {Phys.
  Rev.}\ }\textbf {\bibinfo {volume} {140}},\ \bibinfo {pages} {A1133}
  (\bibinfo {year} {1965})}\BibitemShut {NoStop}%
\bibitem [{\citenamefont {Ceperley}\ and\ \citenamefont
  {Alder}(1980)}]{Ceperley-80}%
  \BibitemOpen
  \bibfield  {author} {\bibinfo {author} {\bibfnamefont {D.~M.}\ \bibnamefont
  {Ceperley}}\ and\ \bibinfo {author} {\bibfnamefont {B.~J.}\ \bibnamefont
  {Alder}},\ }\href@noop {} {\bibfield  {journal} {\bibinfo  {journal} {Phys.
  Rev. Lett.}\ }\textbf {\bibinfo {volume} {45}},\ \bibinfo {pages} {566}
  (\bibinfo {year} {1980})}\BibitemShut {NoStop}%
\bibitem [{\citenamefont {Hohenberg}\ and\ \citenamefont
  {Kohn}(1964)}]{Hohenberg-64}%
  \BibitemOpen
  \bibfield  {author} {\bibinfo {author} {\bibfnamefont {P.}~\bibnamefont
  {Hohenberg}}\ and\ \bibinfo {author} {\bibfnamefont {W.}~\bibnamefont
  {Kohn}},\ }\href@noop {} {\bibfield  {journal} {\bibinfo  {journal} {Phys.
  Rev.}\ }\textbf {\bibinfo {volume} {136}},\ \bibinfo {pages} {B864} (\bibinfo
  {year} {1964})}\BibitemShut {NoStop}%
\bibitem [{Num()}]{Numerov-Gianozzi}%
  \BibitemOpen
  \href@noop {} {}\bibinfo {note} {See the notes by P. Gianozzi at
  http://www.fisica.uniud.it/$\sim$giannozz/Corsi/MQ/LectureNotes/mq-cap1.pdf}\BibitemShut
  {NoStop}%
\bibitem [{\citenamefont {Kleinman}(1981)}]{Kleinman-81}%
  \BibitemOpen
  \bibfield  {author} {\bibinfo {author} {\bibfnamefont {L.}~\bibnamefont
  {Kleinman}},\ }\href@noop {} {\bibfield  {journal} {\bibinfo  {journal}
  {Phys. Rev. B}\ }\textbf {\bibinfo {volume} {24}},\ \bibinfo {pages} {7412}
  (\bibinfo {year} {1981})}\BibitemShut {NoStop}%
\bibitem [{\citenamefont {Tanatar}\ and\ \citenamefont
  {Ceperley}(1989)}]{Tanatar-89}%
  \BibitemOpen
  \bibfield  {author} {\bibinfo {author} {\bibfnamefont {B.}~\bibnamefont
  {Tanatar}}\ and\ \bibinfo {author} {\bibfnamefont {D.~M.}\ \bibnamefont
  {Ceperley}},\ }\href@noop {} {\bibfield  {journal} {\bibinfo  {journal}
  {Phys. Rev. B}\ }\textbf {\bibinfo {volume} {39}},\ \bibinfo {pages} {5005}
  (\bibinfo {year} {1989})}\BibitemShut {NoStop}%
\bibitem [{\citenamefont {Stern}(1974)}]{Stern-74}%
  \BibitemOpen
  \bibfield  {author} {\bibinfo {author} {\bibfnamefont {F.}~\bibnamefont
  {Stern}},\ }\href@noop {} {\bibfield  {journal} {\bibinfo  {journal} {Jpn. J.
  Appl. Phys. Suppl.}\ }\textbf {\bibinfo {volume} {2}},\ \bibinfo {pages}
  {323} (\bibinfo {year} {1974})}\BibitemShut {NoStop}%
\bibitem [{\citenamefont {Skinner}\ and\ \citenamefont
  {Fogler}(2010)}]{Skinner-10}%
  \BibitemOpen
  \bibfield  {author} {\bibinfo {author} {\bibfnamefont {B.}~\bibnamefont
  {Skinner}}\ and\ \bibinfo {author} {\bibfnamefont {M.~M.}\ \bibnamefont
  {Fogler}},\ }\href@noop {} {\bibfield  {journal} {\bibinfo  {journal} {Phys.
  Rev. B}\ }\textbf {\bibinfo {volume} {82}},\ \bibinfo {pages} {201306(R)}
  (\bibinfo {year} {2010})}\BibitemShut {NoStop}%
\bibitem [{\citenamefont {Fu}\ \emph {et~al.}(2015)\citenamefont {Fu},
  \citenamefont {Shklovskii},\ and\ \citenamefont {Skinner}}]{Fu-15}%
  \BibitemOpen
  \bibfield  {author} {\bibinfo {author} {\bibfnamefont {H.}~\bibnamefont
  {Fu}}, \bibinfo {author} {\bibfnamefont {B.~I.}\ \bibnamefont {Shklovskii}},
  \ and\ \bibinfo {author} {\bibfnamefont {B.}~\bibnamefont {Skinner}},\
  }\href@noop {} {\bibfield  {journal} {\bibinfo  {journal} {Phys. Rev. B}\
  }\textbf {\bibinfo {volume} {82}},\ \bibinfo {pages} {155118} (\bibinfo
  {year} {2015})}\BibitemShut {NoStop}%
\bibitem [{\citenamefont {Wu}\ \emph {et~al.}(2016)\citenamefont {Wu},
  \citenamefont {Chen}, \citenamefont {Wu}, \citenamefont {Xu}, \citenamefont
  {Han}, \citenamefont {Lin}, \citenamefont {Skinner}, \citenamefont {Cai},
  \citenamefont {He}, \citenamefont {Cheng},\ and\ \citenamefont
  {Wang}}]{Wu-16}%
  \BibitemOpen
  \bibfield  {author} {\bibinfo {author} {\bibfnamefont {Y.}~\bibnamefont
  {Wu}}, \bibinfo {author} {\bibfnamefont {X.}~\bibnamefont {Chen}}, \bibinfo
  {author} {\bibfnamefont {Z.}~\bibnamefont {Wu}}, \bibinfo {author}
  {\bibfnamefont {S.}~\bibnamefont {Xu}}, \bibinfo {author} {\bibfnamefont
  {T.}~\bibnamefont {Han}}, \bibinfo {author} {\bibfnamefont {J.}~\bibnamefont
  {Lin}}, \bibinfo {author} {\bibfnamefont {B.}~\bibnamefont {Skinner}},
  \bibinfo {author} {\bibfnamefont {Y.}~\bibnamefont {Cai}}, \bibinfo {author}
  {\bibfnamefont {Y.}~\bibnamefont {He}}, \bibinfo {author} {\bibfnamefont
  {C.}~\bibnamefont {Cheng}}, \ and\ \bibinfo {author} {\bibfnamefont
  {N.}~\bibnamefont {Wang}},\ }\href@noop {} {\bibfield  {journal} {\bibinfo
  {journal} {Phys. Rev. B}\ }\textbf {\bibinfo {volume} {93}},\ \bibinfo
  {pages} {035455} (\bibinfo {year} {2016})}\BibitemShut {NoStop}%
\bibitem [{\citenamefont {Go{\~n}i}\ \emph {et~al.}(2002)\citenamefont
  {Go{\~n}i}, \citenamefont {Haboeck}, \citenamefont {Thomsen}, \citenamefont
  {Eberl}, \citenamefont {Reboredo}, \citenamefont {Proetto},\ and\
  \citenamefont {Guinea}}]{Goni-02}%
  \BibitemOpen
  \bibfield  {author} {\bibinfo {author} {\bibfnamefont {A.~R.}\ \bibnamefont
  {Go{\~n}i}}, \bibinfo {author} {\bibfnamefont {U.}~\bibnamefont {Haboeck}},
  \bibinfo {author} {\bibfnamefont {C.}~\bibnamefont {Thomsen}}, \bibinfo
  {author} {\bibfnamefont {K.}~\bibnamefont {Eberl}}, \bibinfo {author}
  {\bibfnamefont {F.~A.}\ \bibnamefont {Reboredo}}, \bibinfo {author}
  {\bibfnamefont {C.~R.}\ \bibnamefont {Proetto}}, \ and\ \bibinfo {author}
  {\bibfnamefont {F.}~\bibnamefont {Guinea}},\ }\href@noop {} {\bibfield
  {journal} {\bibinfo  {journal} {Phys. Rev. B}\ }\textbf {\bibinfo {volume}
  {65}},\ \bibinfo {pages} {121313(R)} (\bibinfo {year} {2002})}\BibitemShut
  {NoStop}%
\bibitem [{\citenamefont {Yadav}\ \emph {et~al.}(2019)\citenamefont {Yadav},
  \citenamefont {Nguyen}, \citenamefont {Hong}, \citenamefont
  {Garc\'{i}a-Fern\'andez}, \citenamefont {Aguado-Puente}, \citenamefont
  {Nelson}, \citenamefont {Das}, \citenamefont {Prasad}, \citenamefont {Kwon},
  \citenamefont {Cheema}, \citenamefont {Khan}, \citenamefont {Hu},
  \citenamefont {\'I{\~n}iguez}, \citenamefont {Junquera}, \citenamefont
  {Chen}, \citenamefont {Muller}, \citenamefont {Ramesh},\ and\ \citenamefont
  {Salahuddin}}]{Yadav-19}%
  \BibitemOpen
  \bibfield  {author} {\bibinfo {author} {\bibfnamefont {A.~K.}\ \bibnamefont
  {Yadav}}, \bibinfo {author} {\bibfnamefont {K.~X.}\ \bibnamefont {Nguyen}},
  \bibinfo {author} {\bibfnamefont {Z.}~\bibnamefont {Hong}}, \bibinfo {author}
  {\bibfnamefont {P.}~\bibnamefont {Garc\'{i}a-Fern\'andez}}, \bibinfo {author}
  {\bibfnamefont {P.}~\bibnamefont {Aguado-Puente}}, \bibinfo {author}
  {\bibfnamefont {C.~T.}\ \bibnamefont {Nelson}}, \bibinfo {author}
  {\bibfnamefont {S.}~\bibnamefont {Das}}, \bibinfo {author} {\bibfnamefont
  {B.}~\bibnamefont {Prasad}}, \bibinfo {author} {\bibfnamefont
  {D.}~\bibnamefont {Kwon}}, \bibinfo {author} {\bibfnamefont {S.}~\bibnamefont
  {Cheema}}, \bibinfo {author} {\bibfnamefont {A.~I.}\ \bibnamefont {Khan}},
  \bibinfo {author} {\bibfnamefont {C.}~\bibnamefont {Hu}}, \bibinfo {author}
  {\bibfnamefont {J.}~\bibnamefont {\'I{\~n}iguez}}, \bibinfo {author}
  {\bibfnamefont {J.}~\bibnamefont {Junquera}}, \bibinfo {author}
  {\bibfnamefont {L.-Q.}\ \bibnamefont {Chen}}, \bibinfo {author}
  {\bibfnamefont {D.~A.}\ \bibnamefont {Muller}}, \bibinfo {author}
  {\bibfnamefont {R.}~\bibnamefont {Ramesh}}, \ and\ \bibinfo {author}
  {\bibfnamefont {S.}~\bibnamefont {Salahuddin}},\ }\href@noop {} {\bibfield
  {journal} {\bibinfo  {journal} {Nature (London)}\ }\textbf {\bibinfo {volume}
  {565}},\ \bibinfo {pages} {468} (\bibinfo {year} {2019})}\BibitemShut
  {NoStop}%
\bibitem [{\citenamefont {Dirac}(1930)}]{Dirac-30}%
  \BibitemOpen
  \bibfield  {author} {\bibinfo {author} {\bibfnamefont {P.~A.~M.}\
  \bibnamefont {Dirac}},\ }\href@noop {} {\bibfield  {journal} {\bibinfo
  {journal} {Proc. Cambridge Phil. Soc.}\ }\textbf {\bibinfo {volume} {26}},\
  \bibinfo {pages} {376} (\bibinfo {year} {1930})}\BibitemShut {NoStop}%
\bibitem [{\citenamefont {Slater}(1951)}]{Slater-51}%
  \BibitemOpen
  \bibfield  {author} {\bibinfo {author} {\bibfnamefont {J.~C.}\ \bibnamefont
  {Slater}},\ }\href@noop {} {\bibfield  {journal} {\bibinfo  {journal} {Phys.
  Rev.}\ }\textbf {\bibinfo {volume} {81}},\ \bibinfo {pages} {385} (\bibinfo
  {year} {1951})}\BibitemShut {NoStop}%
\end{thebibliography}
\end{document}